\pdfoutput=1
\documentclass[numberedappendix,letter]{emulateapj}
\newcommand{\be}{\begin{equation}}
\newcommand{\ee}{\end{equation}}
\newcommand{\ba}{\begin{eqnarray}}
\newcommand{\ea}{\end{eqnarray}}
\def\ltsima{$\; \buildrel < \over \sim \;$}
\def\ltsim{\lower.5ex\hbox{\ltsima}}
\def\gtsima{$\; \buildrel > \over \sim \;$}
\def\gtsim{\lower.5ex\hbox{\gtsima}}

\newcommand{\uk}  
           {\ensuremath{\mu K}}
\usepackage{graphicx}        
\usepackage{bm}              
\usepackage{natbib}
\usepackage{mathrsfs}

\newcommand{\act}    {ACT}                                                                                                         
\def\setsymbol#1#2{\expandafter\def\csname #1\endcsname{#2}}
\def\getsymbol#1{\csname #1\endcsname}

\def\actcosmo#1#2#3#4#5{\getsymbol{ACT:#1:#2:#3:#4:#5}}
\def\actcosmo#1#2#3#4#5{\getsymbol{ACT:#1:#2:#3:#4:#5}}

\setsymbol{ACT:symbol:rchisqfull}{\ensuremath{\chi^2/{\rm dof}}}
\setsymbol{ACT:symbol:rsdvz035}{\ensuremath{r_s(z_d)/D_v(z=0.35)}}
\setsymbol{ACT:symbol:Neff}{\ensuremath{N_{\rm eff}}}
\setsymbol{ACT:symbol:thetastardeg}{\ensuremath{\theta_{*}}}
\setsymbol{ACT:symbol:omegab}{\ensuremath{\Omega_b}}
\setsymbol{ACT:symbol:thetaA}{\ensuremath{\theta_A}}
\setsymbol{ACT:symbol:lA}{\ensuremath{\ell_A}}
\setsymbol{ACT:symbol:omeganuh2}{\ensuremath{\Omega_\nu h^2}}
\setsymbol{ACT:symbol:omegach2}{\ensuremath{\Omega_ch^2}}
\setsymbol{ACT:symbol:1mns}{\ensuremath{1-n_s}}
\setsymbol{ACT:symbol:sdssbias}{\ensuremath{b_{\rm SDSS}}}
\setsymbol{ACT:symbol:dndlnk}{\ensuremath{dn_s/d\ln{k}}}
\setsymbol{ACT:symbol:xe}{\ensuremath{x_e}}
\setsymbol{ACT:symbol:chisqfull}{\ensuremath{\chi^2}}
\setsymbol{ACT:symbol:rsdvz02}{\ensuremath{r_s(z_d)/D_v(z=0.2)}}
\setsymbol{ACT:symbol:bias}{\ensuremath{b}}
\setsymbol{ACT:symbol:sigma8}{\ensuremath{\sigma_8}}
\setsymbol{ACT:symbol:rchisqtt}{\ensuremath{\chi^2/{\rm dof}}}
\setsymbol{ACT:symbol:Aiso2}{\ensuremath{\alpha_{0}}}
\setsymbol{ACT:symbol:treion}{\ensuremath{t_{\rm reion}}}
\setsymbol{ACT:symbol:sigma0}{\ensuremath{\sigma_0}}
\setsymbol{ACT:symbol:omegak2sig}{\ensuremath{\Omega_k}}
\setsymbol{ACT:symbol:2dfbias}{\ensuremath{b_{\rm 2dF}}}
\setsymbol{ACT:symbol:2dfbeta}{\ensuremath{\beta_{\rm 2dF}}}
\setsymbol{ACT:symbol:dzdec}{\ensuremath{\Delta z_{\rm dec}}}
\setsymbol{ACT:symbol:omegam}{\ensuremath{\Omega_m}}
\setsymbol{ACT:symbol:omegak}{\ensuremath{\Omega_k}}
\setsymbol{ACT:symbol:zstar}{\ensuremath{z_{*}}}
\setsymbol{ACT:symbol:nb}{\ensuremath{n_b}}
\setsymbol{ACT:symbol:omegatot2sig}{\ensuremath{\Omega_{\rm tot}}}
\setsymbol{ACT:symbol:omegac}{\ensuremath{\Omega_c}}
\setsymbol{ACT:symbol:logAs}{\ensuremath{\log(10^{10} A_s)}}
\setsymbol{ACT:symbol:pr}{\ensuremath{p_r}}
\setsymbol{ACT:symbol:r}{\ensuremath{r}}
\setsymbol{ACT:symbol:zreion}{\ensuremath{z_{\rm reion}}}
\setsymbol{ACT:symbol:zdec}{\ensuremath{z_{\rm dec}}}
\setsymbol{ACT:symbol:chisqttdof}{\ensuremath{{\rm dof}}}
\setsymbol{ACT:symbol:sdssbeta}{\ensuremath{\beta_{\rm SDSS}}}
\setsymbol{ACT:symbol:tdec}{\ensuremath{t_{\rm dec}}}
\setsymbol{ACT:symbol:lstar}{\ensuremath{\ell_{*}}}
\setsymbol{ACT:symbol:Abaoz035}{\ensuremath{A_{\rm BAO}(z=0.35)}}
\setsymbol{ACT:symbol:omegabh2}{\ensuremath{\Omega_bh^2}}
\setsymbol{ACT:symbol:dtdec}{\ensuremath{\Delta t_{\rm dec}}}
\setsymbol{ACT:symbol:omegal}{\ensuremath{\Omega_\Lambda}}
\setsymbol{ACT:symbol:mnu}{\ensuremath{\sum m_\nu}}
\setsymbol{ACT:symbol:omegatot}{\ensuremath{\Omega_{\rm tot}}}
\setsymbol{ACT:symbol:rsstar}{\ensuremath{r_s(z_*)}}
\setsymbol{ACT:symbol:chisqfulldof}{\ensuremath{{\rm dof}}}
\setsymbol{ACT:symbol:zdrag}{\ensuremath{z_{d}}}
\setsymbol{ACT:symbol:zeq}{\ensuremath{z_{\rm eq}}}
\setsymbol{ACT:symbol:chisqtt}{\ensuremath{\chi^2}}
\setsymbol{ACT:symbol:beta}{\ensuremath{\beta}}
\setsymbol{ACT:symbol:dAeq}{\ensuremath{d_A(z_{\rm eq})}}
\setsymbol{ACT:symbol:thetastar}{\ensuremath{\theta_{*}}}
\setsymbol{ACT:symbol:100omegabh2}{\ensuremath{10^2\Omega_bh^2}}
\setsymbol{ACT:symbol:dw}{\ensuremath{w'}}
\setsymbol{ACT:symbol:shift}{\ensuremath{R}}
\setsymbol{ACT:symbol:c220}{\ensuremath{C_{220}}}
\setsymbol{ACT:symbol:keq}{\ensuremath{k_{\rm eq}}}
\setsymbol{ACT:symbol:baryonfrac}{\ensuremath{\Omega_b/\Omega_m}}
\setsymbol{ACT:symbol:omeganu}{\ensuremath{\Omega_\nu}}
\setsymbol{ACT:symbol:ns}{\ensuremath{n_s}}
\setsymbol{ACT:symbol:sig8omegam0.6}{\ensuremath{\sigma_8\Omega_m^{0.6}}}
\setsymbol{ACT:symbol:rhordec}{\ensuremath{r_{\rm hor}(z_{\rm dec})}}
\setsymbol{ACT:symbol:leq}{\ensuremath{\ell_{\rm eq}}}
\setsymbol{ACT:symbol:h}{\ensuremath{h}}
\setsymbol{ACT:symbol:H0}{\ensuremath{H_0}}
\setsymbol{ACT:symbol:sig8omegam0.5}{\ensuremath{\sigma_8\Omega_m^{0.5}}}
\setsymbol{ACT:symbol:Aiso1}{\ensuremath{\alpha_{-1}}}
\setsymbol{ACT:symbol:t0}{\ensuremath{t_0}}
\setsymbol{ACT:symbol:1pw}{\ensuremath{1+w}}
\setsymbol{ACT:symbol:dA}{\ensuremath{d_A}}
\setsymbol{ACT:symbol:tau}{\ensuremath{\tau}}
\setsymbol{ACT:symbol:tstar}{\ensuremath{t_{*}}}
\setsymbol{ACT:symbol:eta}{\ensuremath{\eta}}
\setsymbol{ACT:symbol:omegamh2}{\ensuremath{\Omega_mh^2}}
\setsymbol{ACT:symbol:w}{\ensuremath{w}}
\setsymbol{ACT:symbol:rsdrag}{\ensuremath{r_s(z_d)}}
\setsymbol{ACT:symbol:DeltaR2}{\ensuremath{\Delta_{\cal R}^2}}
\setsymbol{ACT:symbol:tsz}{\ensuremath{a_{\rm tSZ}}}
\setsymbol{ACT:symbol:ksz}{\ensuremath{a_{\rm kSZ}}}
\setsymbol{ACT:symbol:gamma}{\ensuremath{\Gamma}}
\setsymbol{ACT:symbol:dAstar}{\ensuremath{d_A(z_{*})}}
\setsymbol{ACT:symbol:yhe}{\ensuremath{Y_P}}
\setsymbol{ACT:symbol:ede}{\ensuremath{\Omega_{e}}}
\setsymbol{ACT:symbol:w0}{\ensuremath{w_{0}}}
\setsymbol{ACT:symbol:ad}{\ensuremath{a_d}}
\setsymbol{ACT:symbol:ac}{\ensuremath{a_c}}
\setsymbol{ACT:symbol:betac}{\ensuremath{\beta_c}}
\setsymbol{ACT:symbol:age}{\ensuremath{a_{gs}}}
\setsymbol{ACT:symbol:ags}{\ensuremath{a_{ge}}}
\setsymbol{ACT:symbol:xi}{\ensuremath{\xi}}
\setsymbol{ACT:symbol:ba}{\ensuremath{b_A}}
\setsymbol{ACT:symbol:bs}{\ensuremath{b_S}}
\setsymbol{ACT:symbol:cas1}{\ensuremath{y_{As1}}}
\setsymbol{ACT:symbol:cas2}{\ensuremath{y_{As2}}}
\setsymbol{ACT:symbol:cae1}{\ensuremath{y_{Ae1}}}
\setsymbol{ACT:symbol:cae2}{\ensuremath{y_{Ae2}}}
\setsymbol{ACT:symbol:95}{\ensuremath{c_{95}}}
\setsymbol{ACT:symbol:150}{\ensuremath{c_{150}}}
\setsymbol{ACT:symbol:220}{\ensuremath{c_{220}}}
\setsymbol{ACT:symbol:dalpha}{\ensuremath{d\alpha/\alpha_0}}
\setsymbol{ACT:symbol:gmu}{\ensuremath{G_\mu}}
\setsymbol{ACT:p1:tau:lcdm:wmap7+act:number}{\ensuremath{0.091\pm 0.015}}
\setsymbol{ACT:p1:omegach2:lcdm:wmap7+act:number}{\ensuremath{0.114\pm 0.005}}
\setsymbol{ACT:p1:thetaA:lcdm:wmap7+act:number}{\ensuremath{1.040 \pm 0.002}}
\setsymbol{ACT:p1:sigma8:lcdm:wmap7+act:number}{\ensuremath{0.83\pm 0.03}}
\setsymbol{ACT:p1:omegal:lcdm:wmap7+act:number}{\ensuremath{0.72\pm 0.03}}
\setsymbol{ACT:p1:H0:lcdm:wmap7+act:number}{\ensuremath{70.0\pm2.4}}
\setsymbol{ACT:p1:omegac:lcdm:wmap7+act:number}{\ensuremath{}}
\setsymbol{ACT:p1:ns:lcdm:wmap7+act:number}{\ensuremath{0.972\pm0.012}}
\setsymbol{ACT:p1:omegam:lcdm:wmap7+act:number}{\ensuremath{0.28\pm 0.03}}
\setsymbol{ACT:p1:100omegabh2:lcdm:wmap7+act:number}{\ensuremath{2.251\pm0.047}}
\setsymbol{ACT:p1:109DeltaR2:lcdm:wmap7+act:number}{\ensuremath{3.19\pm 0.04}}
\setsymbol{ACT:p1:tsz:lcdm:wmap7+act:number}{\ensuremath{3.4 \pm 1.4}}
\setsymbol{ACT:p1:ksz:lcdm:wmap7+act:number}{\ensuremath{<8.6}}
\setsymbol{ACT:p1:xi:lcdm:wmap7+act:number}{\ensuremath{<0.2}}
\setsymbol{ACT:p1:ad:lcdm:wmap7+act:number}{\ensuremath{7.0\pm0.5}}
\setsymbol{ACT:p1:ac:lcdm:wmap7+act:number}{\ensuremath{5.0\pm1.0}}
\setsymbol{ACT:p1:as:lcdm:wmap7+act:number}{\ensuremath{3.1\pm0.4}}
\setsymbol{ACT:p1:betac:lcdm:wmap7+act:number}{\ensuremath{2.2\pm 0.1}}
\setsymbol{ACT:p1:ags:lcdm:wmap7+act:number}{\ensuremath{0.4\pm 0.2}}
\setsymbol{ACT:p1:age:lcdm:wmap7+act:number}{\ensuremath{0.9\pm 0.2}}
\setsymbol{ACT:p1:cas1:lcdm:wmap7+act:number}{\ensuremath{1.01\pm 0.01}}
\setsymbol{ACT:p1:cas2:lcdm:wmap7+act:number}{\ensuremath{1.03\pm 0.02}}
\setsymbol{ACT:p1:cae1:lcdm:wmap7+act:number}{\ensuremath{1.00\pm 0.01}}
\setsymbol{ACT:p1:cae2:lcdm:wmap7+act:number}{\ensuremath{1.00\pm 0.01}}
\setsymbol{ACT:p1c220:lcdm:wmap7+act:number}{\ensuremath{}}

\setsymbol{ACT:p1:zreion:lcdm:wmap7+act:number}{\ensuremath{10.9\pm1.2}}

\setsymbol{ACT:p1:tau:lcdm+yhe:wmap7+act:number}{\ensuremath{0.089\pm0.015}}
\setsymbol{ACT:p1:omegabh2:lcdm+yhe:wmap7+act:number}{\ensuremath{0.0223\pm0.0005}}
\setsymbol{ACT:p1:omegach2:lcdm+yhe:wmap7+act:number}{\ensuremath{0.113\pm0.005}}
\setsymbol{ACT:p1:thetaA:lcdm+yhe:wmap7+act:number}{\ensuremath{1.039\pm 0.003}}
\setsymbol{ACT:p1:sigma8:lcdm+yhe:wmap7+act:number}{\ensuremath{0.81\pm0.03}}
\setsymbol{ACT:p1:omegal:lcdm+yhe:wmap7+act:number}{\ensuremath{0.72\pm0.03}}
\setsymbol{ACT:p1:H0:lcdm+yhe:wmap7+act:number}{\ensuremath{70.0\pm2.3}}
\setsymbol{ACT:p1:yhe:lcdm+yhe:wmap7+act:number}{\ensuremath{0.225\pm0.034}}
\setsymbol{ACT:p1:omegac:lcdm+yhe:wmap7+act:number}{\ensuremath{}}
\setsymbol{ACT:p1:ns:lcdm+yhe:wmap7+act:number}{\ensuremath{0.967\pm0.014}}
\setsymbol{ACT:p1:omegam:lcdm+yhe:wmap7+act:number}{\ensuremath{0.28\pm0.03}}
\setsymbol{ACT:p1:100omegabh2:lcdm+yhe:wmap7+act:number}{\ensuremath{2.236\pm0.052}}
\setsymbol{ACT:p1:zreion:lcdm+yhe:wmap7+act:number}{\ensuremath{}}
\setsymbol{ACT:p1:109DeltaR2:lcdm+yhe:wmap7+act:number}{\ensuremath{3.20\pm0.05}}
\setsymbol{ACT:p1:tsz:lcdm+yhe:wmap7+act:number}{\ensuremath{3.5\pm1.4}}
\setsymbol{ACT:p1:ksz:lcdm+yhe:wmap7+act:number}{\ensuremath{<8.2}}
\setsymbol{ACT:p1:xi:lcdm+yhe:wmap7+act:number}{\ensuremath{<0.2}}
\setsymbol{ACT:p1:ad:lcdm+yhe:wmap7+act:number}{\ensuremath{7.1\pm0.5}}
\setsymbol{ACT:p1:ac:lcdm+yhe:wmap7+act:number}{\ensuremath{5.0\pm1.0}}
\setsymbol{ACT:p1:as:lcdm+yhe:wmap7+act:number}{\ensuremath{3.1\pm0.4}}
\setsymbol{ACT:p1:betac:lcdm+yhe:wmap7+act:number}{\ensuremath{2.2\pm0.1}}
\setsymbol{ACT:p1:ags:lcdm+yhe:wmap7+act:number}{\ensuremath{0.4\pm0.2}}
\setsymbol{ACT:p1:age:lcdm+yhe:wmap7+act:number}{\ensuremath{0.9\pm0.2}}
\setsymbol{ACT:p1:cas1:lcdm+yhe:wmap7+act:number}{\ensuremath{1.01 \pm 0.01}}
\setsymbol{ACT:p1:cas2:lcdm+yhe:wmap7+act:number}{\ensuremath{1.03\pm 0.02}}
\setsymbol{ACT:p1:cae1:lcdm+yhe:wmap7+act:number}{\ensuremath{1.01 \pm 0.01}}
\setsymbol{ACT:p1:cae2:lcdm+yhe:wmap7+act:number}{\ensuremath{1.00\pm 0.02}}
\setsymbol{ACT:p1c220:+act:lcdm+yhe:wmap7+act:number}{\ensuremath{<0.93}}

\setsymbol{ACT:p1:tau:lcdm+run:wmap7+act:number}{\ensuremath{0.092\pm 0.017}}
\setsymbol{ACT:p1:omegabh2:lcdm+run:wmap7+act:number}{\ensuremath{0.0225\pm 0.0004}}
\setsymbol{ACT:p1:omegach2:lcdm+run:wmap7+act:number}{\ensuremath{0.115\pm 0.006}}
\setsymbol{ACT:p1:thetaA:lcdm+run:wmap7+act:number}{\ensuremath{1.040\pm 0.002}}
\setsymbol{ACT:p1:sigma8:lcdm+run:wmap7+act:number}{\ensuremath{0.83\pm 0.03}}
\setsymbol{ACT:p1:omegal:lcdm+run:wmap7+act:number}{\ensuremath{0.71\pm 0.03}}
\setsymbol{ACT:p1:H0:lcdm+run:wmap7+act:number}{\ensuremath{69.6\pm2.9}}
\setsymbol{ACT:p1:omegac:lcdm+run:wmap7+act:number}{\ensuremath{}}
\setsymbol{ACT:p1:ns:lcdm+run:wmap7+act:number}{\ensuremath{0.981\pm0.031}}
\setsymbol{ACT:p1:omegam:lcdm+run:wmap7+act:number}{\ensuremath{0.29\pm 0.04}}
\setsymbol{ACT:p1:100omegabh2:lcdm+run:wmap7+act:number}{\ensuremath{2.253\pm 0.051}}
\setsymbol{ACT:p1:zreion:lcdm+run:wmap7+act:number}{\ensuremath{}}
\setsymbol{ACT:p1:109DeltaR2:lcdm+run:wmap7+act:number}{\ensuremath{3.19\pm 0.05}}
\setsymbol{ACT:p1:dndlnk:lcdm+run:wmap7+act:number}{\ensuremath{-0.004\pm 0.012}}
\setsymbol{ACT:p1:tsz:lcdm+run:wmap7+act:number}{\ensuremath{3.3\pm 1.4}}
\setsymbol{ACT:p1:ksz:lcdm+run:wmap7+act:number}{\ensuremath{<8.4}}
\setsymbol{ACT:p1:xi:lcdm+run:wmap7+act:number}{\ensuremath{<0.2}}
\setsymbol{ACT:p1:ad:lcdm+run:wmap7+act:number}{\ensuremath{7.1\pm 0.5}}
\setsymbol{ACT:p1:ac:lcdm+run:wmap7+act:number}{\ensuremath{5.0\pm0.9}}
\setsymbol{ACT:p1:as:lcdm+run:wmap7+act:number}{\ensuremath{3.0\pm0.5}}
\setsymbol{ACT:p1:betac:lcdm+run:wmap7+act:number}{\ensuremath{2.2\pm0.1}}
\setsymbol{ACT:p1:ags:lcdm+run:wmap7+act:number}{\ensuremath{0.4 \pm 0.2}}
\setsymbol{ACT:p1:age:lcdm+run:wmap7+act:number}{\ensuremath{0.9\pm 0.2}}
\setsymbol{ACT:p1:cas1:lcdm+run:wmap7+act:number}{\ensuremath{1.01 \pm 0.01}}
\setsymbol{ACT:p1:cas2:lcdm+run:wmap7+act:number}{\ensuremath{1.03\pm 0.02}}
\setsymbol{ACT:p1:cae1:lcdm+run:wmap7+act:number}{\ensuremath{1.01 \pm 0.01}}
\setsymbol{ACT:p1:cae2:lcdm+run:wmap7+act:number}{\ensuremath{1.00\pm 0.01}}
\setsymbol{ACT:p1c220:+act:lcdm+run:wmap7+act:number}{\ensuremath{<0.93}}

\setsymbol{ACT:p1:tau:lcdm+tens:wmap7+act:number}{\ensuremath{0.092\pm 0.015}}
\setsymbol{ACT:p1:omegabh2:lcdm+tens:wmap7+act:number}{\ensuremath{0.0227\pm 0.0005}}
\setsymbol{ACT:p1:omegach2:lcdm+tens:wmap7+act:number}{\ensuremath{0.111\pm 0.006}}
\setsymbol{ACT:p1:thetaA:lcdm+tens:wmap7+act:number}{\ensuremath{1.040\pm  0.002}}
\setsymbol{ACT:p1:sigma8:lcdm+tens:wmap7+act:number}{\ensuremath{0.82\pm 0.03}}
\setsymbol{ACT:p1:omegal:lcdm+tens:wmap7+act:number}{\ensuremath{0.73\pm 0.03}}
\setsymbol{ACT:p1:H0:lcdm+tens:wmap7+act:number}{\ensuremath{71.1\pm3.1}}
\setsymbol{ACT:p1:omegac:lcdm+tens:wmap7+act:number}{\ensuremath{}}
\setsymbol{ACT:p1:ns:lcdm+tens:wmap7+act:number}{\ensuremath{0.980\pm0.017}}
\setsymbol{ACT:p1:omegam:lcdm+tens:wmap7+act:number}{\ensuremath{0.27\pm 0.03}}
\setsymbol{ACT:p1:100omegabh2:lcdm+tens:wmap7+act:number}{\ensuremath{2.248\pm 0.056}}
\setsymbol{ACT:p1:zreion:lcdm+tens:wmap7+act:number}{\ensuremath{}}
\setsymbol{ACT:p1:109DeltaR2:lcdm+tens:wmap7+act:number}{\ensuremath{3.16\pm 0.07}}
\setsymbol{ACT:p1:r:lcdm+tens:wmap7+act:number}{\ensuremath{<0.30}}
\setsymbol{ACT:p1:tsz:lcdm+tens:wmap7+act:number}{\ensuremath{3.4\pm 1.4}}
\setsymbol{ACT:p1:ksz:lcdm+tens:wmap7+act:number}{\ensuremath{<8.5}}
\setsymbol{ACT:p1:xi:lcdm+tens:wmap7+act:number}{\ensuremath{<0.2}}
\setsymbol{ACT:p1:ad:lcdm+tens:wmap7+act:number}{\ensuremath{7.0\pm 0.5}}
\setsymbol{ACT:p1:ac:lcdm+tens:wmap7+act:number}{\ensuremath{5.0 \pm 1.0}}
\setsymbol{ACT:p1:as:lcdm+tens:wmap7+act:number}{\ensuremath{3.1\pm0.4}}
\setsymbol{ACT:p1:betac:lcdm+tens:wmap7+act:number}{\ensuremath{2.2\pm 0.11}}
\setsymbol{ACT:p1:ags:lcdm+tens:wmap7+act:number}{\ensuremath{0.4\pm 0.2}}
\setsymbol{ACT:p1:age:lcdm+tens:wmap7+act:number}{\ensuremath{0.9\pm 0.2}}
\setsymbol{ACT:p1:cas1:lcdm+tens:wmap7+act:number}{\ensuremath{1.01 \pm 0.01}}
\setsymbol{ACT:p1:cas2:lcdm+tens:wmap7+act:number}{\ensuremath{1.03\pm 0.02}}
\setsymbol{ACT:p1:cae1:lcdm+tens:wmap7+act:number}{\ensuremath{1.01 \pm 0.01}}
\setsymbol{ACT:p1:cae2:lcdm+tens:wmap7+act:number}{\ensuremath{1.00 \pm 0.01}}
\setsymbol{ACT:p1c220:+act:lcdm+tens:wmap7+act:number}{\ensuremath{<0.93}}

\setsymbol{ACT:p1:tau:lcdm+nrel:wmap7+act:number}{\ensuremath{0.091\pm 0.015}}
\setsymbol{ACT:p1:omegabh2:lcdm+nrel:wmap7+act:number}{\ensuremath{0.0224\pm 0.0005}}
\setsymbol{ACT:p1:thetaA:lcdm+nrel:wmap7+act:number}{\ensuremath{1.041\pm  0.003}}
\setsymbol{ACT:p1:tau:lcdm+nrel:wmap7+act:number}{\ensuremath{0.091\pm 0.015}}
\setsymbol{ACT:p1:omegach2:lcdm+nrel:wmap7+act:number}{\ensuremath{0.109\pm 0.011}}
\setsymbol{ACT:p1:sigma8:lcdm+nrel:wmap7+act:number}{\ensuremath{0.81\pm 0.05}}
\setsymbol{ACT:p1:omegal:lcdm+nrel:wmap7+act:number}{\ensuremath{0.72\pm 0.03}}
\setsymbol{ACT:p1:H0:lcdm+nrel:wmap7+act:number}{\ensuremath{68.8\pm3.5}}
\setsymbol{ACT:p1:omegac:lcdm+nrel:wmap7+act:number}{\ensuremath{}}
\setsymbol{ACT:p1:ns:lcdm+nrel:wmap7+act:number}{\ensuremath{0.965\pm0.020}}
\setsymbol{ACT:p1:omegam:lcdm+nrel:wmap7+act:number}{\ensuremath{0.28\pm 0.03}}
\setsymbol{ACT:p1:100omegabh2:lcdm+nrel:wmap7+act:number}{\ensuremath{2.237\pm 0.054}}
\setsymbol{ACT:p1:zreion:lcdm+nrel:wmap7+act:number}{\ensuremath{}}
\setsymbol{ACT:p1:109DeltaR2:lcdm+nrel:wmap7+act:number}{\ensuremath{3.20\pm 0.05}}
\setsymbol{ACT:p1:Neff:lcdm+nrel:wmap7+act:number}{\ensuremath{2.79\pm 0.56}}
\setsymbol{ACT:p1:tsz:lcdm+nrel:wmap7+act:number}{\ensuremath{3.4 \pm 1.4}}
\setsymbol{ACT:p1:ksz:lcdm+nrel:wmap7+act:number}{\ensuremath{<8.4}}
\setsymbol{ACT:p1:xi:lcdm+nrel:wmap7+act:number}{\ensuremath{<0.2}}
\setsymbol{ACT:p1:as:lcdm+nrel:wmap7+act:number}{\ensuremath{3.1\pm0.4}}
\setsymbol{ACT:p1:ad:lcdm+nrel:wmap7+act:number}{\ensuremath{7.0\pm 0.5}}
\setsymbol{ACT:p1:ac:lcdm+nrel:wmap7+act:number}{\ensuremath{5.0\pm 1.0}}
\setsymbol{ACT:p1:betac:lcdm+nrel:wmap7+act:number}{\ensuremath{2.2\pm 0.1}}
\setsymbol{ACT:p1:cas1:lcdm+nrel:wmap7+act:number}{\ensuremath{1.01 \pm 0.01}}
\setsymbol{ACT:p1:cas2:lcdm+nrel:wmap7+act:number}{\ensuremath{1.03\pm 0.02}}
\setsymbol{ACT:p1:cae1:lcdm+nrel:wmap7+act:number}{\ensuremath{1.01 \pm 0.01}}
\setsymbol{ACT:p1:ags:lcdm+nrel:wmap7+act:number}{\ensuremath{0.4 \pm 0.2}}
\setsymbol{ACT:p1:age:lcdm+nrel:wmap7+act:number}{\ensuremath{0.9 \pm 0.2}}
\setsymbol{ACT:p1:cae2:lcdm+nrel:wmap7+act:number}{\ensuremath{1.00 \pm 0.01}}
\setsymbol{ACT:p1c220:+act:lcdm+nrel:wmap7+act:number}{\ensuremath{<0.93}}

\setsymbol{ACT:p1:tau:lcdm+tens:wmap7+act+bao+h0:number}{\ensuremath{0.089\pm 0.014}}
\setsymbol{ACT:p1:omegabh2:lcdm+tens:wmap7+act+bao+h0:number}{\ensuremath{0.0226\pm 0.0004}}
\setsymbol{ACT:p1:omegach2:lcdm+tens:wmap7+act+bao+h0:number}{\ensuremath{0.115\pm 0.003}}
\setsymbol{ACT:p1:thetaA:lcdm+tens:wmap7+act+bao+h0:number}{\ensuremath{1.040\pm  0.002}}
\setsymbol{ACT:p1:sigma8:lcdm+tens:wmap7+act+bao+h0:number}{\ensuremath{0.83\pm 0.02}}
\setsymbol{ACT:p1:omegal:lcdm+tens:wmap7+act+bao+h0:number}{\ensuremath{0.71\pm 0.01}}
\setsymbol{ACT:p1:H0:lcdm+tens:wmap7+act+bao+h0:number}{\ensuremath{69.3\pm1.0}}
\setsymbol{ACT:p1:omegac:lcdm+tens:wmap7+act+bao+h0:number}{\ensuremath{}}
\setsymbol{ACT:p1:ns:lcdm+tens:wmap7+act+bao+h0:number}{\ensuremath{0.972\pm0.010}}
\setsymbol{ACT:p1:omegam:lcdm+tens:wmap7+act+bao+h0:number}{\ensuremath{0.29\pm 0.01}}
\setsymbol{ACT:p1:100omegabh2:lcdm+tens:wmap7+act+bao+h0:number}{\ensuremath{2.256\pm 0.040}}
\setsymbol{ACT:p1:zreion:lcdm+tens:wmap7+act+bao+h0:number}{\ensuremath{}}
\setsymbol{ACT:p1:109DeltaR2:lcdm+tens:wmap7+act+bao+h0:number}{\ensuremath{3.19\pm 0.04}}
\setsymbol{ACT:p1:r:lcdm+tens:wmap7+act+bao+h0:number}{\ensuremath{<0.16}}
\setsymbol{ACT:p1:tsz:lcdm+tens:wmap7+act+bao+h0:number}{\ensuremath{3.4 \pm1.4}}
\setsymbol{ACT:p1:ksz:lcdm+tens:wmap7+act+bao+h0:number}{\ensuremath{<8.2}}
\setsymbol{ACT:p1:xi:lcdm+tens:wmap7+act+bao+h0:number}{\ensuremath{<0.2}}
\setsymbol{ACT:p1:ad:lcdm+tens:wmap7+act+bao+h0:number}{\ensuremath{7.1\pm 0.5}}
\setsymbol{ACT:p1:ac:lcdm+tens:wmap7+act+bao+h0:number}{\ensuremath{5.0 \pm 1.0}}
\setsymbol{ACT:p1:as:lcdm+tens:wmap7+act+bao+h0:number}{\ensuremath{3.1\pm0.4}}
\setsymbol{ACT:p1:betac:lcdm+tens:wmap7+act+bao+h0:number}{\ensuremath{2.2\pm 0.1}}
\setsymbol{ACT:p1:ags:lcdm+tens:wmap7+act+bao+h0:number}{\ensuremath{0.4 \pm 0.2}}
\setsymbol{ACT:p1:age:lcdm+tens:wmap7+act+bao+h0:number}{\ensuremath{0.9\pm 0.2}}
\setsymbol{ACT:p1:cas1:lcdm+tens:wmap7+act+bao+h0:number}{\ensuremath{1.01 \pm 0.01}}
\setsymbol{ACT:p1:cas2:lcdm+tens:wmap7+act+bao+h0:number}{\ensuremath{1.03\pm 0.02}}
\setsymbol{ACT:p1:cae1:lcdm+tens:wmap7+act+bao+h0:number}{\ensuremath{1.01 \pm 0.01}}
\setsymbol{ACT:p1:cae2:lcdm+tens:wmap7+act+bao+h0:number}{\ensuremath{1.00 \pm 0.01}}
\setsymbol{ACT:p1c220:+act:lcdm+tens:wmap7+act+bao+h0:number}{\ensuremath{<0.93}}

\setsymbol{ACT:p1:tau:lcdm+nrel:wmap7+act+bao+h0:number}{\ensuremath{0.092\pm 0.015}}
\setsymbol{ACT:p1:thetaA:lcdm+nrel:wmap7+act+bao+h0:number}{\ensuremath{1.038\pm  0.002}}
\setsymbol{ACT:p1:omegach2:lcdm+nrel:wmap7+act+bao+h0:number}{\ensuremath{0.125\pm 0.008}}
\setsymbol{ACT:p1:sigma8:lcdm+nrel:wmap7+act+bao+h0:number}{\ensuremath{0.86\pm 0.03}}
\setsymbol{ACT:p1:omegal:lcdm+nrel:wmap7+act+bao+h0:number}{\ensuremath{0.71\pm 0.01}}
\setsymbol{ACT:p1:H0:lcdm+nrel:wmap7+act+bao+h0:number}{\ensuremath{71.2\pm2.0}}
\setsymbol{ACT:p1:ns:lcdm+nrel:wmap7+act+bao+h0:number}{\ensuremath{0.982\pm0.014}}
\setsymbol{ACT:p1:omegam:lcdm+nrel:wmap7+act+bao+h0:number}{\ensuremath{0.29\pm 0.01}}
\setsymbol{ACT:p1:100omegabh2:lcdm+nrel:wmap7+act+bao+h0:number}{\ensuremath{2.271\pm 0.046}}
\setsymbol{ACT:p1:zreion:lcdm+nrel:wmap7+act+bao+h0:number}{\ensuremath{}}
\setsymbol{ACT:p1:109DeltaR2:lcdm+nrel:wmap7+act+bao+h0:number}{\ensuremath{3.19\pm 0.03}}
\setsymbol{ACT:p1:Neff:lcdm+nrel:wmap7+act+bao+h0:number}{\ensuremath{3.50\pm 0.42}}
\setsymbol{ACT:p1:tsz:lcdm+nrel:wmap7+act+bao+h0:number}{\ensuremath{3.4\pm1.4}}
\setsymbol{ACT:p1:ksz:lcdm+nrel:wmap7+act+bao+h0:number}{\ensuremath{<8.8}}
\setsymbol{ACT:p1:xi:lcdm+nrel:wmap7+act+bao+h0:number}{\ensuremath{<0.2}}
\setsymbol{ACT:p1:as:lcdm+nrel:wmap7+act+bao+h0:number}{\ensuremath{3.0\pm0.4}}
\setsymbol{ACT:p1:ad:lcdm+nrel:wmap7+act+bao+h0:number}{\ensuremath{7.0\pm 0.5}}
\setsymbol{ACT:p1:ac:lcdm+nrel:wmap7+act+bao+h0:number}{\ensuremath{4.8\pm 1.0}}
\setsymbol{ACT:p1:betac:lcdm+nrel:wmap7+act+bao+h0:number}{\ensuremath{2.2\pm 0.1}}
\setsymbol{ACT:p1:cas1:lcdm+nrel:wmap7+act+bao+h0:number}{\ensuremath{1.01 \pm 0.01}}
\setsymbol{ACT:p1:cas2:lcdm+nrel:wmap7+act+bao+h0:number}{\ensuremath{1.03\pm 0.02}}
\setsymbol{ACT:p1:cae1:lcdm+nrel:wmap7+act+bao+h0:number}{\ensuremath{1.01 \pm 0.01}}
\setsymbol{ACT:p1:ags:lcdm+nrel:wmap7+act+bao+h0:number}{\ensuremath{0.4 \pm 0.2}}
\setsymbol{ACT:p1:age:lcdm+nrel:wmap7+act+bao+h0:number}{\ensuremath{0.9 \pm 0.2}}
\setsymbol{ACT:p1:cae2:lcdm+nrel:wmap7+act+bao+h0:number}{\ensuremath{1.00 \pm 0.01}}
\setsymbol{ACT:p1c220:+act:lcdm+nrel:wmap7+act+bao+h0:number}{\ensuremath{<0.93}}

\setsymbol{ACT:p1:tau:lcdm+mnu:wmap7+act:number}{\ensuremath{0.915 \pm 0.015}}
\setsymbol{ACT:p1:omegabh2:lcdm+mnu:wmap7+act:number}{\ensuremath{0.0224\pm 0.00045}}
\setsymbol{ACT:p1:omegach2:lcdm+mnu:wmap7+act:number}{\ensuremath{0.118\pm 0.006}}
\setsymbol{ACT:p1:thetaA:lcdm+mnu:wmap7+act:number}{\ensuremath{1.040\pm  0.002}}
\setsymbol{ACT:p1:sigma8:lcdm+mnu:wmap7+act:number}{\ensuremath{0.77\pm 0.05}}
\setsymbol{ACT:p1:omegal:lcdm+mnu:wmap7+act:number}{\ensuremath{0.69\pm 0.04}}
\setsymbol{ACT:p1:H0:lcdm+mnu:wmap7+act:number}{\ensuremath{67.2\pm3.1}}
\setsymbol{ACT:p1:omegac:lcdm+mnu:wmap7+act:number}{\ensuremath{}}
\setsymbol{ACT:p1:ns:lcdm+mnu:wmap7+act:number}{\ensuremath{0.970 \pm 0.012}}
\setsymbol{ACT:p1:omegam:lcdm+mnu:wmap7+act:number}{\ensuremath{0.32\pm 0.05}}
\setsymbol{ACT:p1:100omegabh2:lcdm+mnu:wmap7+act:number}{\ensuremath{2.242\pm 0.045}}
\setsymbol{ACT:p1:zreion:lcdm+mnu:wmap7+act:number}{\ensuremath{}}
\setsymbol{ACT:p1:109DeltaR2:lcdm+mnu:wmap7+act:number}{\ensuremath{3.20\pm 0.05}}
\setsymbol{ACT:p1:as:lcdm+mnu:wmap7+act:number}{\ensuremath{3.1\pm0.4}}
\setsymbol{ACT:p1:mnu:lcdm+mnu:wmap7+act:number}{\ensuremath{<0.70}}
\setsymbol{ACT:p1:tsz:lcdm+mnu:wmap7+act:number}{\ensuremath{3.4\pm 1.3}}
\setsymbol{ACT:p1:ksz:lcdm+mnu:wmap7+act:number}{\ensuremath{<8.5}}
\setsymbol{ACT:p1:xi:lcdm+mnu:wmap7+act:number}{\ensuremath{<0.2}}
\setsymbol{ACT:p1:ad:lcdm+mnu:wmap7+act:number}{\ensuremath{7.0\pm 0.5}}
\setsymbol{ACT:p1:ac:lcdm+mnu:wmap7+act:number}{\ensuremath{5.0\pm 1.0}}
\setsymbol{ACT:p1:betac:lcdm+mnu:wmap7+act:number}{\ensuremath{2.2\pm 0.1}}
\setsymbol{ACT:p1:cas1:lcdm+mnu:wmap7+act:number}{\ensuremath{1.01 \pm 0.01}}
\setsymbol{ACT:p1:cas2:lcdm+mnu:wmap7+act:number}{\ensuremath{1.03\pm 0.02}}
\setsymbol{ACT:p1:cae1:lcdm+mnu:wmap7+act:number}{\ensuremath{1.01 \pm 0.01}}
\setsymbol{ACT:p1:cae2:lcdm+mnu:wmap7+act:number}{\ensuremath{1.00 \pm 0.01}}
\setsymbol{ACT:p1:ags:lcdm+mnu:wmap7+act:number}{\ensuremath{0.4\pm 0.2}}
\setsymbol{ACT:p1:age:lcdm+mnu:wmap7+act:number}{\ensuremath{0.9\pm0.2}}

\setsymbol{ACT:p1:tau:lcdm+mnu:wmap7+act+bao+h0:number}{\ensuremath{0.091 \pm 0.014}}
\setsymbol{ACT:p1:omegabh2:lcdm+mnu:wmap7+act+bao+h0:number}{\ensuremath{0.0226\pm 0.00040}}
\setsymbol{ACT:p1:omegach2:lcdm+mnu:wmap7+act+bao+h0:number}{\ensuremath{0.115\pm 0.002}}
\setsymbol{ACT:p1:thetaA:lcdm+mnu:wmap7+act+bao+h0:number}{\ensuremath{1.040\pm  0.002}}
\setsymbol{ACT:p1:sigma8:lcdm+mnu:wmap7+act+bao+h0:number}{\ensuremath{0.79\pm 0.04}}
\setsymbol{ACT:p1:omegal:lcdm+mnu:wmap7+act+bao+h0:number}{\ensuremath{0.71\pm 0.01}}
\setsymbol{ACT:p1:H0:lcdm+mnu:wmap7+act+bao+h0:number}{\ensuremath{68.8\pm1.0}}
\setsymbol{ACT:p1:omegac:lcdm+mnu:wmap7+act+bao+h0:number}{\ensuremath{}}
\setsymbol{ACT:p1:ns:lcdm+mnu:wmap7+act+bao+h0:number}{\ensuremath{0.974 \pm 0.010}}
\setsymbol{ACT:p1:omegam:lcdm+mnu:wmap7+act+bao+h0:number}{\ensuremath{0.29\pm 0.01}}
\setsymbol{ACT:p1:100omegabh2:lcdm+mnu:wmap7+act+bao+h0:number}{\ensuremath{2.259\pm 0.039}}
\setsymbol{ACT:p1:zreion:lcdm+mnu:wmap7+act+bao+h0:number}{\ensuremath{}}
\setsymbol{ACT:p1:109DeltaR2:lcdm+mnu:wmap7+act+bao+h0:number}{\ensuremath{3.19\pm 0.03}}
\setsymbol{ACT:p1:as:lcdm+mnu:wmap7+act+bao+h0:number}{\ensuremath{3.1\pm0.4}}
\setsymbol{ACT:p1:mnu:lcdm+mnu:wmap7+act+bao+h0:number}{\ensuremath{<0.40}}
\setsymbol{ACT:p1:tsz:lcdm+mnu:wmap7+act+bao+h0:number}{\ensuremath{3.4\pm 1.4}}
\setsymbol{ACT:p1:ksz:lcdm+mnu:wmap7+act+bao+h0:number}{\ensuremath{<8.5}}
\setsymbol{ACT:p1:xi:lcdm+mnu:wmap7+act+bao+h0:number}{\ensuremath{<0.2}}
\setsymbol{ACT:p1:ad:lcdm+mnu:wmap7+act+bao+h0:number}{\ensuremath{7.0\pm 0.5}}
\setsymbol{ACT:p1:ac:lcdm+mnu:wmap7+act+bao+h0:number}{\ensuremath{5.0\pm 1.0}}
\setsymbol{ACT:p1:betac:lcdm+mnu:wmap7+act+bao+h0:number}{\ensuremath{2.2\pm 0.1}}
\setsymbol{ACT:p1:cas1:lcdm+mnu:wmap7+act+bao+h0:number}{\ensuremath{1.01 \pm 0.01}}
\setsymbol{ACT:p1:cas2:lcdm+mnu:wmap7+act+bao+h0:number}{\ensuremath{1.03\pm 0.02}}
\setsymbol{ACT:p1:cae1:lcdm+mnu:wmap7+act+bao+h0:number}{\ensuremath{1.01 \pm 0.01}}
\setsymbol{ACT:p1:cae2:lcdm+mnu:wmap7+act+bao+h0:number}{\ensuremath{1.00 \pm 0.01}}
\setsymbol{ACT:p1:ags:lcdm+mnu:wmap7+act+bao+h0:number}{\ensuremath{0.4\pm 0.2}}
\setsymbol{ACT:p1:age:lcdm+mnu:wmap7+act+bao+h0:number}{\ensuremath{0.9\pm0.2}}

\setsymbol{ACT:p1:tau:lcdm+ede:wmap7+act:number}{\ensuremath{0.090 \pm 0.015}}
\setsymbol{ACT:p1:omegabh2:lcdm+ede:wmap7+act:number}{\ensuremath{0.0226\pm 0.00046}}
\setsymbol{ACT:p1:omegach2:lcdm+ede:wmap7+act:number}{\ensuremath{0.117\pm 0.006}}
\setsymbol{ACT:p1:thetaA:lcdm+ede:wmap7+act:number}{\ensuremath{1.059\pm  0.002}}
\setsymbol{ACT:p1:sigma8:lcdm+ede:wmap7+act:number}{\ensuremath{0.72\pm 0.07}}
\setsymbol{ACT:p1:omegal:lcdm+ede:wmap7+act:number}{\ensuremath{0.62\pm 0.07}}
\setsymbol{ACT:p1:H0:lcdm+ede:wmap7+act:number}{\ensuremath{61.5\pm5.0}}
\setsymbol{ACT:p1:omegac:lcdm+ede:wmap7+act:number}{\ensuremath{}}
\setsymbol{ACT:p1:ns:lcdm+ede:wmap7+act:number}{\ensuremath{0.977 \pm 0.012}}
\setsymbol{ACT:p1:omegam:lcdm+ede:wmap7+act:number}{\ensuremath{0.38\pm 0.07}}
\setsymbol{ACT:p1:100omegabh2:lcdm+ede:wmap7+act:number}{\ensuremath{2.257\pm 0.046}}
\setsymbol{ACT:p1:zreion:lcdm+ede:wmap7+act:number}{\ensuremath{}}
\setsymbol{ACT:p1:109DeltaR2:lcdm+ede:wmap7+act:number}{\ensuremath{3.18\pm 0.05}}
\setsymbol{ACT:p1:ede:lcdm+ede:wmap7+act:number}{\ensuremath{<0.03}}
\setsymbol{ACT:p1:w0:lcdm+ede:wmap7+act:number}{\ensuremath{<-0.45}}
\setsymbol{ACT:p1:tsz:lcdm+ede:wmap7+act:number}{\ensuremath{3.5\pm 1.4}}
\setsymbol{ACT:p1:ksz:lcdm+ede:wmap7+act:number}{\ensuremath{<8.4}}
\setsymbol{ACT:p1:xi:lcdm+ede:wmap7+act:number}{\ensuremath{<0.2}}
\setsymbol{ACT:p1:ad:lcdm+ede:wmap7+act:number}{\ensuremath{7.2\pm 0.5}}
\setsymbol{ACT:p1:as:lcdm+ede:wmap7+act:number}{\ensuremath{3.1\pm0.4}}
\setsymbol{ACT:p1:ac:lcdm+ede:wmap7+act:number}{\ensuremath{5.2\pm 1.0}}
\setsymbol{ACT:p1:betac:lcdm+ede:wmap7+act:number}{\ensuremath{2.2\pm 0.1}}
\setsymbol{ACT:p1:ba:lcdm+ede:wmap7+act:number}{\ensuremath{}}
\setsymbol{ACT:p1:bs:lcdm+ede:wmap7+act:number}{\ensuremath{<0.93}}
\setsymbol{ACT:p1:cas1:lcdm+ede:wmap7+act:number}{\ensuremath{1.02 \pm 0.01}}
\setsymbol{ACT:p1:cas2:lcdm+ede:wmap7+act:number}{\ensuremath{1.04\pm 0.02}}
\setsymbol{ACT:p1:cae1:lcdm+ede:wmap7+act:number}{\ensuremath{1.02 \pm 0.01}}
\setsymbol{ACT:p1:cae2:lcdm+ede:wmap7+act:number}{\ensuremath{1.00 \pm 0.02}}
\setsymbol{ACT:p1:ags:lcdm+ede:wmap7+act:number}{\ensuremath{0.4\pm0.2}}
\setsymbol{ACT:p1:age:lcdm+ede:wmap7+act:number}{\ensuremath{0.9\pm0.2}}
\setsymbol{ACT:p1c220:+act:lcdm+ede:wmap7+act:number}{\ensuremath{<0.93}}

\setsymbol{ACT:p1:tau:lcdm+omk:wmap7+act:number}{\ensuremath{0.090\pm 0.015}}
\setsymbol{ACT:p1:omegabh2:lcdm+omk:wmap7+act:number}{\ensuremath{0.0225\pm 0.00044}}
\setsymbol{ACT:p1:omegach2:lcdm+omk:wmap7+act:number}{\ensuremath{0.113\pm0.047}}
\setsymbol{ACT:p1:thetaA:lcdm+omk:wmap7+act:number}{\ensuremath{1.040\pm  0.002}}
\setsymbol{ACT:p1:sigma8:lcdm+omk:wmap7+act:number}{\ensuremath{0.81\pm0.03}}
\setsymbol{ACT:p1:omegal:lcdm+omk:wmap7+act:number}{\ensuremath{0.63\pm0.08}}
\setsymbol{ACT:p1:H0:lcdm+omk:wmap7+act:number}{\ensuremath{59.8\pm8.0}}
\setsymbol{ACT:p1:omegac:lcdm+omk:wmap7+act:number}{\ensuremath{}}
\setsymbol{ACT:p1:ns:lcdm+omk:wmap7+act:number}{\ensuremath{0.970\pm0.011}}
\setsymbol{ACT:p1:omegam:lcdm+omk:wmap7+act:number}{\ensuremath{0.40\pm0.10}}
\setsymbol{ACT:p1:100omegabh2:lcdm+omk:wmap7+act:number}{\ensuremath{2.246\pm0.044}}
\setsymbol{ACT:p1:zreion:lcdm+omk:wmap7+act:number}{\ensuremath{}}
\setsymbol{ACT:p1:109DeltaR2:lcdm+omk:wmap7+act:number}{\ensuremath{3.20\pm0.04}}
\setsymbol{ACT:p1:omk:lcdm+omk:wmap7+act:number}{\ensuremath{-0.031\pm0.026}}
\setsymbol{ACT:p1:tsz:lcdm+omk:wmap7+act:number}{\ensuremath{3.4\pm1.4}}
\setsymbol{ACT:p1:ksz:lcdm+omk:wmap7+act:number}{\ensuremath{<8.1}}
\setsymbol{ACT:p1:xi:lcdm+omk:wmap7+act:number}{\ensuremath{<0.2}}
\setsymbol{ACT:p1:as:lcdm+omk:wmap7+act:number}{\ensuremath{3.1\pm0.4}}
\setsymbol{ACT:p1:ad:lcdm+omk:wmap7+act:number}{\ensuremath{7.0\pm0.5}}
\setsymbol{ACT:p1:ac:lcdm+omk:wmap7+act:number}{\ensuremath{4.9\pm0.9}}
\setsymbol{ACT:p1:betac:lcdm+omk:wmap7+act:number}{\ensuremath{2.2\pm0.1}}
\setsymbol{ACT:p1:cas1:lcdm+omk:wmap7+act:number}{\ensuremath{1.01\pm0.01}}
\setsymbol{ACT:p1:cas2:lcdm+omk:wmap7+act:number}{\ensuremath{1.03\pm0.02}}
\setsymbol{ACT:p1:cae1:lcdm+omk:wmap7+act:number}{\ensuremath{1.01\pm0.01}}
\setsymbol{ACT:p1:ags:lcdm+omk:wmap7+act:number}{\ensuremath{0.4\pm0.2}}
\setsymbol{ACT:p1:age:lcdm+omk:wmap7+act:number}{\ensuremath{0.9\pm0.2}}
\setsymbol{ACT:p1:cae2:lcdm+omk:wmap7+act:number}{\ensuremath{1.00\pm0.01}}
\setsymbol{ACT:p1c220:lcdm+omk:wmap7+act:number}{\ensuremath{<0.93}}

\setsymbol{ACT:p1:tau:lcdm+omk:wmap7+act+bao+h0:number}{\ensuremath{0.092\pm 0.015}}
\setsymbol{ACT:p1:omegabh2:lcdm+omk:wmap7+act+bao+h0:number}{\ensuremath{0.0226\pm 0.00043}}
\setsymbol{ACT:p1:omegach2:lcdm+omk:wmap7+act+bao+h0:number}{\ensuremath{0.114\pm0.005}}
\setsymbol{ACT:p1:thetaA:lcdm+omk:wmap7+act+bao+h0:number}{\ensuremath{1.040\pm  0.002}}
\setsymbol{ACT:p1:sigma8:lcdm+omk:wmap7+act+bao+h0:number}{\ensuremath{0.83\pm0.03}}
\setsymbol{ACT:p1:omegal:lcdm+omk:wmap7+act+bao+h0:number}{\ensuremath{0.72\pm0.01}}
\setsymbol{ACT:p1:H0:lcdm+omk:wmap7+act+bao+h0:number}{\ensuremath{69.0\pm1.1}}
\setsymbol{ACT:p1:omegac:lcdm+omk:wmap7+act+bao+h0:number}{\ensuremath{}}
\setsymbol{ACT:p1:ns:lcdm+omk:wmap7+act+bao+h0:number}{\ensuremath{0.974\pm0.011}}
\setsymbol{ACT:p1:omegam:lcdm+omk:wmap7+act+bao+h0:number}{\ensuremath{0.29\pm0.01}}
\setsymbol{ACT:p1:100omegabh2:lcdm+omk:wmap7+act+bao+h0:number}{\ensuremath{2.261\pm0.044}}
\setsymbol{ACT:p1:zreion:lcdm+omk:wmap7+act+bao+h0:number}{\ensuremath{}}
\setsymbol{ACT:p1:109DeltaR2:lcdm+omk:wmap7+act+bao+h0:number}{\ensuremath{3.19\pm0.04}}
\setsymbol{ACT:p1:omk:lcdm+omk:wmap7+act+bao+h0:number}{\ensuremath{-0.0020\pm0.0047}}
\setsymbol{ACT:p1:tsz:lcdm+omk:wmap7+act+bao+h0:number}{\ensuremath{3.4\pm1.5}}
\setsymbol{ACT:p1:ksz:lcdm+omk:wmap7+act+bao+h0:number}{\ensuremath{<8.5}}
\setsymbol{ACT:p1:xi:lcdm+omk:wmap7+act+bao+h0:number}{\ensuremath{<0.2}}
\setsymbol{ACT:p1:as:lcdm+omk:wmap7+act+bao+h0:number}{\ensuremath{3.1\pm0.4}}
\setsymbol{ACT:p1:ad:lcdm+omk:wmap7+act+bao+h0:number}{\ensuremath{7.0\pm0.6}}
\setsymbol{ACT:p1:ac:lcdm+omk:wmap7+act+bao+h0:number}{\ensuremath{5.0\pm1.0}}
\setsymbol{ACT:p1:betac:lcdm+omk:wmap7+act+bao+h0:number}{\ensuremath{2.2\pm0.1}}
\setsymbol{ACT:p1:cas1:lcdm+omk:wmap7+act+bao+h0:number}{\ensuremath{1.01\pm0.01}}
\setsymbol{ACT:p1:cas2:lcdm+omk:wmap7+act+bao+h0:number}{\ensuremath{1.03\pm0.02}}
\setsymbol{ACT:p1:cae1:lcdm+omk:wmap7+act+bao+h0:number}{\ensuremath{1.00\pm0.01}}
\setsymbol{ACT:p1:ags:lcdm+omk:wmap7+act+bao+h0:number}{\ensuremath{0.4\pm0.2}}
\setsymbol{ACT:p1:age:lcdm+omk:wmap7+act+bao+h0:number}{\ensuremath{0.9\pm0.2}}
\setsymbol{ACT:p1:cae2:lcdm+omk:wmap7+act+bao+h0:number}{\ensuremath{1.00\pm0.01}}
\setsymbol{ACT:p1c220:lcdm+omk:wmap7+act+bao+h0:number}{\ensuremath{<0.93}}

\setsymbol{ACT:p1:tau:lcdm+dalpha:wmap7+act:number}{\ensuremath{0.888 \pm 0.015}}
\setsymbol{ACT:p1:omegabh2:lcdm+dalpha:wmap7+act:number}{\ensuremath{0.02260\pm 0.00045}}
\setsymbol{ACT:p1:omegach2:lcdm+dalpha:wmap7+act:number}{\ensuremath{0.115\pm 0.005}}
\setsymbol{ACT:p1:thetaA:lcdm+dalpha:wmap7+act:number}{\ensuremath{1.060\pm  0.002}}
\setsymbol{ACT:p1:sigma8:lcdm+dalpha:wmap7+act:number}{\ensuremath{0.84\pm 0.03}}
\setsymbol{ACT:p1:omegal:lcdm+dalpha:wmap7+act:number}{\ensuremath{0.73\pm 0.03}}
\setsymbol{ACT:p1:H0:lcdm+dalpha:wmap7+act:number}{\ensuremath{71.2\pm 3.6}}
\setsymbol{ACT:p1:omegac:lcdm+dalpha:wmap7+act:number}{\ensuremath{}}
\setsymbol{ACT:p1:ns:lcdm+dalpha:wmap7+act:number}{\ensuremath{0.970\pm 0.013}}
\setsymbol{ACT:p1:omegam:lcdm+dalpha:wmap7+act:number}{\ensuremath{0.27\pm0.03}}
\setsymbol{ACT:p1:100omegabh2:lcdm+dalpha:wmap7+act:number}{\ensuremath{2.260 \pm 0.045}}
\setsymbol{ACT:p1:zreion:lcdm+dalpha:wmap7+act:number}{\ensuremath{}}
\setsymbol{ACT:p1:109DeltaR2:lcdm+dalpha:wmap7+act:number}{\ensuremath{3.20\pm 0.04}}
\setsymbol{ACT:p1:dalpha:lcdm+dalpha:wmap7+act:number}{\ensuremath{1.004\pm 0.005}}
\setsymbol{ACT:p1:tsz:lcdm+dalpha:wmap7+act:number}{\ensuremath{3.5\pm1.4}}
\setsymbol{ACT:p1:ksz:lcdm+dalpha:wmap7+act:number}{\ensuremath{<8.1}}
\setsymbol{ACT:p1:xi:lcdm+dalpha:wmap7+act:number}{\ensuremath{<0.2}}
\setsymbol{ACT:p1:as:lcdm+dalpha:wmap7+act:number}{\ensuremath{3.1\pm0.4}}
\setsymbol{ACT:p1:ad:lcdm+dalpha:wmap7+act:number}{\ensuremath{7.2 \pm 0.5}}
\setsymbol{ACT:p1:ac:lcdm+dalpha:wmap7+act:number}{\ensuremath{5.1\pm 1.0}}
\setsymbol{ACT:p1:betac:lcdm+dalpha:wmap7+act:number}{\ensuremath{2.2\pm 0.1}}
\setsymbol{ACT:p1:cas1:lcdm+dalpha:wmap7+act:number}{\ensuremath{1.01\pm 0.01}}
\setsymbol{ACT:p1:cas2:lcdm+dalpha:wmap7+act:number}{\ensuremath{1.04\pm 0.02}}
\setsymbol{ACT:p1:cae1:lcdm+dalpha:wmap7+act:number}{\ensuremath{1.02 \pm 0.01}}
\setsymbol{ACT:p1:ags:lcdm+dalpha:wmap7+act:number}{\ensuremath{0.4\pm0.2}}
\setsymbol{ACT:p1:age:lcdm+dalpha:wmap7+act:number}{\ensuremath{0.9\pm0.2}}
\setsymbol{ACT:p1:cae2:lcdm+dalpha:wmap7+act:number}{\ensuremath{1.00 \pm 0.02}}
\setsymbol{ACT:p1c220:lcdm+dalpha:wmap7+act:number}{\ensuremath{<0.93}}
\setsymbol{ACT:p1:tau:lcdm+gmu:wmap7+act:number}{\ensuremath{0.092\pm0.015}}
\setsymbol{ACT:p1:omegabh2:lcdm+gmu:wmap7+act:number}{\ensuremath{0.0228\pm0.0005}}
\setsymbol{ACT:p1:omegach2:lcdm+gmu:wmap7+act:number}{\ensuremath{0.113\pm0.005}}
\setsymbol{ACT:p1:thetaA:lcdm+gmu:wmap7+act:number}{\ensuremath{1.040\pm  0.002}}
\setsymbol{ACT:p1:sigma8:lcdm+gmu:wmap7+act:number}{\ensuremath{0.82\pm0.03}}
\setsymbol{ACT:p1:omegal:lcdm+gmu:wmap7+act:number}{\ensuremath{0.73\pm0.03}}
\setsymbol{ACT:p1:H0:lcdm+gmu:wmap7+act:number}{\ensuremath{70.9\pm2.5}}
\setsymbol{ACT:p1:omegac:lcdm+gmu:wmap7+act:number}{\ensuremath{}}
\setsymbol{ACT:p1:ns:lcdm+gmu:wmap7+act:number}{\ensuremath{0.973\pm0.012}}
\setsymbol{ACT:p1:omegam:lcdm+gmu:wmap7+act:number}{\ensuremath{0.27\pm0.03}}
\setsymbol{ACT:p1:100omegabh2:lcdm+gmu:wmap7+act:number}{\ensuremath{2.279\pm0.052}}
\setsymbol{ACT:p1:109DeltaR2:lcdm+gmu:wmap7+act:number}{\ensuremath{3.18\pm0.05}}
\setsymbol{ACT:p1:gmu:lcdm+gmu:wmap7+act:number}{\ensuremath{<1.7\times 10^{-7}}}
\setsymbol{ACT:p1:tsz:lcdm+gmu:wmap7+act:number}{\ensuremath{3.4\pm1.3}}
\setsymbol{ACT:p1:ksz:lcdm+gmu:wmap7+act:number}{\ensuremath{<6.9}}
\setsymbol{ACT:p1:xi:lcdm+gmu:wmap7+act:number}{\ensuremath{<0.2}}
\setsymbol{ACT:p1:ad:lcdm+gmu:wmap7+act:number}{\ensuremath{7.3\pm0.6}}
\setsymbol{ACT:p1:ac:lcdm+gmu:wmap7+act:number}{\ensuremath{4.9\pm1.0}}
\setsymbol{ACT:p1:betac:lcdm+gmu:wmap7+act:number}{\ensuremath{2.2\pm0.1}}
\setsymbol{ACT:p1:as:lcdm+gmu:wmap7+act:number}{\ensuremath{3.1\pm0.4}}
\setsymbol{ACT:p1:cas1:lcdm+gmu:wmap7+act:number}{\ensuremath{1.02\pm 0.01}}
\setsymbol{ACT:p1:cas2:lcdm+gmu:wmap7+act:number}{\ensuremath{1.04\pm 0.02}}
\setsymbol{ACT:p1:cae1:lcdm+gmu:wmap7+act:number}{\ensuremath{1.01 \pm 0.01}}
\setsymbol{ACT:p1:cae2:lcdm+gmu:wmap7+act:number}{\ensuremath{1.00 \pm 0.02}}
\setsymbol{ACT:p1:ags:lcdm+gmu:wmap7+act:number}{\ensuremath{0.4\pm0.2}}
\setsymbol{ACT:p1:age:lcdm+gmu:wmap7+act:number}{\ensuremath{0.9\pm0.2}}
\setsymbol{ACT:p1c220:+act:lcdm+gmu:wmap7+act:number}{\ensuremath{x}}

\setsymbol{ACT:p1:tau:lcdm:wmap7+act+bao+h0:number}{\ensuremath{0.090\pm 0.014}}
\setsymbol{ACT:p1:omegach2:lcdm:wmap7+act+bao+h0:number}{\ensuremath{0.116\pm 0.003}}
\setsymbol{ACT:p1:thetaA:lcdm:wmap7+act+bao+h0:number}{\ensuremath{1.040 \pm 0.002}}
\setsymbol{ACT:p1:sigma8:lcdm:wmap7+act+bao+h0:number}{\ensuremath{0.83\pm 0.02}}
\setsymbol{ACT:p1:omegal:lcdm:wmap7+act+bao+h0:number}{\ensuremath{0.71\pm 0.01}}
\setsymbol{ACT:p1:omegac:lcdm:wmap7+act+bao+h0:number}{\ensuremath{}}
\setsymbol{ACT:p1:ns:lcdm:wmap7+act+bao+h0:number}{\ensuremath{0.971\pm0.009}}
\setsymbol{ACT:p1:omegam:lcdm:wmap7+act+bao+h0:number}{\ensuremath{0.29\pm 0.01}}
\setsymbol{ACT:p1:100omegabh2:lcdm:wmap7+act+bao+h0:number}{\ensuremath{2.225\pm0.038}}
\setsymbol{ACT:p1:109DeltaR2:lcdm:wmap7+act+bao+h0:number}{\ensuremath{3.20\pm 0.03}}
\setsymbol{ACT:p1:tsz:lcdm:wmap7+act+bao+h0:number}{\ensuremath{3.4 \pm 1.4}}
\setsymbol{ACT:p1:ksz:lcdm:wmap7+act+bao+h0:number}{\ensuremath{<8.3}}
\setsymbol{ACT:p1:xi:lcdm:wmap7+act+bao+h0:number}{\ensuremath{<0.2}}
\setsymbol{ACT:p1:ad:lcdm:wmap7+act+bao+h0:number}{\ensuremath{7.0\pm0.5}}
\setsymbol{ACT:p1:ac:lcdm:wmap7+act+bao+h0:number}{\ensuremath{5.0\pm1.0}}
\setsymbol{ACT:p1:as:lcdm:wmap7+act+bao+h0:number}{\ensuremath{3.1\pm0.4}}
\setsymbol{ACT:p1:betac:lcdm:wmap7+act+bao+h0:number}{\ensuremath{2.2\pm 0.1}}
\setsymbol{ACT:p1:ags:lcdm:wmap7+act+bao+h0:number}{\ensuremath{0.4\pm 0.2}}
\setsymbol{ACT:p1:age:lcdm:wmap7+act+bao+h0:number}{\ensuremath{0.9\pm 0.2}}
\setsymbol{ACT:p1:cas1:lcdm:wmap7+act+bao+h0:number}{\ensuremath{1.01\pm 0.01}}
\setsymbol{ACT:p1:cas2:lcdm:wmap7+act+bao+h0:number}{\ensuremath{1.03\pm 0.02}}
\setsymbol{ACT:p1:cae1:lcdm:wmap7+act+bao+h0:number}{\ensuremath{1.01\pm 0.01}}
\setsymbol{ACT:p1:cae2:lcdm:wmap7+act+bao+h0:number}{\ensuremath{1.00\pm 0.02}}
\setsymbol{ACT:p1c220:lcdm:wmap7+act+bao+h0:number}{\ensuremath{}}
\setsymbol{ACT:p1:H0:lcdm:wmap7+act+bao+h0:number}{\ensuremath{69.3\pm1.0}}
\setsymbol{ACT:p1:zreion:lcdm:wmap7+act+bao+h0:number}{\ensuremath{10.9\pm1.2}}
\begin{document}
\bibliographystyle{yahapj} 
\graphicspath{{./}{./Figs/}}

\author{Jonathan~L.~Sievers\altaffilmark{1,2},
Ren\'ee~A.~Hlozek\altaffilmark{3},
Michael~R.~Nolta\altaffilmark{2},
Viviana~Acquaviva\altaffilmark{4},
Graeme~E.~Addison\altaffilmark{5,6},
Peter~A.~R.~Ade\altaffilmark{7},
Paula~Aguirre\altaffilmark{8},
Mandana~Amiri\altaffilmark{5},
John~William~Appel\altaffilmark{1},
L.~Felipe~Barrientos\altaffilmark{8},
Elia~S.~Battistelli\altaffilmark{5,9},
Nick~Battaglia\altaffilmark{2,10},
J.~Richard~Bond\altaffilmark{2},
Ben~Brown\altaffilmark{11},
Bryce~Burger\altaffilmark{5},
Erminia~Calabrese\altaffilmark{6},
Jay~Chervenak\altaffilmark{12},
Devin~Crichton\altaffilmark{13},
Sudeep~Das\altaffilmark{15,14},
Mark~J.~Devlin\altaffilmark{16},
Simon~R.~Dicker\altaffilmark{16},
W.~Bertrand~Doriese\altaffilmark{17},
Joanna~Dunkley\altaffilmark{6},
Rolando~D\"{u}nner\altaffilmark{8},
Thomas~Essinger-Hileman\altaffilmark{1},
David~Faber\altaffilmark{1},
Ryan~P.~Fisher\altaffilmark{1},
Joseph~W.~Fowler\altaffilmark{1,17},
Patricio~Gallardo\altaffilmark{8},
Michael~S.~Gordon\altaffilmark{3},
Megan~B.~Gralla\altaffilmark{13},
Amir~Hajian\altaffilmark{2,3,1},
Mark~Halpern\altaffilmark{5},
Matthew~Hasselfield\altaffilmark{3,5},
Carlos~Hern\'andez-Monteagudo\altaffilmark{18},
J.~Colin~Hill\altaffilmark{3},
Gene~C.~Hilton\altaffilmark{17},
Matt~Hilton\altaffilmark{19,20},
Adam~D.~Hincks\altaffilmark{1,2},
Dave~Holtz\altaffilmark{1},
Kevin~M.~Huffenberger\altaffilmark{21},
David~H.~Hughes\altaffilmark{22},
John~P.~Hughes\altaffilmark{23},
Leopoldo~Infante\altaffilmark{8},
Kent~D.~Irwin\altaffilmark{17},
David~R.~Jacobson\altaffilmark{16},
Brittany~Johnstone\altaffilmark{25},
Jean~Baptiste~Juin\altaffilmark{8},
Madhuri~Kaul\altaffilmark{16},
Jeff~Klein\altaffilmark{16},
Arthur~Kosowsky\altaffilmark{11},
Judy~M~Lau\altaffilmark{1},
Michele~Limon\altaffilmark{26,16,1},
Yen-Ting~Lin\altaffilmark{27,28,3},
Thibaut~Louis\altaffilmark{6},
Robert~H.~Lupton\altaffilmark{3},
Tobias~A.~Marriage\altaffilmark{13,3,1},
Danica~Marsden\altaffilmark{24,16},
Krista~Martocci\altaffilmark{1},
Phil~Mauskopf\altaffilmark{7,30},
Michael~McLaren\altaffilmark{16},
Felipe~Menanteau\altaffilmark{23},
Kavilan~Moodley\altaffilmark{20},
Harvey~Moseley\altaffilmark{12},
Calvin~B~Netterfield\altaffilmark{31},
Michael~D.~Niemack\altaffilmark{1,17,32},
Lyman~A.~Page\altaffilmark{1},
William~A.~Page\altaffilmark{1},
Lucas~Parker\altaffilmark{1},
Bruce~Partridge\altaffilmark{33},
Reed~Plimpton\altaffilmark{16},
Hernan~Quintana\altaffilmark{8},
Erik~D.~Reese\altaffilmark{16},
Beth~Reid\altaffilmark{1},
Felipe~Rojas\altaffilmark{8},
Neelima~Sehgal\altaffilmark{34,1},
Blake~D.~Sherwin\altaffilmark{1},
Benjamin~L.~Schmitt\altaffilmark{16},
David~N.~Spergel\altaffilmark{3},
Suzanne~T.~Staggs\altaffilmark{1},
Omelan~Stryzak\altaffilmark{1},
Daniel~S.~Swetz\altaffilmark{16,17},
Eric~R.~Switzer\altaffilmark{1,2,29},
Robert~Thornton\altaffilmark{16,25},
Hy~Trac\altaffilmark{10},
Carole~Tucker\altaffilmark{7},
Masao~Uehara\altaffilmark{1},
Katerina~Visnjic\altaffilmark{1},
Ryan~Warne\altaffilmark{20},
Grant~Wilson\altaffilmark{35},
Ed~Wollack\altaffilmark{12},
Yue~Zhao\altaffilmark{1},
Caroline~Zunckel\altaffilmark{36}}
\altaffiltext{1}{Joseph Henry Laboratories of Physics, Jadwin Hall, Princeton University, Princeton, NJ 08544, USA}  
\altaffiltext{2}{Canadian Institute for Theoretical Astrophysics, University of Toronto, Toronto, ON M5S 3H8, Canada}
\altaffiltext{3}{Department of Astrophysical Sciences, Peyton Hall, Princeton University, Princeton, NJ 08544, USA}
\altaffiltext{4}{New York City College of Technology, 300 Jay Street ¥ Brooklyn, NY 11201, USA}
\altaffiltext{5}{Department of Physics and Astronomy, University of British Columbia, Vancouver, BC V6T 1Z4, Canada}
\altaffiltext{6}{Department of Astrophysics, Oxford University, Oxford OX1 3RH, UK}
\altaffiltext{7}{School of Physics and Astronomy, Cardiff University, The Parade, Cardiff, Wales CF24 3AA, UK}
\altaffiltext{8}{Departamento de Astronom{\'{i}}a y Astrof{\'{i}}sica, Facultad de F{\'{i}}sica, Pontificia Universidad Cat\'{o}lica de Chile, Casilla 306, Santiago 22, Chile}
\altaffiltext{9}{Department of Physics, University of Rome ``La Sapienza'', Piazzale Aldo Moro 5, I-00185 Rome, Italy}
\altaffiltext{10}{Department of Astronomy, Carnegie Mellon University, 100 Allen Hall, 3941 O'Hara Street, Pittsburgh, PA 15260, USA}
\altaffiltext{11}{Department of Physics and Astronomy, University of Pittsburgh, Pittsburgh, PA 15260, USA}
\altaffiltext{12}{Code 553/665, NASA/Goddard Space Flight Center, Greenbelt, MD 20771, USA}
\altaffiltext{13}{Department of Physics and Astronomy, The Johns Hopkins University, 3400 N. Charles St., Baltimore, MD 21218-2686, USA}
\altaffiltext{14}{Berkeley Center for Cosmological Physics, LBL and Department of Physics, University of California, Berkeley, CA 94720, USA}
\altaffiltext{15}{High Energy Physics Division, Argonne National Laboratory, 9700 S Cass Avenue, Lemont IL 60439 USA}
\altaffiltext{16}{Department of Physics and Astronomy, University of Pennsylvania, 209 South 33rd Street, Philadelphia, PA 19104, USA}
\altaffiltext{17}{NIST Quantum Devices Group, 325 Broadway Mailcode 817.03, Boulder, CO 80305, USA}
\altaffiltext{18}{Max Planck Institut f\"ur Astrophysik, Postfach 1317, D-85741 Garching bei M\"unchen, Germany}
\altaffiltext{19}{Centre for Astronomy \& Particle Theory, School of Physics \& Astronomy, University of Nottingham, Nottingham NG7 2RD, UK}
\altaffiltext{20}{Astrophysics and Cosmology Research Unit, School of Mathematical Sciences, University of KwaZulu-Natal, Durban, 4041, South Africa}
\altaffiltext{21}{Department of Physics, University of Miami, Coral Gables, FL 33124, USA}
\altaffiltext{22}{Instituto Nacional de Astrof\'isica, \'Optica y Electr\'onica (INAOE), Tonantzintla, Puebla, Mexico}
\altaffiltext{23}{Department of Physics and Astronomy, Rutgers, The State University of New Jersey, Piscataway, NJ 08854-8019, USA}
\altaffiltext{24}{Department of Physics, University of California Santa Barbara, CA 93106, USA}
\altaffiltext{25}{Department of Physics, West Chester University of Pennsylvania, West Chester, PA 19383, USA}
\altaffiltext{26}{Columbia Astrophysics Laboratory, 550 W. 120th St. Mail Code 5247, New York, NY 10027, USA}
\altaffiltext{27}{Institute for the Physics and Mathematics of the Universe, The University of Tokyo, Kashiwa, Chiba 277-8568, Japan}
\altaffiltext{28}{Institute of Astronomy \& Astrophysics, Academia Sinica, Taipei, Taiwan}
\altaffiltext{29}{Kavli Institute for Cosmological Physics, Laboratory for Astrophysics and Space Research, 5620 South Ellis Ave., Chicago, IL 60637, USA}
\altaffiltext{30}{Arizona State University, 4701 Thunderbird Road, Glendale, AZ 85306, USA}
\altaffiltext{31}{Department of Physics, University of Toronto, 60 St. George Street, Toronto, ON M5S 1A7, Canada}
\altaffiltext{32}{Department of Physics, Cornell University, Ithaca, NY, USA 14853}
\altaffiltext{33}{Department of Physics and Astronomy, Haverford College, Haverford, PA 19041, USA}
\altaffiltext{34}{Physics and Astronomy Department, Stony Brook University, Stony Brook, NY 11794-3800, USA}
\altaffiltext{35}{Department of Astronomy, University of Massachusetts, Amherst, MA 01003, USA}
\altaffiltext{36}{Department of Chemistry and Physics, University of Kwa-Zulu Natal, University Road. Westville. Postal Address: Private Bag X01 Scottsville 3209, South Africa.}
\title{The Atacama Cosmology Telescope: Cosmological parameters from three seasons of data.}
\begin{abstract}
We present constraints on cosmological and astrophysical parameters from high-resolution microwave background maps at 148 GHz and 218 GHz made by the Atacama Cosmology Telescope (ACT) in three seasons of observations from 2008 to 2010. A model of primary cosmological and secondary foreground parameters is fit to the map power spectra and lensing deflection power spectrum, including contributions from both the thermal Sunyaev-ZelÕdovich (tSZ) effect and the kinematic Sunyaev-ZelÕdovich (kSZ) effect, Poisson and correlated anisotropy from unresolved infrared sources, radio sources, and the correlation between the tSZ effect and infrared sources. The power $\ell^2 C_\ell/2\pi$ of the thermal SZ power spectrum at 148 GHz is measured to be $3.4\pm 1.4\,\,\mu{\rm K}^2$ at $\ell=3000,$ while the corresponding amplitude of the kinematic SZ power spectrum has a 95\% confidence level upper limit of $8.6\,\,\mu{\rm K}^2$. Combining ACT power spectra with the WMAP 7-year temperature and polarization power spectra, we find excellent consistency with the LCDM model.  We constrain the number of effective relativistic degrees of freedom in the early universe to be $N_{\rm eff}=2.79\pm 0.56$, in agreement with the canonical value of $N_{\rm eff}=3.046$ for three massless neutrinos. We constrain the sum of the neutrino masses to be $\Sigma m_\nu < 0.39$ eV at 95\% confidence when combining ACT and WMAP 7-year data with BAO and Hubble constant measurements. We constrain the amount of primordial helium to be $Y_p = 0.225 \pm 0.034$, and measure no variation in the fine structure constant $\alpha$ since recombination, with $\alpha/\alpha_0 = 1.004 \pm 0.005$. We also find no evidence for any running of the scalar spectral index, $dn_s/d\ln k = -0.004\pm 0.012$. 
\end{abstract}

\maketitle

\section{Introduction}
Studies of the cosmic microwave background (CMB) have dramatically progressed over the past two decades \citep[e.g.,][]{smoot/etal:1992, cheng/etal:1997, baker/etal:1999, miller/etal:1999, deBernardis/etal:2000, knox/page:2000, hanany/etal:2000, lee/etal:2001, romeo/etal:2001, netterfield/etal:2002,                                                                                                           
halverson/etal:2002, kovac/etal:2002, carlstrom/etal:2003, pearson/etal:2003, scott/etal:2003, benoit/etal:2003b,spergel/etal:2003, johnson/etal:2007, chiang/etal:2010}. The current $\Lambda$CDM cosmological model provides an excellent fit to the CMB data across a wide range of angular scales, and is supported by complementary observations of large scale structure \citep[e.g.,][]{reid/etal:2012, ho/etal:2012}, Baryon Acoustic Oscillations \citep[BAO e.g.,][]{blake/etal:2011, anderson/etal:2012,  busca/etal:2012}, Type Ia supernovae \citep[e.g.,][]{hicken/etal:2009,kessler/etal:2009, conley/etal:2011}, galaxy cluster measurements \citep[e.g.,][]{vikhlinin/etal:2009a,mantz/etal:2010,rozo/etal:2010, tinker/etal:2011} and observations of gravitational lensing \citep[e.g.,][]{massey/etal:2007,fu/etal:2008,schrabback/etal:2010,suyu/etal:2010, heymans/etal:2012, kilbinger/etal:2012}.  While the CMB on angular scales of greater than $0.3^\circ$ has been definitively measured by the Wilkinson Microwave Anisotropy Probe \citep[WMAP,][]{bennett/etal:2012}, a wealth of information in the CMB on smaller angular scales continues to be probed with ever-increasing precision \citep[e.g.,][]{hedman/etal:2002, kuo/etal:2007, pryke/etal:2009, reichardt/etal:2009, sievers/etal:prep, bischoff/etal:2010, reichardt/etal:2012, das/etal:2011, keisler/etal:2011, araujio/etal:2012}. As this paper was being finalized the SPT collaboration released a new set of papers \citep{story/etal:2012, hou/etal:2013}
and the WMAP team released its final 9-year results \citep{bennett/etal:2012,hinshaw/etal:2012}. Our analysis does not incorporate these results although in a few places we make direct comparisons. While this paper was under peer review, the Planck satellite released its first cosmological results. We leave a final combined ACT+Planck analysis to future work.

The Atacama Cosmology Telescope (ACT) complements measurements from WMAP by observing from $\ell \simeq 300$ to $\ell=10000$. This widens the range of data available to constrain both cosmological parameters through the Silk damping tail of the primary CMB \citep{silk:1968} and the residual power from secondary sources between us and the surface of last scattering. These sources include galaxy clusters, which are detectable at microwave frequencies through the Sunyaev-Zel'dovich (SZ) effect \citep{zeldovich/sunyaev:1969, sunyaev/zeldovich:1970}. The thermal SZ (tSZ) effect describes the spectral distortion due to the inverse-Compton scattering of CMB photons to higher frequencies by the hot gas in clusters, while the kinematic SZ (kSZ) effect measures the corresponding temperature shift due to the bulk peculiar motion of the clusters. While these effects produce a diffuse signal at small scales from unresolved clusters, their influence has recently been detected directly through the cross-correlation of the ACT temperature maps with other tracers \citep{hand/etal:2011, hand/etal:2012}. In addition to emission from clusters via the SZ effect, radio galaxies and dusty star-forming galaxies also contribute to the power on small scales, and indeed dominate the cosmological signal for multipoles $\ell > 3000$ and frequencies $\nu \ge 90$ GHz. Gravitational lensing by structures along the line of sight also generates a microwave background signal, and distorts the primordial CMB.

\begin{figure*}[htbp!]
\begin{center}
$\begin{array}{@{\hspace{-0.0in}}l}
\includegraphics[scale=0.6]{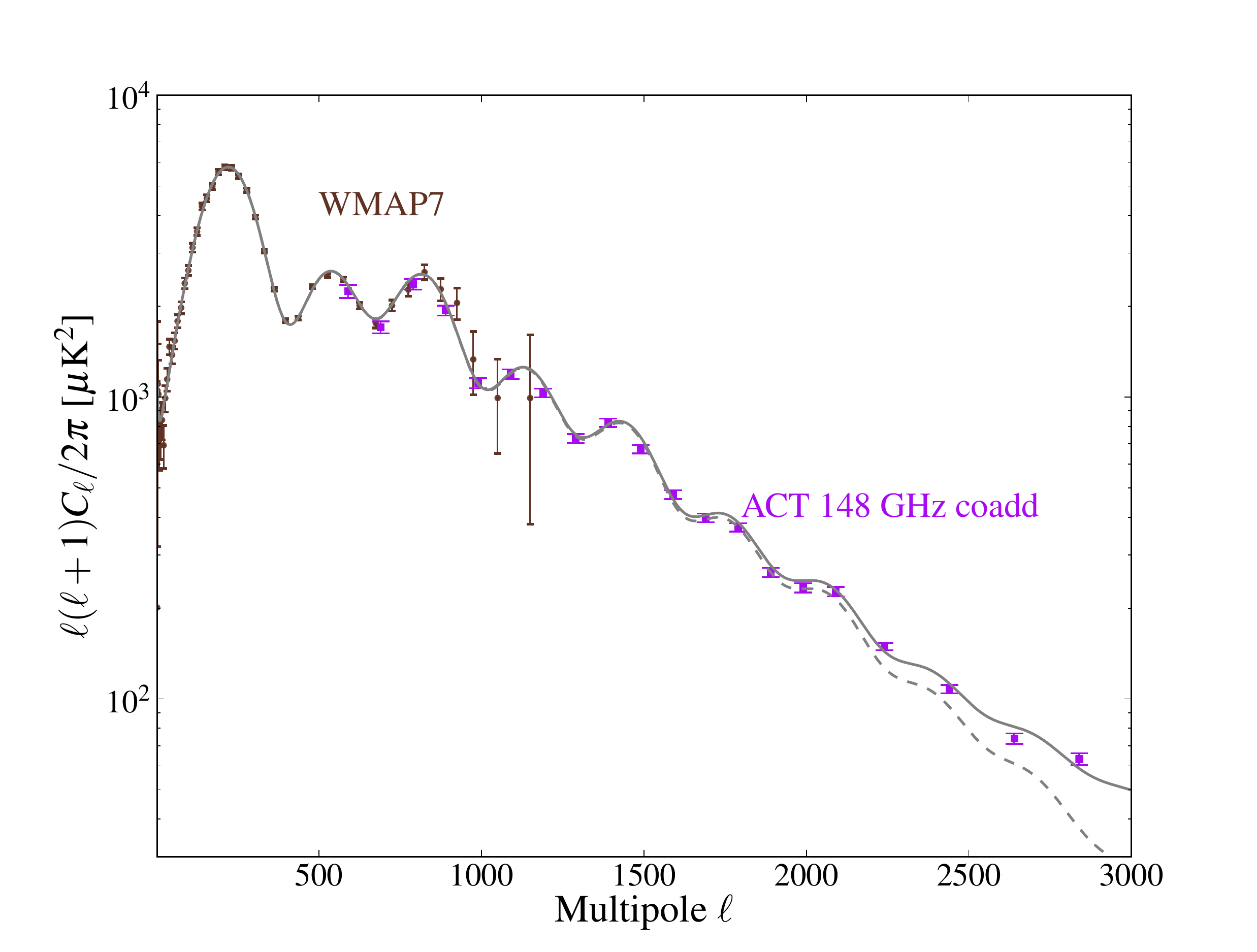}\\[0.0cm]
 \end{array}$
\caption{Data used in this cosmological analysis. The data from the WMAP 7-year data release \citep{larson/etal:2011,komatsu/etal:2011} are combined with the ACT data. In this figure we show a weighted co-added spectrum from the equatorial and southern patches at 148 GHz. The full ACT likelihood, however, considers independent southern and equatorial spectra for both 148 GHz and 218 GHz \citep[given in Table 3 of][and online on LAMBDA]{das/etal:2012prep} and their cross-frequency spectrum, which are shown in Figure~\ref{fig:multispec}. The solid line indicates the best-fit cosmological model including foreground emission, while the dashed line shows the best-fit primordial CMB spectrum. This binning was selected for the common analysis of the equatorial and southern data. See \citet{das/etal:2012prep} for alternative binnings. \label{fig:cmbdata} }
 \end{center}
 \end{figure*}

This paper forms part of a set of papers presenting the 3-year analysis of the ACT data; the ACT temperature and deflection power spectra are presented in \citet{das/etal:2012prep}, while \citet{dunkley/etal:preplike} presents the likelihood used in this analysis. \citet{hasselfield/etal:2013a} presents a catalogue of SZ-detected clusters from the ACT data, and interprets them. This paper contains the parameter estimation of both primary (cosmological) and secondary (foreground) parameters. We outline the data used in this analysis in Section~\ref{sec:data} and describe the methodology and likelihood in Section~\ref{sec:method}. We present constraints on primary cosmological parameters in Section~\ref{sec:cosmo} and on secondary parameters in Section~\ref{sec:secondaries}. We conclude in Section~\ref{sec:conclusions}, after which we provide an appendix of analysis tests.
\section{Data}
\label{sec:data}
This paper presents results from a combination of observations at two frequencies, 148 GHz and 218 GHz, of multiple fields taken over three years. The southern fields (ACT-S) varied over different seasons, with the 2008 season containing a 292 deg$^2$ patch and the 2009 and 2010 seasons focusing on a smaller 146 deg$^2$ footprint. Equatorial data (ACT-E) were taken only over the 2009 and 2010 seasons; we use a 300 deg$^2$ patch for the 3-year analysis. 
We follow a similar procedure for going from maps to the temperature power spectra as was used in \citet{das/etal:2011}, using the power spectrum estimation procedure presented in \citet{das/hajian/spergel:2009}. 
 
The data and map-making procedure are described in \citet{dunner/etal:2012}; the power spectrum method and systematic tests are presented in \citet{das/etal:2012prep}.
Using the spectrum presented in \citet{das/etal:2012prep}, we have constructed two likelihoods; these likelihoods that are also presented in \citet{dunkley/etal:preplike}. The multi-frequency likelihood parameterizes the foreground emission using additional parameters which we list in Section~\ref{sec:method}, while the CMB-only likelihood marginalizes over these foregrounds. For the analysis presented here we use the multi-frequency likelihood; however we show in Appendix~\ref{likeconsistency} that the two likelihoods give equivalent results.

The data used in this analysis are the multi-frequency temperature spectra estimated from the ACT maps with combined spectra from both ACT-S and ACT-E including their covariance. The spectra are presented in \citet{das/etal:2012prep}. For the results presented in Figure~\ref{fig:marg_cls} we use the marginalized CMB-only likelihood. In addition, we include the measurement of the power spectrum of the lensing deflection angle, ${\ell^2}/{4} {C_\ell}^{d d}$  \citep{das/etal:2012prep}.

The lensing spectrum is estimated from the ACT temperature maps using an optimal quadratic estimator \citep{hu/okamoto:2002}. Only data from the ACT equatorial patches are used to measure the deflection power, as the signal-to-noise of the southern patch was much lower than that of the equatorial data. The covariance between the lensing power spectrum and the temperature power spectrum is small. When adding in the deflection data, we will use the abbreviation `ACTDefl'. 

The maps are cross-correlated with WMAP7 maps in order to obtain a calibration factor in multipole space. The cross-correlation calibration method is described in \citet{hajian/etal:2012}; details of the calibration of the ACT 3-year data are given in \citet{das/etal:2012prep}.

 \subsection{Beam and calibration errors}
Understanding the beam profiles is essential for interpreting the high-$\ell$ aspects of the power spectrum. At $\ell=3000$ and 148 GHz, the window function is $\approx 0.5$ its value at $\ell=200$. The beams are estimated independently for each array and season \citep{dunner/etal:2012} from observations of Saturn and Uranus. The beams vary slightly with season due to changes in the telescope focus.  We include a contribution in the likelihood to the full covariance matrix from the covariance of the beams for each season. The beam error includes contributions from the uncertainty in the pointing variation of the telescope.
At $\ell=700$, the pivot point for the beam and calibration uncertainties, the effective calibration error is 2\% for the 148 GHz maps and 2.6\% for the 218 GHz maps. Different seasons have somewhat different calibration uncertainties and so these numbers should be considered as representative of the effective combined calibration. While the absolute calibration is performed from cross-correlations to WMAP data, the telescope pointing solution, beams, and detector responsivity are characterized independently in each observing season and thus the calibration uncertainties of ACT-E and ACT-S are relatively independent.

In the cosmological analysis, we apply a calibration prior for the 148 GHz spectra obtained from the WMAP-ACT cross-correlation calibration procedure described above. In the chains, we allow for a small error in the overall calibration of the spectrum by marginalizing over independent calibration factors for the south and equatorial spectra, at both 148 GHz and 218 GHz. This extra calibration allows for the overall spectra to adjust themselves at the 1\% level and has a negligible effect on the cosmological parameters, as discussed in Appendix~\ref{beam_error}. Similarly, in the same appendix, we test for the dependence of the cosmological parameters on the beam's assumed uncertainty, and find that beam error has a negligible effect on the parameters of interest.\\
 \subsection{Additional data}
We use temperature and polarization data from the seven year data release of the WMAP\ satellite \citep{larson/etal:2011,komatsu/etal:2011} in addition to the measurements of the microwave temperature from ACT. We include measurements of the Baryon Acoustic Oscillations (BAO) from the Six-degree Field Galaxy Redshift Survey  \citep[6dFGRS,][]{beutler/etal:2011} and the Sloan Digital Sky Survey Data Releases 7 \citep[SDSS DR7,][]{percival/etal:2009} and 9 \citep[SDSS DR9,][]{anderson/etal:2012}, measured at redshifts: $z = 0.106, 0.35$ and $0.57.$
In addition, we supplement our data with a measurement of the Hubble constant of $H_0 = 73.8 \pm2.4~\mathrm{km}~\mathrm{s}^{-1}~\mathrm{Mpc}^{-1}$ \citep{riess/etal:2011}, although \citet{freedman/etal:2012} find  $H_0 = 74.3 \pm2.1~\mathrm{km}~\mathrm{s}^{-1}~\mathrm{Mpc}^{-1}.$ In some cases we also include a prior on $\sigma_8$ from skewness measurements of the tSZ effect from ACT \citep{wilson/etal:2012}. 

\begin{figure}[htbp!]
\begin{center}
$\begin{array}{@{\hspace{-0.3in}}l}
\includegraphics[scale=0.5]{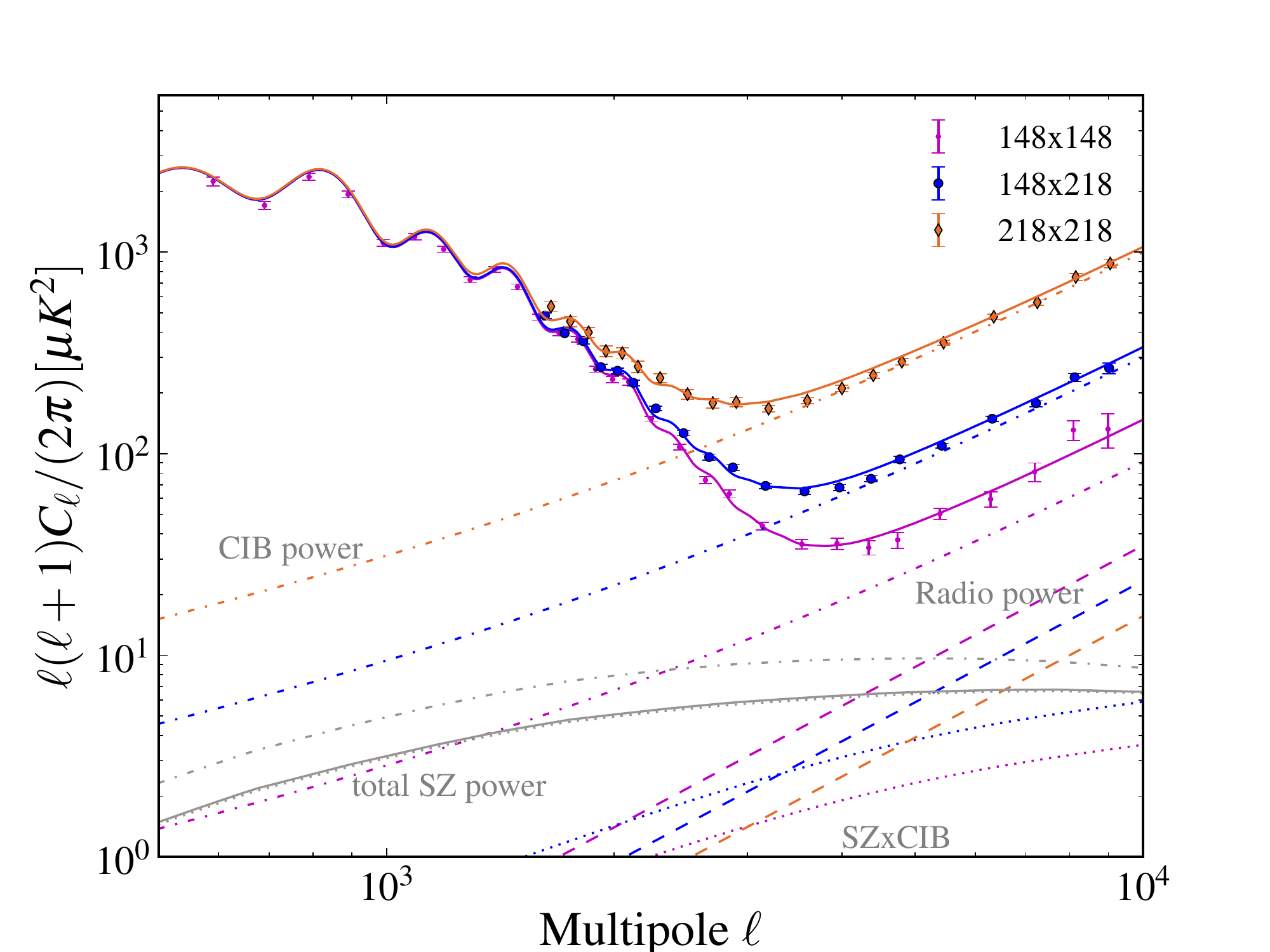}\\ [0.0cm]
 \end{array}$
 \caption{The ACT multi-frequency data. Each solid line indicates the best-fit total model for each cross spectrum, while the dashed lines show the radio power; the dot-dashed line indicates the power from CIB sources (both Poisson and correlated CIB); the gray lines show the total SZ power. The solid and dotted gray lines are for those spectra including 218 GHz and consist of only the kSZ contribution, while the dashed gray line is for the 148x148 GHz spectrum. The dotted lines show the (negative of) the (SZ-CIB) correlation. In all cases the models are for the best fit in a $\Lambda$CDM scenario. \label{fig:multispec}}
\end{center}
 \end{figure}

Unless explicitly specified, the ACT 3-year data are combined with WMAP7 data. For some model constraints, such as those on $N_\mathrm{eff}$ and the secondary parameters, we also include the `low-$\ell$' and `high-$\ell$' spectrum measurements from SPT \citep{keisler/etal:2011, reichardt/etal:2012}, following the prescription in \citet{dunkley/etal:preplike}. We show the ACT and WMAP7 data in Figure~\ref{fig:cmbdata}. The best-fit model for the various frequency components is shown in Figure~\ref{fig:multispec}.

\section{Methodology}
\label{sec:method}
We use Markov Chain Monte Carlo (MCMC) methods to determine parameters associated with a variety of models.
The basic cosmological  $\Lambda$CDM model consists of 6 parameters describing a flat universe. These include the physical baryon density 
$\Omega_bh^2$ (where $h = H_0/100~\mathrm{km}~\mathrm{s}^{-1}~\mathrm{Mpc}^{-1}$ is the dimensionless Hubble parameter), cold dark matter (CDM) density 
$\Omega_ch^2$, and $\theta_A = r_a/d_A$, the ratio of the acoustic horizon to the angular diameter distance at decoupling. This parameter is sensitive to the dark energy density, $\Omega_\Lambda,$ but less degenerate with other parameters \citep{kosowsky/etal:2002}. The value of $\Omega_\Lambda$ is then a derived parameter.
We assume the primordial perturbations to be scalar, adiabatic,
and Gaussian and parametrize them via a spectral
tilt $n_s$, and amplitude $\Delta^2_\mathcal{R}$, defined at pivot scale
$k_0 = 0.002$ Mpc$^{-1}$. We assume that the universe transitioned from a neutral to an ionized state over a small redshift range, $\Delta z = 0.5,$ with optical depth $\tau.$ The reionization history of the universe can be probed by small-scale CMB measurements through the impact of reionization on the kSZ effect \citep{ostriker/vishniac:1986, gruzinov/hu:1998, knox/etal:1998, zahn/etal:2012}, although care must be taken to allow for correlations between the tSZ effect and the microwave emission from unresolved dusty galaxies \citep{mesinger/etal:2012, addison/dunkley/spergel:2012}.

We express the $\Lambda$CDM set of parameters as
\begin{equation}
\{\Omega_bh^2,\Omega_ch^2, \theta_A,\Delta^2_\mathcal{R}, n_s, \tau\}.\\
\end{equation}
In this $\Lambda$CDM model, the number of effective relativistic degrees of freedom is assumed to be $N_\mathrm{eff} = 3.046,$ with the abundance of primordial helium fixed at $Y_p = 0.24.$ 
The likelihood used in the ACT analysis is described in \citet{dunkley/etal:preplike}, which we briefly summarize here. We fit a model of secondary emission to the ACT multi-frequency power spectra that includes an additional nine parameters when considering ACT data in combination with WMAP7 data. For the 148 GHz data we use modes $500< \ell < 10000,$ while for the 218 GHz data we restrict ourselves to $1500 < \ell < 10000.$
The theoretical spectrum for frequency bands $i$ and $j$ is
\be
{\cal B}_\ell^{\rm{th,ij}} = {\cal B}_\ell^{\rm{CMB}} + {\cal B}_\ell^{\rm{sec,ij}}, 
\ee
where  ${\cal B}_\ell = \ell(\ell+1)C_\ell/2\pi$ and $C^{\rm{CMB}}_\ell$ is the lensed primary CMB  power spectrum. The secondary spectra components are modeled as
\begin{eqnarray}
\label{eq:model}
{\cal B}_\ell^{\rm{sec,ij}}  &=&  {\cal B}_\ell^{\rm{tSZ,ij}} 
+ {\cal B}_\ell^{\rm{kSZ,ij}}  \nonumber \\
&&+ {\cal B}_\ell^{\rm{CIB,ij}}
+ {\cal B}_\ell^{\rm{tSZ-CIB,ij}}
+  {\cal B}_\ell^{\rm{rad,ij}}
+ {\cal B}_\ell^{\rm{Gal,ij}},
\label{eqn:spectra_th}
\end{eqnarray}
with contributions from the tSZ and kSZ effects, CIB sources, the cross-correlation between the tSZ and CIB  signals (tSZ-CIB), radio galaxies (rad), and residual Galactic dust (Gal). In addition to the six primary parameters, we add the following nine parameters
\begin{equation}
\{a_\mathrm{tSZ}, a_\mathrm{kSZ}, a_{p},a_{c},  a_{gs}, a_{ge}, \beta_c, a_s, \xi  \}.
\end{equation}
The parameter $a_\mathrm{tSZ}$ parameterizes the amplitude of the tSZ power; $a_\mathrm{kSZ}$ the kSZ amplitude; $a_p$ and $a_c$ the Poisson and clustered Cosmic Infrared Background (CIB) power and $a_{gs}, a_{ge}$ model the residual Galactic dust anisotropy in the southern and equatorial survey regions. All the parameter amplitudes are dimensionless, and are defined for a template spectrum normalized to $1~\mu\mathrm{K}^2$ at $\ell_0 = 3000$, and frequency $\nu = 150$ GHz. The frequency dependence of the correlated and Poisson CIB power is given in flux density units by the product of modified blackbodies with effective temperature 9.7~K and emissivity index $\beta_c$, following \citet{addison/etal:2012}, and as described in Equations (8) and (9) of \citet{dunkley/etal:preplike}. The radio source power has an amplitude $a_s$ and a spectral index $\alpha_s$ fixed to $-0.5$, based on the assumption that the index obtained for brighter sources from ACT and SPT source catalogs \citep{vieira/etal:2010, marriage/etal:2011a} holds for fainter sources.

The clustered and Poisson (both CIB and radio) templates vary with scale as $\ell(\ell+1)C_{\ell}/2\pi\propto\ell^{0.8}$ and $\propto\ell^2$, respectively. 
We allow for a correlation between the tSZ effect and CIB sources, with scale dependence given by the template calculated by \citet{addison/dunkley/spergel:2012}, and a frequency-independent correlation coefficient, $\xi$, which is defined in Equation (11) of \citet{dunkley/etal:preplike}, and is restricted to lie in the range $0<\xi<0.2$. The parameters in our foreground model are summarized in Table~\ref{table:secondaries}.
 \begin{table*}[htbp!] 
\caption{\small{Secondary parameters in the ACT 3-year foreground model.} }
\begin{center}
\begin{tabular}{c|c}
\hline
\hline
Symbol & Description \\
\hline
$a_\mathrm{tSZ}$\tablenotemark{a} & Thermal Sunyaev-Zel'dovich (SZ) power amplitude. \\
$a_\mathrm{kSZ}$\tablenotemark{a} & Kinematic SZ power amplitude. \\
$a_p$\tablenotemark{a} & Poisson Cosmic Infrared Background (CIB) power amplitude. \\
$a_c$\tablenotemark{a} & Clustered CIB power amplitude. \\
$a_{gs}$\tablenotemark{a} & Residual galactic emission amplitude for the ACT-S spectrum.\\
$a_{ge}$\tablenotemark{a} & Residual galactic emission amplitude for the ACT-E spectrum.\\
$\beta_c$ & Emissivity index of the clustered CIB power. \\
$a_s$\tablenotemark{a} & Radio Poisson power amplitude. \\
$\xi$\tablenotemark{a} & tSZ-CIB correlation amplitude.\\
\hline
\end{tabular}
\label{table:secondaries}
\tablenotetext{1}{Amplitudes relative to templates normalized to $\ell(\ell+1)C_\ell/2\pi = 1\mu\mathrm{K}^2$ at $\ell_0 = 3000$ and 150 GHz. See \citet{dunkley/etal:preplike} for a complete description.}
\end{center}
\end{table*}

The spatially variable Galactic emission has been masked, leaving only a small residual component. To determine the amplitude and spectrum of the residual component we cross correlate with the IRIS map \citep{iris:2005} as described in detail in \citet{das/etal:2012prep}. We then marginalize over this dust component in the likelihood with separate amplitudes in the equatorial and southern regions. There is roughly double the amount of dust in the equatorial region than in the south. In addition, there are some bright clouds. Investigations were performed to identify possible residual dusty clouds that were missed by our treatment, but no clear evidence for them was found. The likelihood marginalization described in \citet{dunkley/etal:preplike} was found to be the most general and parsimonious treatment.

We mask all sources above a detection threshold of 15 mJy. The source detection algorithm is described in \citet{marriage/etal:2011a}.  The 148 and 218 GHz source samples in the southern map will be presented in \citet{marsden/etal:prep}, and source catalogs for the full data set will be presented in \citet{gralla/etal:prep2013}. Sources are masked to a flux level of 6.4 mJy in the SPT analysis presented in \citet{reichardt/etal:2012}, resulting in a source amplitude at 150 GHz of $a'_s = 1.3\pm0.2.$ Sources are masked to a level of 50 mJy in \citet{keisler/etal:2011}. The estimated difference in unresolved radio source power is $9.2~\mu \mathrm{K}^2.$ Hence we include a separate amplitude for radio emission measured by the South Pole telescope, $a'_s,$ when including the SPT data in our analysis, whereas the other parameters in the secondary model, apart from the Galactic dust parameters, are common to both data sets. Thus we first subtract an amplitude of radio Poisson power of $9.2~\mu\mathrm{K}^2$ from the \citet{keisler/etal:2011} `low-$\ell$' spectrum.

For the parameter analysis, we use the publicly available CosmoMC code, which includes version 1.5 of the Recfast code \citep{seager/sasselov/scott:1999, seager/sasselov/scott:2000, wong/moss/scott:2008, switzer/hirata:2008, alihaimoud/hirata:2011, chluba/thomas:2011}.

 We combine measurements of the three independent lensed cross spectra: 148x148 GHz, 148x218 GHz and 218x218 GHz made from the ACT-S and ACT-E fields described in Section~\ref{sec:data}. In addition, we use the measurement of the lensing deflection field by ACT, presented in \citep{das/etal:2012prep}.
We investigate four types of fits in this analysis:
\begin{itemize}
\item{}We apply the full ACT likelihood to the data in combination with WMAP7 data and obtain constraints on both the primary and secondary parameters, with a full 15, 16 or 17 parameter model (and four additional calibration nuisance parameters).
\item{}We estimate $C_\ell$ bandpowers marginalized over the secondary foregrounds, and obtain constraints on primary parameters based on these marginalized bandpowers.
\item{}We combine ACT and SPT to check for consistency between these small-scale experiments.
\item{}We consider the ACT data alone without WMAP7 data, while fixing (or placing priors on) the spectral index $n_s$ and the optical depth $\tau$.
\end{itemize}

The models obtained from fitting only the ACT-S and ACT-E spectra are consistent with the models fit to the combined data, the best-fit spectra from each region agreeing to within 4\%. Moreover, the best-fit theoretical spectrum from the current 3-year ACT data agrees with the spectrum derived from the 1-year ACT data to the 1\% level. We discuss the consistency of the ACT spectrum in Appendix~\ref{dataconsistency}.
\begin{figure*}[htbp!]
\begin{center}
$\begin{array}{@{\hspace{-0.25in}}l}
\includegraphics[scale=0.75,trim = 0mm 0mm 0mm 0mm, clip]{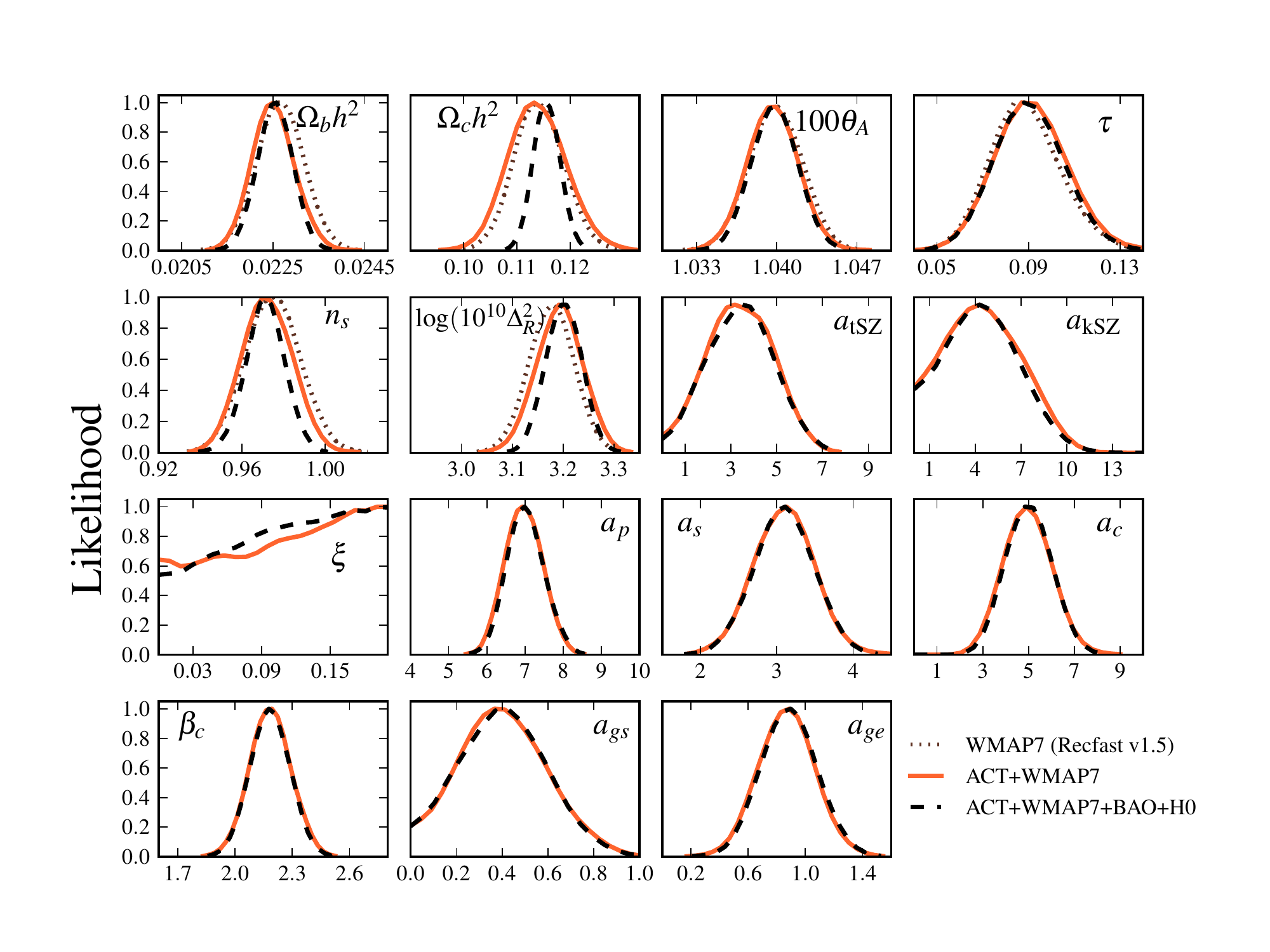}\\ [0.0cm]
 \end{array}$
 \caption{Parameter constraints on the $\Lambda$CDM model for the combined ACT and WMAP 7-year
power spectra. For each parameter, the full likelihood is marginalized over the other
parameters. Dotted curves for the primary parameters are for WMAP7 data only. The
solid curves are for ACT plus WMAP7, and the dashed lines add priors from measurements
of the baryon acoustic peak \citep{anderson/etal:2012, percival/etal:2009, beutler/etal:2011} and a direct measurement of $H_0$ \citep{riess/etal:2011}. 
The constraints are summarized in Tables~\ref{table:wa_damp_params},\ref{table:wa_params} and \ref{table:wabh_params}. Since the $\xi$ parameter is unconstrained and set only by the prior, $0< \xi < 0.2,$ it is not shown in the tables.\label{fig:lcdm_params}}. 
\end{center}
 \end{figure*}
 
\section{Constraints on primordial parameters}
\label{sec:cosmo}
The $\Lambda$CDM model continues to fit the ACT data well, when combined with the independent WMAP7 data.
Figure~\ref{fig:lcdm_params} illustrates the constraints on the $\Lambda$CDM model with the additional secondary parameters for the ACT+WMAP7 data combination. In addition, we plot the constraints from the WMAP7 power spectrum alone. ACT extends the angular range measured by WMAP, but the parameters from the joint fit are consistent with those from WMAP alone. In addition, the plot shows that the six parameters are robust to the presence of low levels of foreground emission that can be identified and extracted by ACT because of its higher resolution and different frequency coverage.

We start by constraining the parameters in the $\Lambda$CDM model. Our constraint on the scalar spectral index is $n_s = 0.971 \pm 0.009,$ using ACT data in combination with WMAP7 data, BAO and $H_0$ measurements. We discuss the constraints on $n_s$ in Section~\ref{sec:inflation}. In addition, we improve the constraints on the baryon density $\Omega_bh^2,$ using only CMB data, to $\Omega_bh^2 = 0.0225 \pm 0.00047,$ which is due to the fact that the ACT spectrum places tight limits on the positions of the higher order peaks below  $\ell \approx 3000.$ The models are summarized in Tables~\ref{table:wa_damp_params} and \ref{table:wa_params} for the ACT data in combination with WMAP7. Table~\ref{table:wabh_params} shows the constraints when adding BAO and $H_0$ measurements to the ACT power spectrum data. In Appendix~\ref{likeconsistency} we show that the $\Lambda$CDM parameters derived from the CMB-only likelihood agree with the full likelihood to within $0.1\sigma$.
\footnote{Updated beams are presented in \citep{hasselfield/etal:2013b}. We have checked that using these beams (rather than the ones on which this analysis is based) results in less than a $0.2\sigma$ shift in parameters, as described in Appendix~\ref{beam_error}.}
 \begin{figure*}[htbp!]
\begin{center}
$\begin{array}{@{\hspace{-0.0in}}l}
\includegraphics[scale=0.65,trim = 10mm 40mm 0mm 40mm, clip]{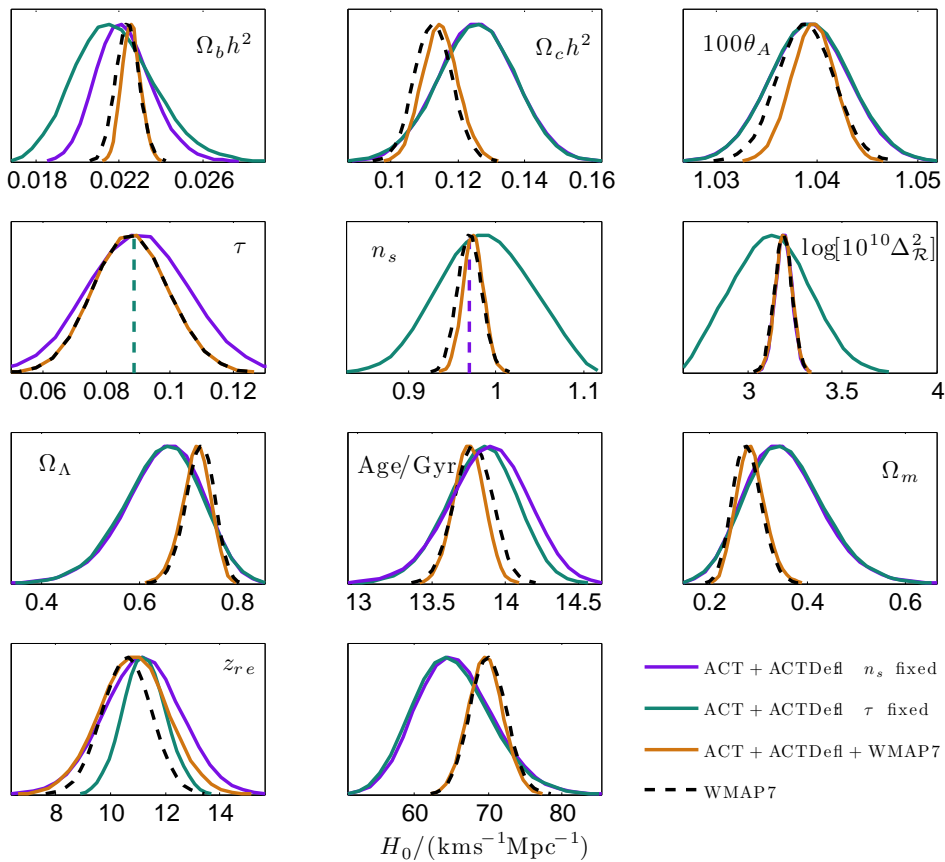}\\ [0.0cm]
 \end{array}$
 \caption{Constraints on the $\Lambda$CDM cosmological model using the likelihood of marginalized CMB bandpowers using ACT+ACTDefl data, compared to the constraints from ACT+ACTDefl+WMAP7, where ACTDefl is the reconstructed deflection power spectrum from ACT data, and WMAP7 alone. The top six panels are fitted parameters, while the bottom two rows are derived parameters. For the ACT-only parameters, two cases are shown: one where we fix the optical depth $\tau$, and another where the scalar spectral index $n_s$ is fixed. Fixing $n_s$ tightens the constraint on the amplitude of scalar perturbations (the ACT+ACTDefl, $n_s$ fixed curve lies on the ACT+ACTDefl+WMAP7 curve), but does not cause significant shifts in the other primary parameters, relative to leaving $n_s$ free. The independent CMB data sets, ACT and WMAP7, are consistent with the $\Lambda$CDM model.\label{fig:marg_cls}} 
\end{center}
 \end{figure*}

The greatest power of ACT comes when quantifying models beyond the standard cosmological model because the temperature power spectrum
contains little additional information on the simple $\Lambda$CDM parameters at angular scales
smaller than the third acoustic peak, $\ell \approx 800$ \citep[e.g.,][]{kosowsky/etal:2002}. The damping of the higher-order acoustic peaks relative to the baseline model \citep{komatsu/etal:2009} provides constraints on a variety of non-standard models. The ACT 1-year spectrum data \citep{das/etal:2011} showed a slight excess of damping at small scales relative to the baseline model. Evidence for this slight excess is not seen in the ACT 3-year data set. More data resulted in a spectrum with smaller error bars, providing tighter constraints on parameters such as the baryon density, while the best-fit theoretical spectra are consistent between the two results at the 1\% level. The consistency between the results presented here and those presented in \citet{dunkley/etal:2011} is discussed in Section~\ref{likeconsistency} of the appendix. 
 \begin{figure*}[htbp!]
\begin{center}
$\begin{array}{@{\hspace{-0.0in}}l@{\hspace{-0.0in}}l}
\includegraphics[scale=0.4]{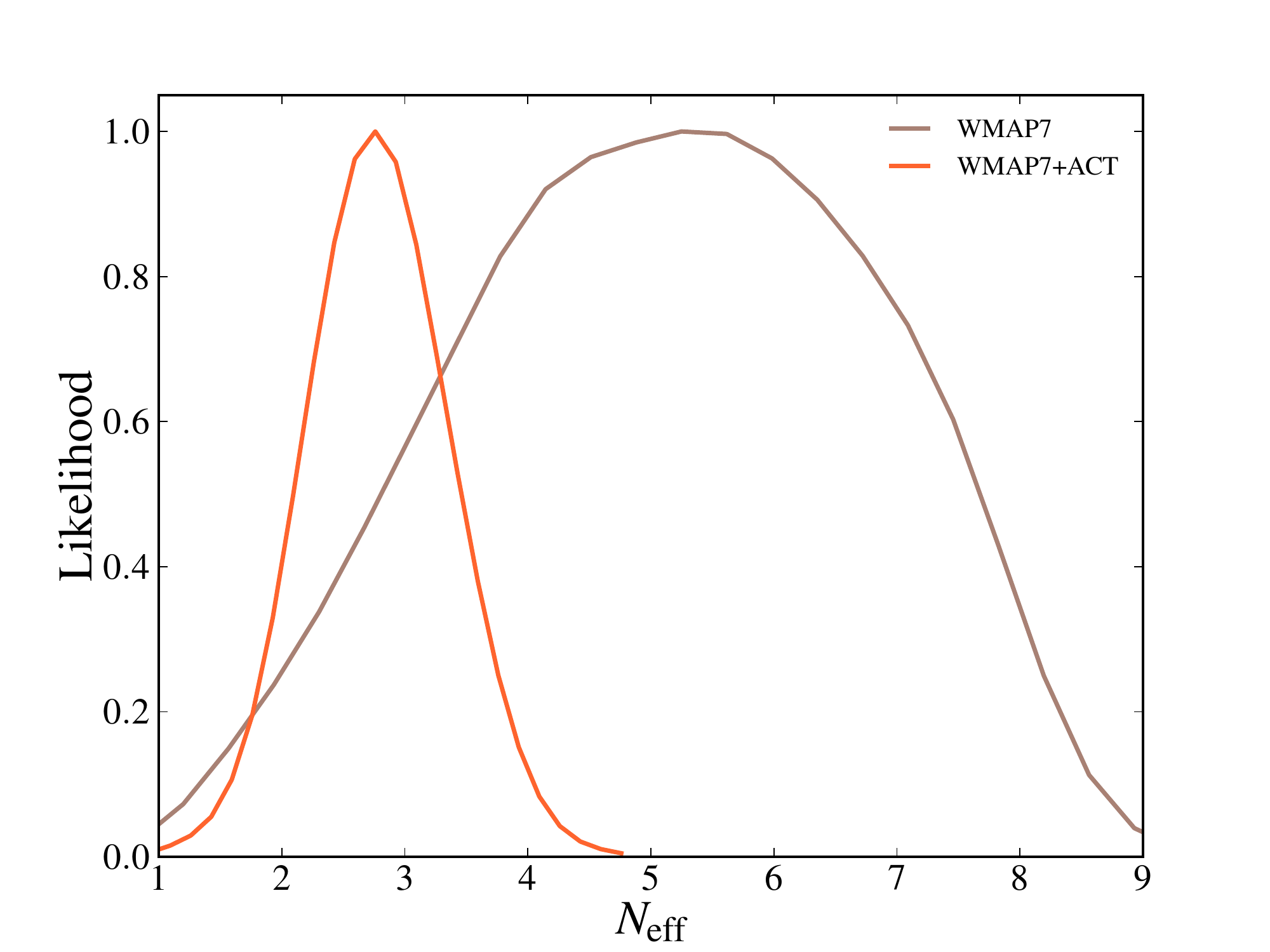}&
\includegraphics[scale=0.4]{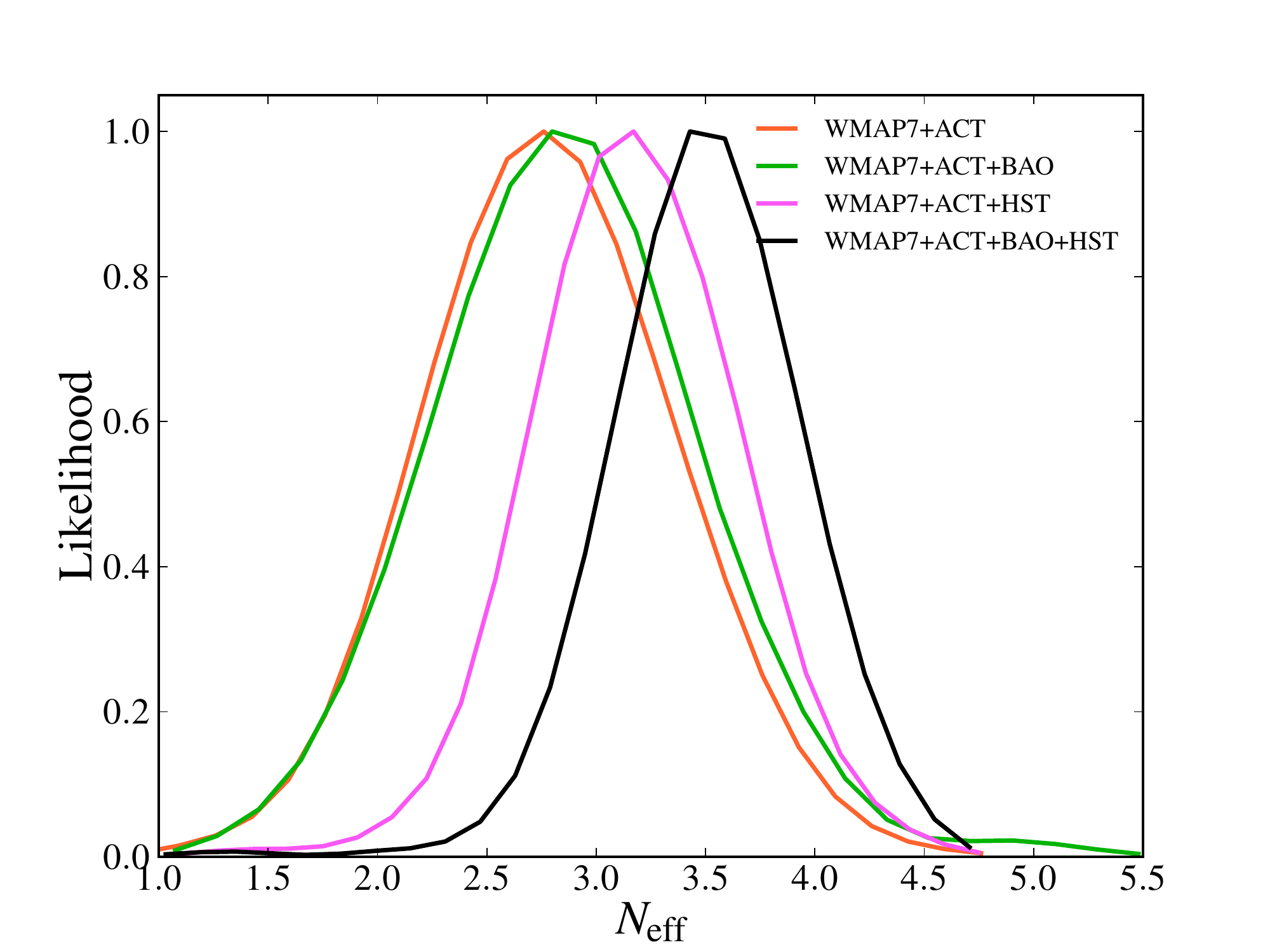}\\ [0.06cm]
 \end{array}$
 \caption{Marginalized one-dimensional likelihood of the $N_\mathrm{eff}$ for ACT data in combination with other probes. {\textit Left panel:} The improvement in constraints of the ACT in combination with WMAP7 data relative to WMAP7 data on its own. {\textit Right panel:} The $N_\mathrm{eff}$ likelihood for WMAP7+ACT data and a variety of other probes. The constraints are consistent with $N_\mathrm{eff} = 3.046$, the value in the model with three neutrino species. The slight shifts in the central value can be understood in the context of the changes in the value of the Hubble parameter, as illustrated in Figure~\ref{fig:neff_h0}\label{fig:neff_like}.}
\end{center}
 \end{figure*}
 
\begin{figure}[htbp!]
\begin{center}
$\begin{array}{@{\hspace{-0.0in}}l}
\includegraphics[scale=0.4]{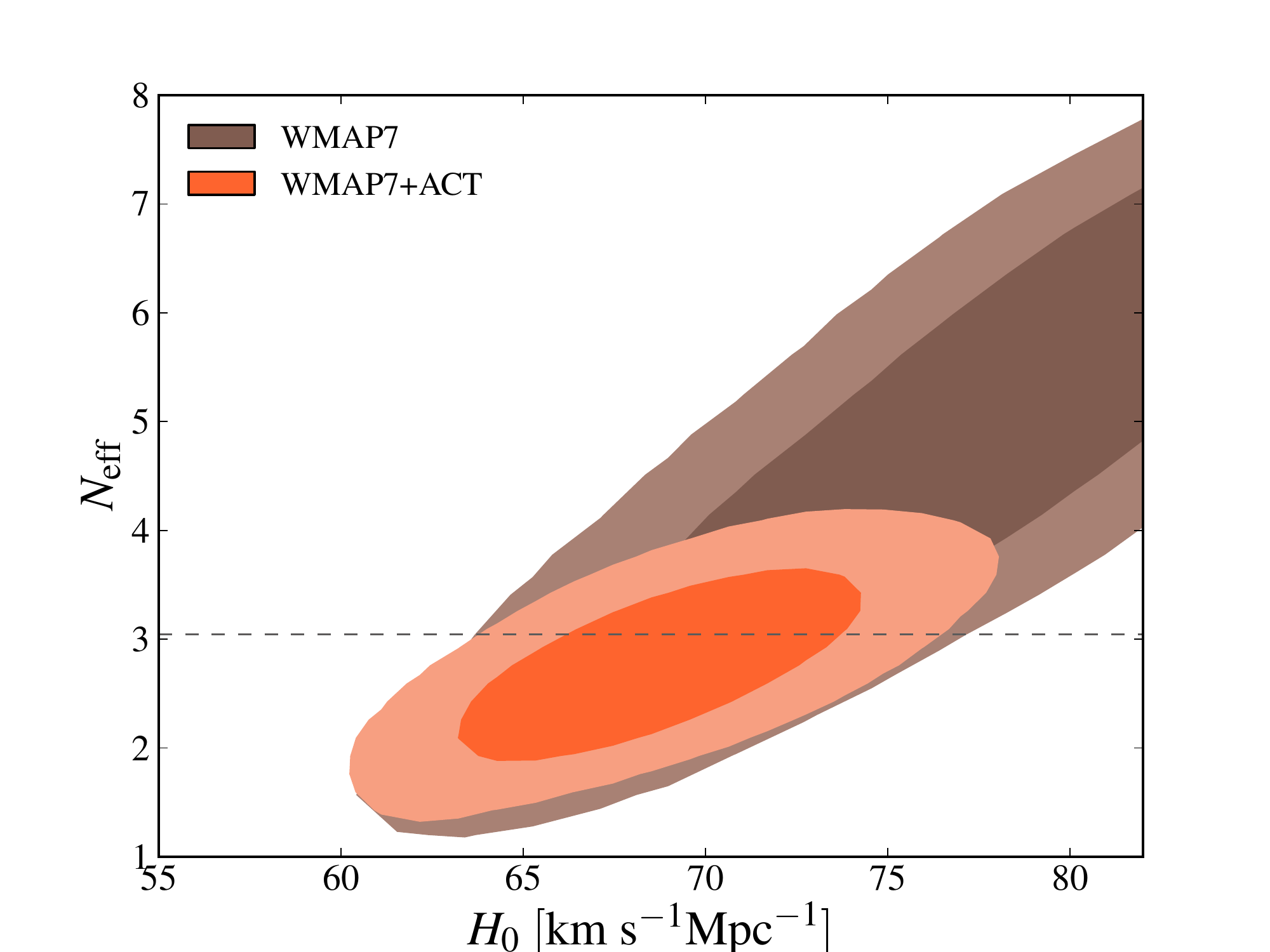}\\ [-0.45cm]
\includegraphics[scale=0.4]{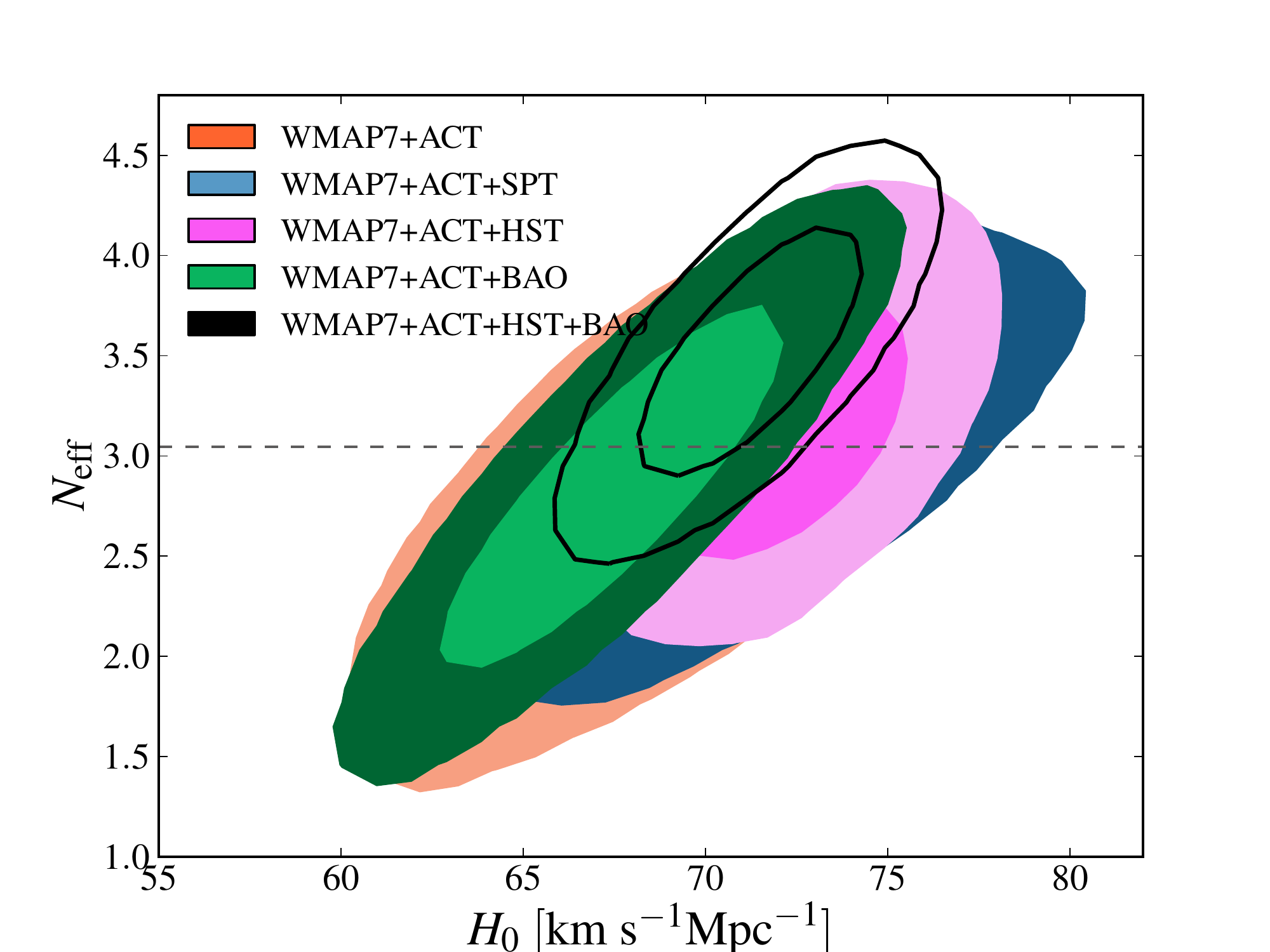}\\ [0.0cm]
 \end{array}$
 \caption{Marginalized two-dimensional ($68, 95\%$) contours in the $N_\mathrm{eff}-H_0$ plane. The two parameters are highly correlated, with larger values of $H_0$ leading to higher values of the effective number of relativistic degrees of freedom. Here BAO refers to the combined 6dF and SDSS DR9 BAO measurements, while HST is the $H_0 = 73.8 \pm2.4~\mathrm{km}~\mathrm{s}^{-1}~\mathrm{Mpc}^{-1}$ measurement from \citet{riess/etal:2011}. In contrast, the analysis of \citet{chen/ratra:2011} prefers $H_0 = 65.5 \pm5.5~\mathrm{km}~\mathrm{s}^{-1}~\mathrm{Mpc}^{-1},$ where the 95\% error bar includes both systematic and statistical errors. When combining both BAO and HST measurements, $N_\mathrm{eff}$ is pushed to higher values than in either of the individual cases, as $N_\mathrm{eff}$ tries to reconcile the differences in the distance ratio $r_s/D_V$ (where $r_s$ is the co-moving sound horizon at recombination and $D_V$ is an effective geometric distance) between the two probes. A similar trend is seen for the matter density, $\Omega_ch^2$. \label{fig:neff_h0}}
\end{center}
 \end{figure}

Various authors have explored possible models which lead to excess damping of the Silk damping tail \citep[e.g.,][]{galli/etal:2011, calabrese/etal:2011a, calabrese/etal:2011b, hasenkamp:2012,hamann:2011,menestrina/scherrer:2011,foot/etal:2011,abazajian/etal:2011,menegoni/etal:2012, farhang/etal:2012}. In this analysis we broaden our standard picture with other parameters and interpret the damping tail of the ACT data in this context. 
\bigskip
\\
\subsection{ACT data alone}
In Figure~\ref{fig:marg_cls}, we compare the ACT and WMAP7 data both separately and together with the constraints on  the $\Lambda$CDM model. This provides an important cross-check of the consistency of the two data sets. A key parameter which is primarily constrained with polarization data on the largest scales, such as those probed by WMAP, is the optical depth $\tau.$ Hence, when considering the constraints from the ACT data alone, without including WMAP7, we impose a prior on the optical depth, $\tau = 0.089 \pm 0.015.$ In Figure~\ref{fig:marg_cls} we show two cases, one in which the scalar spectral index is fixed at $n_s = 0.967$ and one in which $n_s$ is allowed to vary freely. Fixing $n_s$ tightens the bound on the amplitude of fluctuations, while other parameters are largely insensitive to the effect.
The agreement shows that the same model that describes the WMAP7 data for $\ell < 1000$ independently fits the damping tail measurement from ACT of $\ell > 500.$ The same behavior is observed in \citet{story/etal:2012}.

\begin{figure}[htbp!]
\begin{center}
$\begin{array}{@{\hspace{-0.0in}}l}
\includegraphics[scale=0.4]{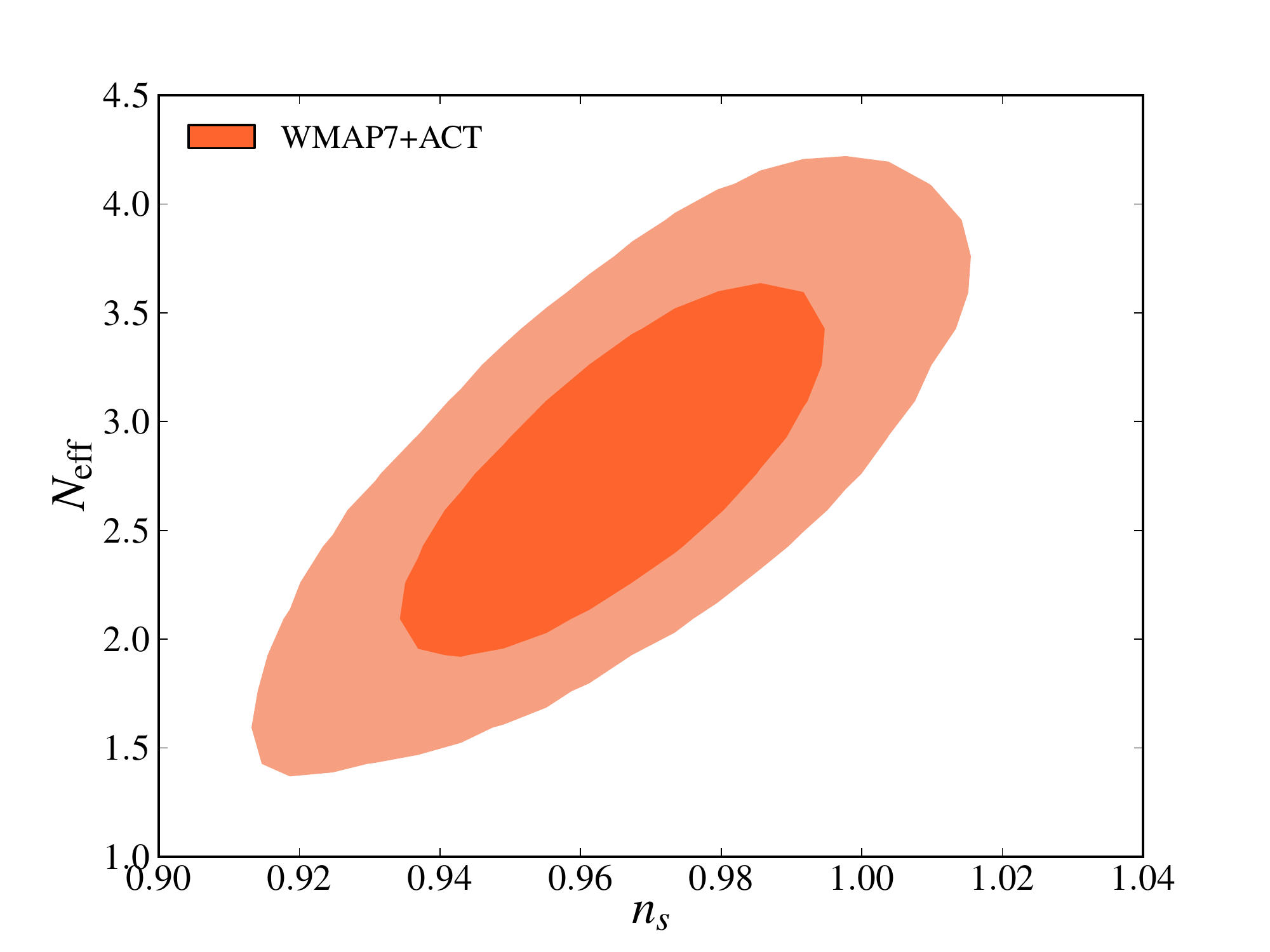}
 \end{array}$
 \caption{Marginalized two-dimensional ($68, 95\%$) contours in the $N_\mathrm{eff}-n_s$ plane. The two parameters are strongly correlated, with larger values of $n_s$ leading to higher values of the effective number of relativistic degrees of freedom. This figure is based on the 16 parameter fit of the standard cosmology and including variations in $N_\mathrm{eff}$. \label{fig:neff_ns}}
\end{center}
 \end{figure}

\subsection{Effective number of relativistic species}
\label{sec:neff}
The standard cosmological model has three neutrino species, all of which have negligible mass and contribute to $N_\mathrm{eff}$, the effective number of relativistic species at
recombination.\footnote{The effective number of relativistic species due to three neutrinos is slightly larger than three even in the standard scenario due to heating caused by the injection of the entropy from the $e^+/e^-$ annihilation \citep[e.g.,][]{dicus/etal:1982,rana:1991, dodelson/etal:1992, dolgov/fukugita:1992,hannestad/madsen:1995,dolgov/etal:1997, gnedin/eta:1997, lopez/etal:1998}.} Precision electroweak measurements place tight constraints on the number of light neutrino species with standard-model couplings  \citep{aleph:2005_zboson} through Z production in $e^+e^{-}$ collisions:
\begin{equation}
N_\mathrm{eff} = 2.984 \pm 0.008.
\end{equation}

Relativistic species (whether neutrinos or other early relativistic species) change the expansion rate of the universe through their energy density and impact the perturbations in the early universe, affecting the damping tail of the primary CMB spectrum \citep{bowen/etal:2002, bashinsky/seljak:2004,hou/etal:2011}. In the case of neutrinos, the energy density $\rho_\nu$ is lower than that of photons by a factor 
\be
\rho_\nu/\rho_\gamma = (7/8) (4/11)^{4/3} N_\mathrm{eff} \,,
\ee where $N_\mathrm{eff}$ is 3.046 in the standard $\Lambda$CDM model. 

Extra relativistic energy density damps the small-scale CMB power -- see \citet{hou/etal:2011} for a concise recent review, and the discussions in \citet{hu/dodelson:2002, hu/etal:2001, bashinsky/seljak:2004, tegmark:2005,lesgourgues/pastor:2006,hannestad:2010}.

Figure~\ref{fig:neff_like} and Table~\ref{table:wa_damp_params} illustrate the constraints on the $N_\mathrm{eff}$ from ACT in combination with various probes. Previous analyses \citep{dunkley/etal:2011, keisler/etal:2011} suggested a slight excess in the $N_\mathrm{eff}$. This preference for more damping from extra relativistic degrees of freedom is no longer present when analyzing the ACT 3-year data in combination with WMAP7 data. The change is consistent with the improved statistics of the ACT 3-year data. We find
\be
N_\mathrm{eff} = 2.79 \pm 0.56 ~~(\mathrm{WMAP7+ACT}).
\ee
The improvement in this value relative to the WMAP-only constraints is shown in the top panel of Figure~\ref{fig:neff_h0}.

The result in this analysis was obtained by imposing the consistency relation between the primordial helium fraction at Big Bang Nucleosynthesis (BBN) and the number of effective relativistic degrees of freedom \citep[see Section~\ref{sec:helium}]{trotta/hansen:2004, kneller/steigman:2004,steigman:2007,simha/steigman:2008}:
\begin{eqnarray}
Y_P &= &0.2485+0.0016[(273.9\Omega_bh^2- 6)+100(S -1)]; \nonumber \\
&&S =\sqrt{1 + (7/43)(N_\mathrm{eff} -3}). \nonumber \\
\label{eq:bbn_consistency}
\end{eqnarray}
Hence, in the present analysis, the helium fraction is a determined parameter given $N_\mathrm{eff}$ and the baryon density, rather than remaining fixed at the standard value of $Y_P = 0.24.$
The value presented in \citet{dunkley/etal:2011} was higher at $N_\mathrm{eff} = 5.3 \pm 1.3,$ as we did not impose this relation;  imposing this constraint on the previous ACT-S data would yield a modified value of $N_\mathrm{eff} = 4.3 \pm1.3,$ consistent at $1\sigma$ with the value of 3.046 expected in standard $\Lambda$CDM.

The value of $N_\mathrm{eff}$ obtained when using only the ACT-S data is closer to the value reported in \citet{dunkley/etal:2011}. The ACT-E data prefer a lower value for $N_\mathrm{eff}$, leading to a combined result which is $\approx 1\sigma$ lower than presented in \citet{dunkley/etal:2011}.

Including the recent BAO data does not change the constraints on the relativistic species, while adding in the SPT data shifts the mean value to slightly higher values of $N_\mathrm{eff},$ but still consistent with the WMAP7+ACT data. The highest value for $N_\mathrm{eff}$ comes from the addition of the BAO and Hubble constant data, yielding
 \be N_\mathrm{eff} = 3.50\pm0.42 ~~(\mathrm{WMAP7+ACT+BAO+HST}).\ee
This is due to the degeneracy between $\Omega_ch^2$ (and therefore the Hubble constant in a flat universe) and the relativistic species, shown in Figure~\ref{fig:neff_h0}. In a flat universe, higher $H_0$ leads to lower $\Omega_ch^2,$ which increases the power on medium to small scales $\ell \gtsim 200$ (as the radiation driving of the acoustic oscillations is reduced), leading in turn to a larger value of $N_\mathrm{eff}$ needed to damp power in the tail of the spectrum. The mild tension between the $H_0$ inferred through BAO distance measurements and the Hubble constant measurements leads to a value of $N_\mathrm{eff}$ which is higher than the constraint when only the Hubble constant is added to the CMB data. A similar trend was seen in the recently released SPT results \citep{hou/etal:2013}. Smaller values of the Hubble constant prefer lower values of $N_\mathrm{eff}$ \citep[e.g.,][]{chen/ratra:2011,calabrese/etal:2012}. The results are summarized in Table~\ref{table:wa_damp_params_datasets}.

In addition, the correlation between the scalar spectral index and the $N_\mathrm{eff}$ is shown in Figure~\ref{fig:neff_ns}. Decreasing power at small scales (through increasing $N_\mathrm{eff})$ is compensated by increasing the scalar spectral index, which increases small-scale power. 

Finally, we also consider an additional constraint from the recent ACT measurement of the skewness induced by the tSZ effect \citep{wilson/etal:2012}.  The tSZ skewness signal is more sensitive to $\sigma_8$ than any other cosmological parameter (scaling approximately as $\sigma_8^{11}$), allowing for a tight constraint with few degeneracies.  Using theoretical calculations similar to those in \citet{wilson/etal:2012}, we find that the most significant degeneracy is with $\Omega_b h^2$, for which the tSZ skewness scales approximately as $(\Omega_b h^2)^{3.3}$.  Thus, we include the $\sigma_8$ constraint from \citet{wilson/etal:2012} in the following form
\begin{equation}
\sigma_8 (\Omega_b h^2/0.0226)^{0.3} = 0.79 \pm 0.03, \label{eq:skewness}
\end{equation}
where $0.30  =  3.3/11.1,$ and 11.1 is the fiducial scaling of the tSZ skewness value with $\sigma_8.$
In addition to its correlation with $n_s,$ the effective number of relativistic degrees of freedom is strongly correlated with $\sigma_8$: as $N_\mathrm{eff}$ increases, so does $\sigma_8,$ hence this prior lowers the effective number of relativistic degrees of freedom to $N_\mathrm{eff} =2.55 \pm 0.40.$ 

The fact that ACT resolves the higher order peaks of the CMB spectrum allows for comparison with models that allow for departure from pure free-streaming \citep[e.g.,][]{cyr-racine/sigurdson:2012}. The effect on the small-scale power of a model with dark photons which are initially coupled to dark matter and hence only start free-streaming after they decouple implies that the phase shift and amplitude suppression associated with the free-streaming
of radiation will not be uniform across all multipoles. We leave the testing of such models to future work.

\subsection{Massive neutrinos}
In the previous subsection, we estimated the number of effective relativistic species, $N_\mathrm{eff}$. If $N_\mathrm{eff}$ is assumed to arise solely from neutrinos impacting the CMB, these are also assumed to be massless in our fiducial model. However, the CMB is sensitive to the sum of neutrino masses, which is related to the energy density of massive neutrinos via \begin{equation}
\Omega_\nu h^2= \frac{\Sigma m_\nu}{93.14~\mathrm{eV}}.\end{equation}
In general there will be a degeneracy between the mass of a neutrino species and increased relativistic degrees of freedom at early times.
\begin{figure}[htbp!]
\begin{center}
$\begin{array}{@{\hspace{-0.0in}}l}
\includegraphics[scale=0.4]{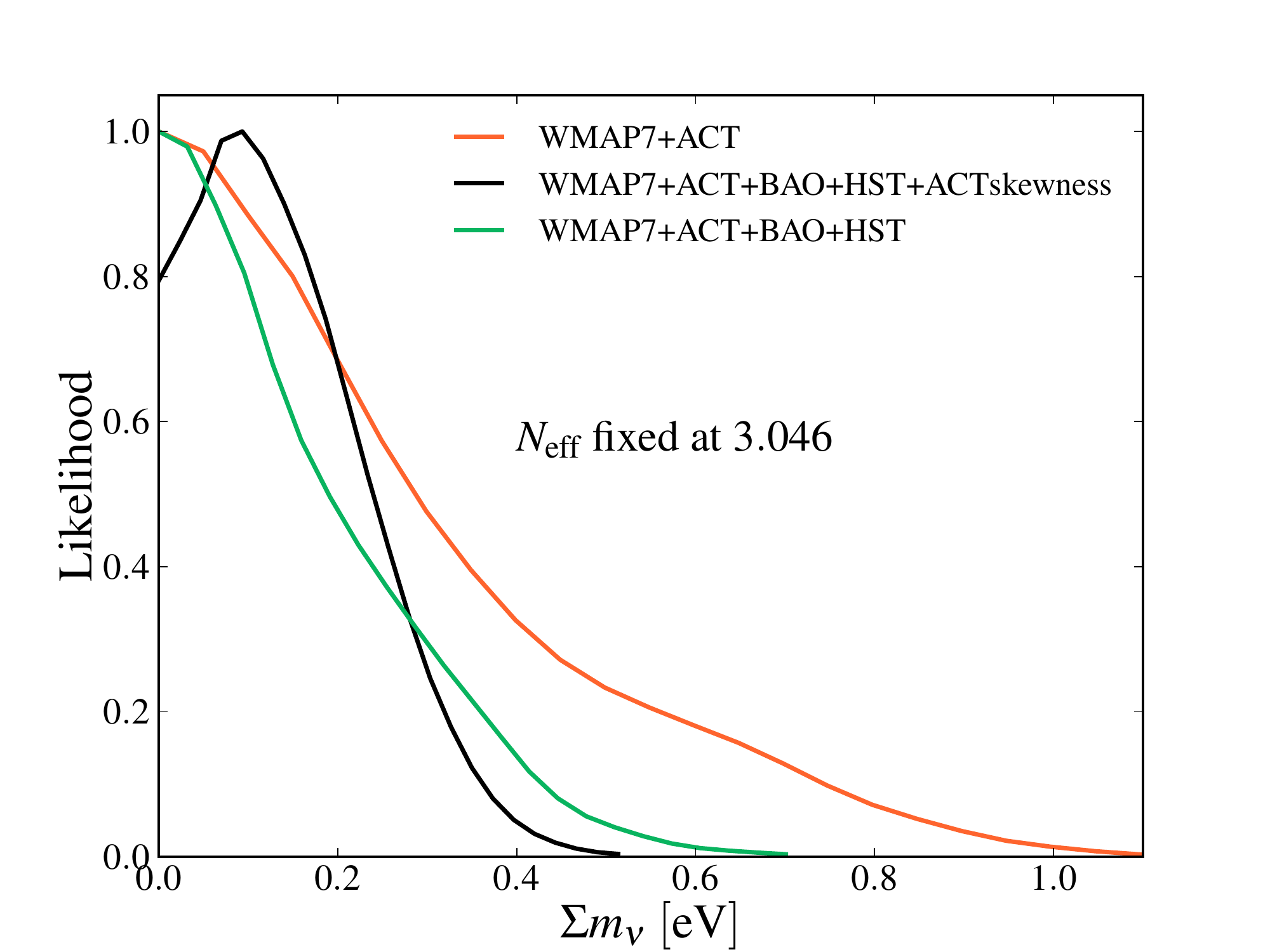}\\ [0.0cm]
 \end{array}$
 \caption{Marginalized one-dimensional likelihood for the sum of the neutrino mass, $\Sigma m_\nu$ in the case of the ACT+WMAP7 data, and with the addition of the BAO and $H_0$ constraint. In addition, we add a constraint on $\sigma_8$ from ACT skewness measurements \citep{wilson/etal:2012}. \label{fig:mnu}}
\end{center}
 \end{figure}
After the neutrinos become non-relativistic, neutrino free streaming washes out structure on small scales.  In Figure~\ref{fig:mnu} we show the constraints on the sum of the neutrino masses assuming $N_\mathrm{eff}$ is fixed to 3.046. For the ACT in combination with WMAP7 data (and keeping the number of neutrinos fixed at $N_\mathrm{eff} = 3.046$),  we find
\be
\Sigma m_\nu < 0.70~\mathrm{eV}~\mathrm{(WMAP7+ACT, 95\%~\mathrm{CL})} \label{eq:mnu_act}.
\ee
Adding in distance measurements through the BAO and Hubble constant prior (which breaks the degeneracy between $\Sigma m_\nu$ and $H_0$) improves the constraint to 
\begin{equation}
\Sigma m_\nu < 0.39~~\mathrm{eV}~(\mathrm{WMAP7+ACT+BAO}+H_0, 95\%~\mathrm{CL}) \label{eq:mnu_act_bh}.
\end{equation}
Imposing the constraint on $\sigma_8$ from the ACT skewness measurement given in Eq.~\ref{eq:skewness} yields
\begin{eqnarray}
\Sigma m_\nu&<&0.30~\mathrm{eV}\nonumber \\
&&~\mathrm{(WMAP7+ACT+ACTSkewness+BAO+}H_0, \nonumber \\
&&95\%~\mathrm{CL}).\nonumber
\\
\end{eqnarray}

Equations~(\ref{eq:mnu_act}) and~(\ref{eq:mnu_act_bh}) show that adding the ACT data strengthens the constraints by factors of 1.5 and 1.8, respectively, compared to \citet{komatsu/etal:2011}. 

\citet{hou/etal:2013} find a 3$\sigma$ preference for non-zero neutrino mass, $\Sigma m_{\nu}=0.32\pm0.11~$eV when combining WMAP7 and SPT CMB data with BAO, HST and SPT cluster constraints. This preference is driven by a combination of factors, including mild tension between the SPT and WMAP7 values of $n_s$, between SPT and BAO constraints, and between the CMB and SPT cluster constraints on $\sigma_8$, in the massless neutrino case, which reduces the error by a factor of 1.6 without changing the central value. The lack of evidence for non-zero neutrino mass in our analysis reflects the agreement between the ACT CMB, WMAP7, BAO and ACT cluster measurements when neutrino mass is fixed to zero. A more quantitative comparison is deferred to further work in light of recent Planck results. The constraint on $\sigma_8$ from ACT clusters alone in combination with WMAP7 data (and BAO+$H_0$ data) for the $\Lambda$CDM+$\Sigma m_\nu$ model \citep{hasselfield/etal:2013a} is $\Sigma m_\nu < 0.29~\mathrm{eV}$, for which $\sigma_8 = 0.802\pm0.031.$

\begin{table*} [t] 
\caption{\small{$\Lambda$CDM and extended `damping tail' parameters confidence limits from the CMB data alone: \act\ data combined with seven-year WMAP data.}}
\begin{center}
\begin{tabular}{llccccc}
\hline
\hline
&Parameter\tablenotemark{a}  & $\Lambda$CDM & $\Lambda$CDM & $\Lambda$CDM & $\Lambda$CDM  & $\Lambda$CDM \\
&   &  & + $N_{\rm eff}$ & +$
 m_\nu$&+ $\Omega_{e}$  & + $\alpha/\alpha_0$\\
\hline
\hline
Primary &$100\Omega_b h^2$ 
& \actcosmo{p1}{100omegabh2}{lcdm}{wmap7+act}{number} 
& \actcosmo{p1}{100omegabh2}{lcdm+nrel}{wmap7+act}{number} 
& \actcosmo{p1}{100omegabh2}{lcdm+mnu}{wmap7+act}{number} 
& \actcosmo{p1}{100omegabh2}{lcdm+ede}{wmap7+act}{number}
& \actcosmo{p1}{100omegabh2}{lcdm+dalpha}{wmap7+act}{number}
\\
$\Lambda$CDM&$\Omega_c h^2$ 
& \actcosmo{p1}{omegach2}{lcdm}{wmap7+act}{number} 
& \actcosmo{p1}{omegach2}{lcdm+nrel}{wmap7+act}{number} 
& \actcosmo{p1}{omegach2}{lcdm+mnu}{wmap7+act}{number} 
& \actcosmo{p1}{omegach2}{lcdm+ede}{wmap7+act}{number}
& \actcosmo{p1}{omegach2}{lcdm+dalpha}{wmap7+act}{number}
\\
&$100\theta_A$ 
& \actcosmo{p1}{thetaA}{lcdm}{wmap7+act}{number} 
& \actcosmo{p1}{thetaA}{lcdm+nrel}{wmap7+act}{number} 
& \actcosmo{p1}{thetaA}{lcdm+mnu}{wmap7+act}{number} 
& \actcosmo{p1}{thetaA}{lcdm+ede}{wmap7+act}{number}
& \actcosmo{p1}{thetaA}{lcdm+dalpha}{wmap7+act}{number}
\\
&$n_s$ 
& \actcosmo{p1}{ns}{lcdm}{wmap7+act}{number} 
& \actcosmo{p1}{ns}{lcdm+nrel}{wmap7+act}{number} 
& \actcosmo{p1}{ns}{lcdm+mnu}{wmap7+act}{number} 
& \actcosmo{p1}{ns}{lcdm+ede}{wmap7+act}{number}
& \actcosmo{p1}{ns}{lcdm+dalpha}{wmap7+act}{number}
\\
&$\tau$ 
& \actcosmo{p1}{tau}{lcdm}{wmap7+act}{number} 
& \actcosmo{p1}{tau}{lcdm+nrel}{wmap7+act}{number} 
& \actcosmo{p1}{tau}{lcdm+mnu}{wmap7+act}{number} 
& \actcosmo{p1}{tau}{lcdm+ede}{wmap7+act}{number}
& \actcosmo{p1}{tau}{lcdm+dalpha}{wmap7+act}{number}
\\
&$\log(10^{10}\Delta_{\cal R}^2)$ 
& \actcosmo{p1}{109DeltaR2}{lcdm}{wmap7+act}{number} 
& \actcosmo{p1}{109DeltaR2}{lcdm+nrel}{wmap7+act}{number}
& \actcosmo{p1}{109DeltaR2}{lcdm+mnu}{wmap7+act}{number}
& \actcosmo{p1}{109DeltaR2}{lcdm+ede}{wmap7+act}{number}
& \actcosmo{p1}{109DeltaR2}{lcdm+dalpha}{wmap7+act}{number}\\
\hline
Extended& $N_\mathrm{eff}$  & &\actcosmo{p1}{Neff}{lcdm+nrel}{wmap7+act}{number}&&& \\
&   $\Sigma m_\nu~(\mathrm{eV})$
 &&&\actcosmo{p1}{mnu}{lcdm+mnu}{wmap7+act}{number}&\\
&$\Omega_{e},w_0$ &&&&\actcosmo{p1}{ede}{lcdm+ede}{wmap7+act}{number} \\
&$w_0$ &&&&\actcosmo{p1}{w0}{lcdm+ede}{wmap7+act}{number} \\
&$\alpha/\alpha_0$ &&&&&\actcosmo{p1}{dalpha}{lcdm+dalpha}{wmap7+act}{number} \\
\hline
Derived & $\sigma_8$ 
& \actcosmo{p1}{sigma8}{lcdm}{wmap7+act}{number} 
& \actcosmo{p1}{sigma8}{lcdm+nrel}{wmap7+act}{number} 
& \actcosmo{p1}{sigma8}{lcdm+mnu}{wmap7+act}{number} 
& \actcosmo{p1}{sigma8}{lcdm+ede}{wmap7+act}{number}
& \actcosmo{p1}{sigma8}{lcdm+dalpha}{wmap7+act}{number}\\
&$\Omega_\Lambda$ 
& \actcosmo{p1}{omegal}{lcdm}{wmap7+act}{number} 
& \actcosmo{p1}{omegal}{lcdm+nrel}{wmap7+act}{number} 
& \actcosmo{p1}{omegal}{lcdm+mnu}{wmap7+act}{number} 
& \actcosmo{p1}{omegal}{lcdm+ede}{wmap7+act}{number}
& \actcosmo{p1}{omegal}{lcdm+dalpha}{wmap7+act}{number}
\\
&$\Omega_m$ 
& \actcosmo{p1}{omegam}{lcdm}{wmap7+act}{number} 
& \actcosmo{p1}{omegam}{lcdm+nrel}{wmap7+act}{number}
& \actcosmo{p1}{omegam}{lcdm+mnu}{wmap7+act}{number}
& \actcosmo{p1}{omegam}{lcdm+ede}{wmap7+act}{number}
& \actcosmo{p1}{omegam}{lcdm+dalpha}{wmap7+act}{number}\\
&$H_0~(\mathrm{km}~\mathrm{s}^{-1}~\mathrm{Mpc}^{-1})$ 
& \actcosmo{p1}{H0}{lcdm}{wmap7+act}{number} 
& \actcosmo{p1}{H0}{lcdm+nrel}{wmap7+act}{number} 
& \actcosmo{p1}{H0}{lcdm+mnu}{wmap7+act}{number} 
& \actcosmo{p1}{H0}{lcdm+ede}{wmap7+act}{number}
& \actcosmo{p1}{H0}{lcdm+dalpha}{wmap7+act}{number}\\
\hline
Secondary&$a_{\rm tSZ}$
& \actcosmo{p1}{tsz}{lcdm}{wmap7+act}{number} 
& \actcosmo{p1}{tsz}{lcdm+nrel}{wmap7+act}{number}
& \actcosmo{p1}{tsz}{lcdm+mnu}{wmap7+act}{number}
& \actcosmo{p1}{tsz}{lcdm+ede}{wmap7+act}{number}
& \actcosmo{p1}{tsz}{lcdm+dalpha}{wmap7+act}{number}\\
&$a_{\rm kSZ}$
& \actcosmo{p1}{ksz}{lcdm}{wmap7+act}{number} 
& \actcosmo{p1}{ksz}{lcdm+nrel}{wmap7+act}{number}
& \actcosmo{p1}{ksz}{lcdm+mnu}{wmap7+act}{number}
& \actcosmo{p1}{ksz}{lcdm+ede}{wmap7+act}{number}
& \actcosmo{p1}{ksz}{lcdm+dalpha}{wmap7+act}{number}\\
&$a_p$
& \actcosmo{p1}{ad}{lcdm}{wmap7+act}{number} 
& \actcosmo{p1}{ad}{lcdm+nrel}{wmap7+act}{number}
& \actcosmo{p1}{ad}{lcdm+mnu}{wmap7+act}{number}
& \actcosmo{p1}{ad}{lcdm+ede}{wmap7+act}{number}
& \actcosmo{p1}{ad}{lcdm+dalpha}{wmap7+act}{number}\\
&$a_c$  
& \actcosmo{p1}{ac}{lcdm}{wmap7+act}{number} 
& \actcosmo{p1}{ac}{lcdm+nrel}{wmap7+act}{number}
& \actcosmo{p1}{ac}{lcdm+mnu}{wmap7+act}{number}
 & \actcosmo{p1}{ac}{lcdm+ede}{wmap7+act}{number}
& \actcosmo{p1}{ac}{lcdm+dalpha}{wmap7+act}{number}\\
&$a_s$ 
& \actcosmo{p1}{as}{lcdm}{wmap7+act}{number} 
& \actcosmo{p1}{as}{lcdm+nrel}{wmap7+act}{number}
& \actcosmo{p1}{as}{lcdm+mnu}{wmap7+act}{number}
 & \actcosmo{p1}{as}{lcdm+ede}{wmap7+act}{number}
& \actcosmo{p1}{as}{lcdm+dalpha}{wmap7+act}{number}\\
&$\beta_c$ 
& \actcosmo{p1}{betac}{lcdm}{wmap7+act}{number} 
& \actcosmo{p1}{betac}{lcdm+nrel}{wmap7+act}{number}
& \actcosmo{p1}{betac}{lcdm+mnu}{wmap7+act}{number}
 & \actcosmo{p1}{betac}{lcdm+ede}{wmap7+act}{number}
& \actcosmo{p1}{betac}{lcdm+dalpha}{wmap7+act}{number}\\
&$a_{ge} $
& \actcosmo{p1}{age}{lcdm}{wmap7+act}{number} 
& \actcosmo{p1}{age}{lcdm+nrel}{wmap7+act}{number}
& \actcosmo{p1}{age}{lcdm+mnu}{wmap7+act}{number}
 & \actcosmo{p1}{age}{lcdm+ede}{wmap7+act}{number}
& \actcosmo{p1}{age}{lcdm+dalpha}{wmap7+act}{number}\\
&$a_{gs}$
& \actcosmo{p1}{ags}{lcdm}{wmap7+act}{number} 
& \actcosmo{p1}{ags}{lcdm+nrel}{wmap7+act}{number}
& \actcosmo{p1}{ags}{lcdm+mnu}{wmap7+act}{number}
 & \actcosmo{p1}{ags}{lcdm+ede}{wmap7+act}{number}
& \actcosmo{p1}{ags}{lcdm+dalpha}{wmap7+act}{number}\\
Calibration&$y{1s}$ 
& \actcosmo{p1}{cas1}{lcdm}{wmap7+act}{number} 
& \actcosmo{p1}{cas1}{lcdm+nrel}{wmap7+act}{number}
& \actcosmo{p1}{cas1}{lcdm+mnu}{wmap7+act}{number}
 & \actcosmo{p1}{cas1}{lcdm+ede}{wmap7+act}{number}
& \actcosmo{p1}{cas1}{lcdm+dalpha}{wmap7+act}{number}\\
&$y{2s}$ 
& \actcosmo{p1}{cas2}{lcdm}{wmap7+act}{number} 
& \actcosmo{p1}{cas2}{lcdm+nrel}{wmap7+act}{number}
& \actcosmo{p1}{cas2}{lcdm+mnu}{wmap7+act}{number}
 & \actcosmo{p1}{cas2}{lcdm+ede}{wmap7+act}{number}
& \actcosmo{p1}{cas2}{lcdm+dalpha}{wmap7+act}{number}\\
&$y{1e}$ 
& \actcosmo{p1}{cae1}{lcdm}{wmap7+act}{number} 
& \actcosmo{p1}{cae1}{lcdm+mnu}{wmap7+act}{number}
& \actcosmo{p1}{cae1}{lcdm+nrel}{wmap7+act}{number}
 & \actcosmo{p1}{cae1}{lcdm+ede}{wmap7+act}{number}
& \actcosmo{p1}{cae1}{lcdm+dalpha}{wmap7+act}{number}\\
&$y{2e}$ 
& \actcosmo{p1}{cae2}{lcdm}{wmap7+act}{number} 
& \actcosmo{p1}{cae2}{lcdm+nrel}{wmap7+act}{number}
& \actcosmo{p1}{cae2}{lcdm+mnu}{wmap7+act}{number}
 & \actcosmo{p1}{cae2}{lcdm+ede}{wmap7+act}{number}
& \actcosmo{p1}{cae2}{lcdm+dalpha}{wmap7+act}{number}\\
\hline
&$-2\ln {\mathscr L}$ \tablenotemark{b}
&8147
&8147
&8148
&8147
&8146
\\
\hline
\hline
\end{tabular}
\label{table:wa_damp_params}
\tablenotetext{1}{For one-tailed distributions, the upper 95\% CL is given. For two-tailed distributions the 68\% CL are shown.}
\tablenotetext{2}{The $-2\ln {\mathscr L}$ is based on the contribution from ACT (710 data points) and the WMAP7 likelihood (whose best-fit gives $-2\ln {\mathscr L}$=7478), including priors on the calibration parameters. The total degrees of freedom are therefore 8169 or 8168 if we consider the 19 or 20 parameter models.}
\tablenotetext{3}{In Table 2 of \citet{dunkley/etal:preplike}, the secondary parameters, including calibrations, are reported assuming a best-fitting cosmological model. Here we estimate the calibration factors simultaneously with the $\Lambda$CDM parameters.}
\end{center}
\end{table*}

\begin{figure}[htbp!]
\begin{center}
$\begin{array}{@{\hspace{-0.0in}}l}
\includegraphics[scale=0.4]{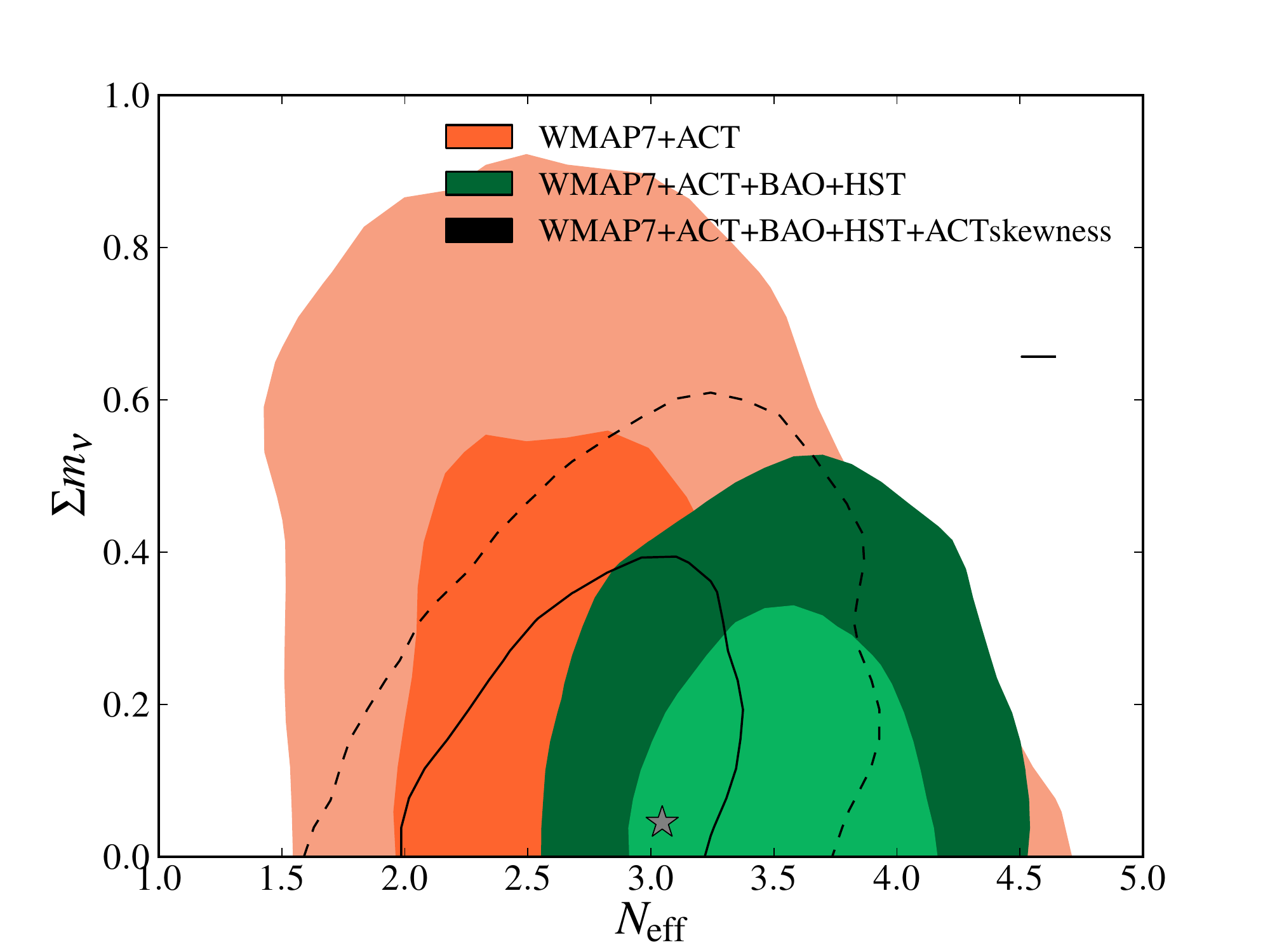}\\ [0.0cm]
 \end{array}$
 \caption{Marginalized two-dimensional ($68, 95\%$) contours in the $N_\mathrm{eff}-\Sigma m_\nu$ plane. The number of effective relativistic degrees of freedom is not strongly correlated with neutrino mass using CMB data alone; however, adding in complementary data sets helps tighten constraints. The star indicates the standard point $(N_\mathrm{eff}, \Sigma m_\nu) = (3.046, 0.045~\mathrm{eV}),$ where 0.045 eV is the minimum neutrino mass. \label{fig:neff_mnu}}
\end{center}
 \end{figure}
Figure~\ref{fig:neff_mnu} illustrates the marginalized $68\%$ and $95\%$ contours for the sum of the neutrino masses and the $N_\mathrm{eff}$ when both are varied simultaneously. Adding in the Hubble constant and BAO data pushes $N_\mathrm{eff}$ to slightly higher values, while the prior on $\sigma_8$ from ACT tSZ skewness measurements lowers the effective number of degrees of freedom. In all cases the sum of the neutrino masses is consistent with zero.
\subsection{Early dark energy}
The small-scale damping seen in the ACT data can also be interpreted as arising from a non-negligible amount of dark energy at decoupling. The early dark energy (EDE) component may be specified through its density parameter (relative to the energy required for a flat universe) $\Omega_\mathrm{de}(a)$ and an equation of state $w(a),$ given by  \citet{wetterich:1988, doran/robbers:2006} 
\ba
\Omega_\mathrm{de}(a) = \frac{\Omega_\mathrm{de}(0) - \Omega_\mathrm{e}(1-a^{-3w_0})}{\Omega_\mathrm{de}(0) + \Omega_m(0)a^{3w_0}} + \Omega_\mathrm{e}(1-a^{-3w_0}); \nonumber \\
w(a) = -\frac{1}{3(1-\Omega_\mathrm{de}(a))}\frac{d\ln \Omega_\mathrm{de}(a)}{d\ln a} + \frac{a_\mathrm{eq}}{3(a+a_\mathrm{eq})}. \nonumber \\
\ea
\begin{table*} [t] 
\caption{\small{Constraints on $N_\mathrm{eff}$ from the \act\ data combined with seven-year WMAP data and other data.}}
\begin{center}
\begin{tabular}{c|c|c|c}
\hline
\hline
Data set & $N_\mathrm{eff}$&$H_0$&$\Omega_ch^2$ \\
\hline
WMAP7+ACT& $2.79\pm0.56$&$68.8\pm3.5$&$0.109\pm0.011$\\
WMAP7+ACT+SPT&$2.98\pm 0.47$&$71.2\pm3.4$&$0.110\pm0.010$\\
WMAP7+ACT+BAO&$2.87\pm0.60$&$67.7\pm3.1$&$0.114\pm0.011$ \\
WMAP7+ACT+HST&$3.18\pm0.46$&$72.2\pm2.2$&$0.113\pm0.011$\\
WMAP7+ACT+BAO+HST&$3.50\pm0.42$&$71.2\pm2.1$&$0.124\pm0.009$\\
\hline
\end{tabular}
\label{table:wa_damp_params_datasets}
\end{center}
\end{table*}

\begin{figure}[htbp!]
\begin{center}
$\begin{array}{@{\hspace{-0.0in}}l}
\includegraphics[scale=0.4]{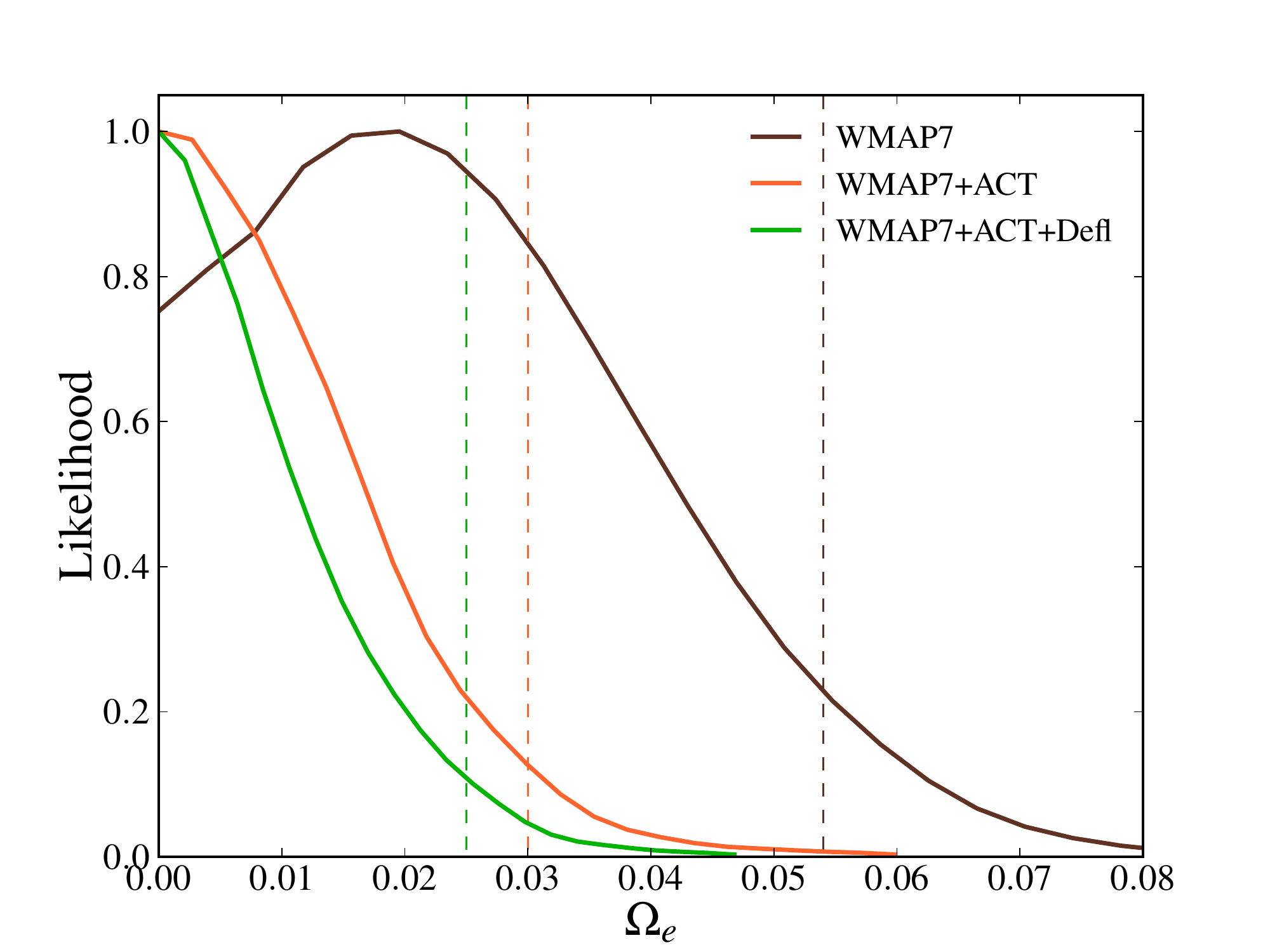}\\ [0.0cm]
 \end{array}$
 \caption{One-dimensional marginalized likelihood for the early dark energy density, $\Omega_e$. Adding in deflection measurements (shown by the inner-most curve) tightens the constraints on the allowed energy density in early dark energy. \label{fig:ede}}
\end{center}
 \end{figure}
The amounts of dark energy and matter today are given by $\Omega_\mathrm{de}(0)\equiv\Omega_\Lambda, \Omega_m(0)\equiv \Omega_c+\Omega_b$ respectively, and $a_\mathrm{eq}$ is the scale factor at matter-radiation equality. $\Omega_\mathrm{e}$ is the fraction of dark energy allowed at early times, while the present value of the equation of state of this early dark energy is expressed as $w_0$. The scaling behavior of the equation of state, tracking the dominant component at each cosmic era, gives rise to a radiation-like component at high redshifts, enhancing the damping of the small-scale CMB tail. As the amount of EDE at early times tends to zero, this model approximates the standard $\Lambda$CDM cosmological model. We place a prior that $w(a) > -1.$ This EDE model has been considered by many authors \citep{dePutter/etal:2009, hollenstein/etal:2009, dePutter/etal:2010, calabrese/etal:2010, calabrese/etal:2011a, calabrese/etal:2011b, reichardt/etal:2012} showing however that current CMB observations combined with large scale structure data have no preference for a non-zero EDE density. We include dark energy clustering as discussed in \citet{calabrese/etal:2010}, but we fix the dark energy sound speed and viscosity parameters to one and zero, respectively, as expected for a perfect fluid. Prior to this work, the most recent constraint on this model is from \citet{reichardt/etal:2012} who report an upper limit of $\Omega_\mathrm{e} < 0.018$ at $95\%$ confidence from CMB only data, combining WMAP and SPT. 
In our analysis of the ACT data in combination with WMAP7 data and the ACT deflection measurement, we obtain the upper bound 
\be 
\Omega_e < 0.025~(\mathrm{WMAP7+ACT+ACTDefl}, 95\%~\mathrm{CL}).
\ee \\
with the bound on the equation of state found to be $w_0 < -0.45$ (see Figure~\ref{fig:ede}). We do not include the combination of early dark energy and $N_\mathrm{eff}$, as the combination of these two parameters is largely unconstrained using the current small-scale CMB data.
\begin{table*} [t] 
\caption{\small{$\Lambda$CDM and extended single-model inflationary and damping parameters from CMB data alone: \act\ data combined with seven-year WMAP data.}}
\begin{center}
\begin{tabular}{llcccccc}
\hline
\hline
&Parameter\tablenotemark{a}  & $\Lambda$CDM & $\Lambda$CDM & $\Lambda$CDM & $\Lambda$CDM & $\Lambda$CDM  & $\Lambda$CDM \\
& &  & + $dn_s/d\ln k$ & + $r$ & + $Y_p$ & + $\Omega_k$\tablenotemark{b}  & + $G\mu$\\
\hline
\hline
Primary &$100\Omega_b h^2$ 
& \actcosmo{p1}{100omegabh2}{lcdm}{wmap7+act}{number} 
& \actcosmo{p1}{100omegabh2}{lcdm+run}{wmap7+act}{number} 
& \actcosmo{p1}{100omegabh2}{lcdm+tens}{wmap7+act}{number}
& \actcosmo{p1}{100omegabh2}{lcdm+yhe}{wmap7+act}{number} 
& \actcosmo{p1}{100omegabh2}{lcdm+omk}{wmap7+act}{number}
& \actcosmo{p1}{100omegabh2}{lcdm+gmu}{wmap7+act}{number}
\\
$\Lambda$CDM&$\Omega_c h^2$ 
& \actcosmo{p1}{omegach2}{lcdm}{wmap7+act}{number} 
& \actcosmo{p1}{omegach2}{lcdm+run}{wmap7+act}{number} 
& \actcosmo{p1}{omegach2}{lcdm+tens}{wmap7+act}{number}
& \actcosmo{p1}{omegach2}{lcdm+yhe}{wmap7+act}{number}
& \actcosmo{p1}{omegach2}{lcdm+omk}{wmap7+act}{number}
& \actcosmo{p1}{omegach2}{lcdm+gmu}{wmap7+act}{number}
\\
&$100\theta_A$ 
& \actcosmo{p1}{thetaA}{lcdm}{wmap7+act}{number} 
& \actcosmo{p1}{thetaA}{lcdm+run}{wmap7+act}{number} 
& \actcosmo{p1}{thetaA}{lcdm+tens}{wmap7+act}{number}
& \actcosmo{p1}{thetaA}{lcdm+yhe}{wmap7+act}{number} 
& \actcosmo{p1}{thetaA}{lcdm+omk}{wmap7+act}{number}
& \actcosmo{p1}{thetaA}{lcdm+gmu}{wmap7+act}{number}
\\
&$n_s$ 
& \actcosmo{p1}{ns}{lcdm}{wmap7+act}{number} 
& \actcosmo{p1}{ns}{lcdm+run}{wmap7+act}{number} 
& \actcosmo{p1}{ns}{lcdm+tens}{wmap7+act}{number}
& \actcosmo{p1}{ns}{lcdm+yhe}{wmap7+act}{number} 
& \actcosmo{p1}{ns}{lcdm+omk}{wmap7+act}{number}
& \actcosmo{p1}{ns}{lcdm+gmu}{wmap7+act}{number}
\\
&$\tau$ 
& \actcosmo{p1}{tau}{lcdm}{wmap7+act}{number} 
& \actcosmo{p1}{tau}{lcdm+run}{wmap7+act}{number} 
& \actcosmo{p1}{tau}{lcdm+tens}{wmap7+act}{number}
& \actcosmo{p1}{tau}{lcdm+yhe}{wmap7+act}{number} 
& \actcosmo{p1}{tau}{lcdm+omk}{wmap7+act}{number}
& \actcosmo{p1}{tau}{lcdm+gmu}{wmap7+act}{number}
\\
&$\log(10^{10}\Delta_{\cal R}^2)$ 
& \actcosmo{p1}{109DeltaR2}{lcdm}{wmap7+act}{number} 
& \actcosmo{p1}{109DeltaR2}{lcdm+run}{wmap7+act}{number} 
& \actcosmo{p1}{109DeltaR2}{lcdm+tens}{wmap7+act}{number}
& \actcosmo{p1}{109DeltaR2}{lcdm+yhe}{wmap7+act}{number}
& \actcosmo{p1}{109DeltaR2}{lcdm+omk}{wmap7+act}{number}
& \actcosmo{p1}{109DeltaR2}{lcdm+gmu}{wmap7+act}{number}\\
\hline
Extended &$dn_s/d\ln k$ & &\actcosmo{p1}{dndlnk}{lcdm+run}{wmap7+act}{number} &&& \\
&$r$ & & &\actcosmo{p1}{r}{lcdm+tens}{wmap7+act}{number}&& \\
&$Y_{p}$ & && &\actcosmo{p1}{yhe}{lcdm+yhe}{wmap7+act}{number}&\\
&$\Omega_{k}$ &&&&&\actcosmo{p1}{omk}{lcdm+omk}{wmap7+act}{number} \\
&$G\mu$ &&&&&&\actcosmo{p1}{gmu}{lcdm+gmu}{wmap7+act}{number} \\
\hline
Derived & $\sigma_8$ 
& \actcosmo{p1}{sigma8}{lcdm}{wmap7+act}{number} 
& \actcosmo{p1}{sigma8}{lcdm+run}{wmap7+act}{number} 
& \actcosmo{p1}{sigma8}{lcdm+tens}{wmap7+act}{number}
& \actcosmo{p1}{sigma8}{lcdm+yhe}{wmap7+act}{number} 
& \actcosmo{p1}{sigma8}{lcdm+omk}{wmap7+act}{number}
& \actcosmo{p1}{sigma8}{lcdm+gmu}{wmap7+act}{number}\\
&$\Omega_\Lambda$ 
& \actcosmo{p1}{omegal}{lcdm}{wmap7+act}{number} 
& \actcosmo{p1}{omegal}{lcdm+run}{wmap7+act}{number} 
& \actcosmo{p1}{omegal}{lcdm+tens}{wmap7+act}{number}
& \actcosmo{p1}{omegal}{lcdm+yhe}{wmap7+act}{number} 
& \actcosmo{p1}{omegal}{lcdm+omk}{wmap7+act}{number}
& \actcosmo{p1}{omegal}{lcdm+gmu}{wmap7+act}{number}\\
&$\Omega_m$ 
& \actcosmo{p1}{omegam}{lcdm}{wmap7+act}{number} 
& \actcosmo{p1}{omegam}{lcdm+run}{wmap7+act}{number} 
& \actcosmo{p1}{omegam}{lcdm+tens}{wmap7+act}{number}
& \actcosmo{p1}{omegam}{lcdm+yhe}{wmap7+act}{number}
& \actcosmo{p1}{omegam}{lcdm+omk}{wmap7+act}{number}
& \actcosmo{p1}{omegam}{lcdm+gmu}{wmap7+act}{number}\\
&$H_0~(\mathrm{km}~\mathrm{s}^{-1}~\mathrm{Mpc}^{-1})$ 
& \actcosmo{p1}{H0}{lcdm}{wmap7+act}{number} 
& \actcosmo{p1}{H0}{lcdm+run}{wmap7+act}{number} 
& \actcosmo{p1}{H0}{lcdm+tens}{wmap7+act}{number}
& \actcosmo{p1}{H0}{lcdm+yhe}{wmap7+act}{number} 
& \actcosmo{p1}{H0}{lcdm+omk}{wmap7+act}{number}
& \actcosmo{p1}{H0}{lcdm+gmu}{wmap7+act}{number}\\
\hline
Secondary&$a_{\rm tSZ}$
& \actcosmo{p1}{tsz}{lcdm}{wmap7+act}{number} 
& \actcosmo{p1}{tsz}{lcdm+run}{wmap7+act}{number} 
& \actcosmo{p1}{tsz}{lcdm+tens}{wmap7+act}{number}
& \actcosmo{p1}{tsz}{lcdm+yhe}{wmap7+act}{number}
& \actcosmo{p1}{tsz}{lcdm+omk}{wmap7+act}{number}
& \actcosmo{p1}{tsz}{lcdm+gmu}{wmap7+act}{number}\\
&$a_{\rm kSZ}$ 
& \actcosmo{p1}{ksz}{lcdm}{wmap7+act}{number} 
& \actcosmo{p1}{ksz}{lcdm+run}{wmap7+act}{number} 
& \actcosmo{p1}{ksz}{lcdm+tens}{wmap7+act}{number}
& \actcosmo{p1}{ksz}{lcdm+yhe}{wmap7+act}{number}
& \actcosmo{p1}{ksz}{lcdm+omk}{wmap7+act}{number}
& \actcosmo{p1}{ksz}{lcdm+gmu}{wmap7+act}{number}\\
&$a_p$
& \actcosmo{p1}{ad}{lcdm}{wmap7+act}{number} 
& \actcosmo{p1}{ad}{lcdm+run}{wmap7+act}{number} 
& \actcosmo{p1}{ad}{lcdm+tens}{wmap7+act}{number}
& \actcosmo{p1}{ad}{lcdm+yhe}{wmap7+act}{number}
& \actcosmo{p1}{ad}{lcdm+omk}{wmap7+act}{number}
& \actcosmo{p1}{ad}{lcdm+gmu}{wmap7+act}{number}\\
&$a_c$
& \actcosmo{p1}{ac}{lcdm}{wmap7+act}{number} 
& \actcosmo{p1}{ac}{lcdm+run}{wmap7+act}{number} 
& \actcosmo{p1}{ac}{lcdm+tens}{wmap7+act}{number}
& \actcosmo{p1}{ac}{lcdm+yhe}{wmap7+act}{number}
 & \actcosmo{p1}{ac}{lcdm+omk}{wmap7+act}{number}
& \actcosmo{p1}{ac}{lcdm+gmu}{wmap7+act}{number}\\
&$a_s$
& \actcosmo{p1}{as}{lcdm}{wmap7+act}{number} 
& \actcosmo{p1}{as}{lcdm+run}{wmap7+act}{number} 
& \actcosmo{p1}{as}{lcdm+tens}{wmap7+act}{number}
& \actcosmo{p1}{as}{lcdm+yhe}{wmap7+act}{number}
 & \actcosmo{p1}{as}{lcdm+omk}{wmap7+act}{number}
& \actcosmo{p1}{as}{lcdm+gmu}{wmap7+act}{number}\\
&$\beta_c$ 
& \actcosmo{p1}{betac}{lcdm}{wmap7+act}{number} 
& \actcosmo{p1}{betac}{lcdm+run}{wmap7+act}{number} 
& \actcosmo{p1}{betac}{lcdm+tens}{wmap7+act}{number}
& \actcosmo{p1}{betac}{lcdm+yhe}{wmap7+act}{number}
 & \actcosmo{p1}{betac}{lcdm+omk}{wmap7+act}{number}
& \actcosmo{p1}{betac}{lcdm+gmu}{wmap7+act}{number}\\
&$a_{ge}$
& \actcosmo{p1}{age}{lcdm}{wmap7+act}{number} 
& \actcosmo{p1}{age}{lcdm+run}{wmap7+act}{number} 
& \actcosmo{p1}{age}{lcdm+tens}{wmap7+act}{number}
& \actcosmo{p1}{age}{lcdm+yhe}{wmap7+act}{number}
 & \actcosmo{p1}{age}{lcdm+omk}{wmap7+act}{number}
& \actcosmo{p1}{age}{lcdm+gmu}{wmap7+act}{number}\\
&$a_{gs}$
& \actcosmo{p1}{ags}{lcdm}{wmap7+act}{number} 
& \actcosmo{p1}{ags}{lcdm+run}{wmap7+act}{number} 
& \actcosmo{p1}{ags}{lcdm+tens}{wmap7+act}{number}
& \actcosmo{p1}{ags}{lcdm+yhe}{wmap7+act}{number}
 & \actcosmo{p1}{ags}{lcdm+omk}{wmap7+act}{number}
& \actcosmo{p1}{ags}{lcdm+gmu}{wmap7+act}{number}\\
Calibration&$y{1s}$ 
& \actcosmo{p1}{cas1}{lcdm}{wmap7+act}{number} 
& \actcosmo{p1}{cas1}{lcdm+run}{wmap7+act}{number} 
& \actcosmo{p1}{cas1}{lcdm+tens}{wmap7+act}{number}
& \actcosmo{p1}{cas1}{lcdm+yhe}{wmap7+act}{number}
 & \actcosmo{p1}{cas1}{lcdm+omk}{wmap7+act}{number}
& \actcosmo{p1}{cas1}{lcdm+gmu}{wmap7+act}{number}\\
&$y{2s}$ 
& \actcosmo{p1}{cas2}{lcdm}{wmap7+act}{number} 
& \actcosmo{p1}{cas2}{lcdm+run}{wmap7+act}{number} 
& \actcosmo{p1}{cas2}{lcdm+tens}{wmap7+act}{number}
& \actcosmo{p1}{cas2}{lcdm+yhe}{wmap7+act}{number}
 & \actcosmo{p1}{cas2}{lcdm+omk}{wmap7+act}{number}
& \actcosmo{p1}{cas2}{lcdm+gmu}{wmap7+act}{number}\\
&$y{1e}$ 
& \actcosmo{p1}{cae1}{lcdm}{wmap7+act}{number} 
& \actcosmo{p1}{cae1}{lcdm+run}{wmap7+act}{number} 
& \actcosmo{p1}{cae1}{lcdm+tens}{wmap7+act}{number}
& \actcosmo{p1}{cae1}{lcdm+yhe}{wmap7+act}{number}
 & \actcosmo{p1}{cae1}{lcdm+omk}{wmap7+act}{number}
& \actcosmo{p1}{cae1}{lcdm+gmu}{wmap7+act}{number}\\
&$y{2e}$ 
& \actcosmo{p1}{cae2}{lcdm}{wmap7+act}{number} 
& \actcosmo{p1}{cae2}{lcdm+run}{wmap7+act}{number} 
& \actcosmo{p1}{cae2}{lcdm+tens}{wmap7+act}{number}
& \actcosmo{p1}{cae2}{lcdm+yhe}{wmap7+act}{number}
 & \actcosmo{p1}{cae2}{lcdm+omk}{wmap7+act}{number}
& \actcosmo{p1}{cae2}{lcdm+gmu}{wmap7+act}{number}\\
\hline
&$-2\ln {\mathscr L}\tablenotemark{b}$ 
&8147
&8147
&8148
&8147
&8148\tablenotemark{c}
&8149
\\
\hline
\hline
\end{tabular}
\tablenotetext{1}{For one-tailed distributions, the upper 95\% CL is given. For two-tailed distributions the 68\% CL are shown.}
\tablenotetext{2}{The $-2\ln {\mathscr L}$ is based on the contribution from ACT (710 data points) and the WMAP7 likelihood (whose best-fit gives $-2\ln {\mathscr L}$=7478), including priors on the calibration parameters. The total degrees of freedom are therefore 8169 or 8168 if we consider the 19 or 20 parameter models. }
\tablenotetext{3}{In the case of the $\Lambda$CDM + $\Omega_k$ model, both ACT power spectrum and deflection measurements are used, which increases the $-2\mathscr{L}$ value.}
\label{table:wa_params}
\end{center}
\end{table*}

\begin{table*} [t] 
\caption{\small{Extended model parameters confidence limits from the \act\ data combined with seven-year WMAP data, BAO and $H_0$ measurements.}}
\begin{center}
\begin{tabular}{llccccc}
\hline
\hline
&Parameter\tablenotemark{a}  & $\Lambda$CDM & $\Lambda$CDM & $\Lambda$CDM & $\Lambda$CDM  & $\Lambda$CDM \\
&   &  & + $N_{\rm eff}$ & +$
 m_\nu$&$r$  & + $\Omega_k$\\
\hline
\hline
Primary &$100\Omega_b h^2$ 
& \actcosmo{p1}{100omegabh2}{lcdm}{wmap7+act+bao+h0}{number} 
& \actcosmo{p1}{100omegabh2}{lcdm+nrel}{wmap7+act+bao+h0}{number} 
& \actcosmo{p1}{100omegabh2}{lcdm+mnu}{wmap7+act+bao+h0}{number} 
& \actcosmo{p1}{100omegabh2}{lcdm+tens}{wmap7+act+bao+h0}{number}
& \actcosmo{p1}{100omegabh2}{lcdm+omk}{wmap7+act+bao+h0}{number}
\\
$\Lambda$CDM&$\Omega_c h^2$ 
& \actcosmo{p1}{omegach2}{lcdm}{wmap7+act+bao+h0}{number} 
& \actcosmo{p1}{omegach2}{lcdm+nrel}{wmap7+act+bao+h0}{number} 
& \actcosmo{p1}{omegach2}{lcdm+mnu}{wmap7+act+bao+h0}{number} 
& \actcosmo{p1}{omegach2}{lcdm+tens}{wmap7+act+bao+h0}{number}
& \actcosmo{p1}{omegach2}{lcdm+omk}{wmap7+act+bao+h0}{number}
\\
&$100\theta_A$ 
& \actcosmo{p1}{thetaA}{lcdm}{wmap7+act+bao+h0}{number} 
& \actcosmo{p1}{thetaA}{lcdm+nrel}{wmap7+act+bao+h0}{number} 
& \actcosmo{p1}{thetaA}{lcdm+mnu}{wmap7+act+bao+h0}{number} 
& \actcosmo{p1}{thetaA}{lcdm+tens}{wmap7+act+bao+h0}{number}
& \actcosmo{p1}{thetaA}{lcdm+omk}{wmap7+act+bao+h0}{number}
\\
&$n_s$ 
& \actcosmo{p1}{ns}{lcdm}{wmap7+act+bao+h0}{number} 
& \actcosmo{p1}{ns}{lcdm+nrel}{wmap7+act+bao+h0}{number} 
& \actcosmo{p1}{ns}{lcdm+mnu}{wmap7+act+bao+h0}{number} 
& \actcosmo{p1}{ns}{lcdm+tens}{wmap7+act+bao+h0}{number}
& \actcosmo{p1}{ns}{lcdm+omk}{wmap7+act+bao+h0}{number}
\\
&$\tau$ 
& \actcosmo{p1}{tau}{lcdm}{wmap7+act+bao+h0}{number} 
& \actcosmo{p1}{tau}{lcdm+nrel}{wmap7+act+bao+h0}{number} 
& \actcosmo{p1}{tau}{lcdm+mnu}{wmap7+act+bao+h0}{number} 
& \actcosmo{p1}{tau}{lcdm+tens}{wmap7+act+bao+h0}{number}
& \actcosmo{p1}{tau}{lcdm+omk}{wmap7+act+bao+h0}{number}
\\
&$\log(10^{10}\Delta_{\cal R}^2)$ 
& \actcosmo{p1}{109DeltaR2}{lcdm}{wmap7+act+bao+h0}{number} 
& \actcosmo{p1}{109DeltaR2}{lcdm+nrel}{wmap7+act+bao+h0}{number}
& \actcosmo{p1}{109DeltaR2}{lcdm+mnu}{wmap7+act+bao+h0}{number}
& \actcosmo{p1}{109DeltaR2}{lcdm+tens}{wmap7+act+bao+h0}{number}
& \actcosmo{p1}{109DeltaR2}{lcdm+omk}{wmap7+act+bao+h0}{number}\\
\hline
Extended& $N_\mathrm{eff}$  & &\actcosmo{p1}{Neff}{lcdm+nrel}{wmap7+act+bao+h0}{number}&&& \\
&   $\Sigma m_\nu~(\mathrm{eV})$
 &&&\actcosmo{p1}{mnu}{lcdm+mnu}{wmap7+act+bao+h0}{number}&\\
&$r$ &&&&\actcosmo{p1}{r}{lcdm+tens}{wmap7+act+bao+h0}{number} \\
&$\Omega_k$ &&&&&\actcosmo{p1}{omk}{lcdm+omk}{wmap7+act+bao+h0}{number} \\
\hline
Derived & $\sigma_8$ 
& \actcosmo{p1}{sigma8}{lcdm}{wmap7+act+bao+h0}{number} 
& \actcosmo{p1}{sigma8}{lcdm+nrel}{wmap7+act+bao+h0}{number} 
& \actcosmo{p1}{sigma8}{lcdm+mnu}{wmap7+act+bao+h0}{number} 
& \actcosmo{p1}{sigma8}{lcdm+tens}{wmap7+act+bao+h0}{number}
& \actcosmo{p1}{sigma8}{lcdm+omk}{wmap7+act+bao+h0}{number}\\
&$\Omega_\Lambda$ 
& \actcosmo{p1}{omegal}{lcdm}{wmap7+act+bao+h0}{number} 
& \actcosmo{p1}{omegal}{lcdm+nrel}{wmap7+act+bao+h0}{number} 
& \actcosmo{p1}{omegal}{lcdm+mnu}{wmap7+act+bao+h0}{number} 
& \actcosmo{p1}{omegal}{lcdm+tens}{wmap7+act+bao+h0}{number}
& \actcosmo{p1}{omegal}{lcdm+omk}{wmap7+act+bao+h0}{number}
\\
&$\Omega_m$ 
& \actcosmo{p1}{omegam}{lcdm}{wmap7+act+bao+h0}{number} 
& \actcosmo{p1}{omegam}{lcdm+nrel}{wmap7+act+bao+h0}{number}
& \actcosmo{p1}{omegam}{lcdm+mnu}{wmap7+act+bao+h0}{number}
& \actcosmo{p1}{omegam}{lcdm+tens}{wmap7+act+bao+h0}{number}
& \actcosmo{p1}{omegam}{lcdm+omk}{wmap7+act+bao+h0}{number}\\
&$H_0~(\mathrm{km}~\mathrm{s}^{-1}~\mathrm{Mpc}^{-1})$ 
& \actcosmo{p1}{H0}{lcdm}{wmap7+act+bao+h0}{number} 
& \actcosmo{p1}{H0}{lcdm+nrel}{wmap7+act+bao+h0}{number} 
& \actcosmo{p1}{H0}{lcdm+mnu}{wmap7+act+bao+h0}{number} 
& \actcosmo{p1}{H0}{lcdm+tens}{wmap7+act+bao+h0}{number}
& \actcosmo{p1}{H0}{lcdm+omk}{wmap7+act+bao+h0}{number}\\
\hline
Secondary&$a_{\rm tSZ}$
& \actcosmo{p1}{tsz}{lcdm}{wmap7+act+bao+h0}{number} 
& \actcosmo{p1}{tsz}{lcdm+nrel}{wmap7+act+bao+h0}{number}
& \actcosmo{p1}{tsz}{lcdm+mnu}{wmap7+act+bao+h0}{number}
& \actcosmo{p1}{tsz}{lcdm+tens}{wmap7+act+bao+h0}{number}
& \actcosmo{p1}{tsz}{lcdm+omk}{wmap7+act+bao+h0}{number}\\
&$a_{\rm kSZ}$
& \actcosmo{p1}{ksz}{lcdm}{wmap7+act+bao+h0}{number} 
& \actcosmo{p1}{ksz}{lcdm+nrel}{wmap7+act+bao+h0}{number}
& \actcosmo{p1}{ksz}{lcdm+mnu}{wmap7+act+bao+h0}{number}
& \actcosmo{p1}{ksz}{lcdm+tens}{wmap7+act+bao+h0}{number}
& \actcosmo{p1}{ksz}{lcdm+omk}{wmap7+act+bao+h0}{number}\\
&$a_p$
& \actcosmo{p1}{ad}{lcdm}{wmap7+act+bao+h0}{number} 
& \actcosmo{p1}{ad}{lcdm+nrel}{wmap7+act+bao+h0}{number}
& \actcosmo{p1}{ad}{lcdm+mnu}{wmap7+act+bao+h0}{number}
& \actcosmo{p1}{ad}{lcdm+tens}{wmap7+act+bao+h0}{number}
& \actcosmo{p1}{ad}{lcdm+omk}{wmap7+act+bao+h0}{number}\\
&$a_c$  
& \actcosmo{p1}{ac}{lcdm}{wmap7+act+bao+h0}{number} 
& \actcosmo{p1}{ac}{lcdm+nrel}{wmap7+act+bao+h0}{number}
& \actcosmo{p1}{ac}{lcdm+mnu}{wmap7+act+bao+h0}{number}
 & \actcosmo{p1}{ac}{lcdm+tens}{wmap7+act+bao+h0}{number}
& \actcosmo{p1}{ac}{lcdm+omk}{wmap7+act+bao+h0}{number}\\
&$a_s$ 
& \actcosmo{p1}{as}{lcdm}{wmap7+act+bao+h0}{number} 
& \actcosmo{p1}{as}{lcdm+nrel}{wmap7+act+bao+h0}{number}
& \actcosmo{p1}{as}{lcdm+mnu}{wmap7+act+bao+h0}{number}
 & \actcosmo{p1}{as}{lcdm+tens}{wmap7+act+bao+h0}{number}
& \actcosmo{p1}{as}{lcdm+omk}{wmap7+act+bao+h0}{number}\\
&$\beta_c$ 
& \actcosmo{p1}{betac}{lcdm}{wmap7+act+bao+h0}{number} 
& \actcosmo{p1}{betac}{lcdm+nrel}{wmap7+act+bao+h0}{number}
& \actcosmo{p1}{betac}{lcdm+mnu}{wmap7+act+bao+h0}{number}
 & \actcosmo{p1}{betac}{lcdm+tens}{wmap7+act+bao+h0}{number}
& \actcosmo{p1}{betac}{lcdm+omk}{wmap7+act+bao+h0}{number}\\
&$a_{ge} $
& \actcosmo{p1}{age}{lcdm}{wmap7+act+bao+h0}{number} 
& \actcosmo{p1}{age}{lcdm+nrel}{wmap7+act+bao+h0}{number}
& \actcosmo{p1}{age}{lcdm+mnu}{wmap7+act+bao+h0}{number}
 & \actcosmo{p1}{age}{lcdm+tens}{wmap7+act+bao+h0}{number}
& \actcosmo{p1}{age}{lcdm+omk}{wmap7+act+bao+h0}{number}\\
&$a_{gs}$
& \actcosmo{p1}{ags}{lcdm}{wmap7+act+bao+h0}{number} 
& \actcosmo{p1}{ags}{lcdm+nrel}{wmap7+act+bao+h0}{number}
& \actcosmo{p1}{ags}{lcdm+mnu}{wmap7+act+bao+h0}{number}
 & \actcosmo{p1}{ags}{lcdm+tens}{wmap7+act+bao+h0}{number}
& \actcosmo{p1}{ags}{lcdm+omk}{wmap7+act+bao+h0}{number}\\
Calibration&$y{1s}$ 
& \actcosmo{p1}{cas1}{lcdm}{wmap7+act+bao+h0}{number} 
& \actcosmo{p1}{cas1}{lcdm+nrel}{wmap7+act+bao+h0}{number}
& \actcosmo{p1}{cas1}{lcdm+mnu}{wmap7+act+bao+h0}{number}
 & \actcosmo{p1}{cas1}{lcdm+tens}{wmap7+act+bao+h0}{number}
& \actcosmo{p1}{cas1}{lcdm+omk}{wmap7+act+bao+h0}{number}\\
&$y{2s}$ 
& \actcosmo{p1}{cas2}{lcdm}{wmap7+act+bao+h0}{number} 
& \actcosmo{p1}{cas2}{lcdm+nrel}{wmap7+act+bao+h0}{number}
& \actcosmo{p1}{cas2}{lcdm+mnu}{wmap7+act+bao+h0}{number}
 & \actcosmo{p1}{cas2}{lcdm+tens}{wmap7+act+bao+h0}{number}
& \actcosmo{p1}{cas2}{lcdm+omk}{wmap7+act+bao+h0}{number}\\
&$y{1e}$ 
& \actcosmo{p1}{cae1}{lcdm}{wmap7+act+bao+h0}{number} 
& \actcosmo{p1}{cae1}{lcdm+mnu}{wmap7+act+bao+h0}{number}
& \actcosmo{p1}{cae1}{lcdm+nrel}{wmap7+act+bao+h0}{number}
 & \actcosmo{p1}{cae1}{lcdm+tens}{wmap7+act+bao+h0}{number}
& \actcosmo{p1}{cae1}{lcdm+omk}{wmap7+act+bao+h0}{number}\\
&$y{2e}$ 
& \actcosmo{p1}{cae2}{lcdm}{wmap7+act+bao+h0}{number} 
& \actcosmo{p1}{cae2}{lcdm+nrel}{wmap7+act+bao+h0}{number}
& \actcosmo{p1}{cae2}{lcdm+mnu}{wmap7+act+bao+h0}{number}
 & \actcosmo{p1}{cae2}{lcdm+tens}{wmap7+act+bao+h0}{number}
& \actcosmo{p1}{cae2}{lcdm+omk}{wmap7+act+bao+h0}{number}\\
\hline
&$-2\ln {\mathscr L}$ \tablenotemark{b}
&8153
&8152
&8153
&8153
&8152
\\
\hline
\hline
\end{tabular}
\label{table:wabh_params}
\tablenotetext{1}{For one-tailed distributions, the upper 95\% CL is given. For two-tailed distributions the 68\% CL are shown.}
\tablenotetext{2}{The $-2\ln {\mathscr L}$ is based on the contribution from ACT (710 data points) and the WMAP7 likelihood (whose best-fit gives $-2\ln {\mathscr L}$=7478), including priors on the calibration parameters. The total degrees of freedom are therefore 8169 or 8168 if we consider the 19 or 20 parameter models.}
\end{center}
\end{table*}
\subsection{Inflationary parameters}
\label{sec:inflation}
Inflation provides a mechanism for the seeding of cosmological structure through small fluctuations in the early universe. We constrain the spectral index of the initial spectrum of scalar density fluctuations through the measurement of the high$-\ell$ tail of the angular power spectrum. The scalar spectrum of curvature perturbations is parameterized via \citep{kosowsky/turner:1995}
\begin{equation}
\Delta^2_\mathcal{R}(k) = \Delta^2_\mathcal{R}(k_0) \left( \frac{k}{k_0}\right)_.^{n_s (k_0)- 1 + \frac{1}{2}\ln(k/k_0)d n_s/d\ln k}
\end{equation}
We constrain the amplitude $ \Delta^2_\mathcal{R}(k_0)$, spectral index $n_s (k_0)$ and `running' of the spectrum $d n_s/d\ln k,$ defined at a pivot point $k=0.002{\rm Mpc}^{-1}.$
The amplitude of scalar perturbations is found to be \be \log(10^{10}\Delta^2_\mathcal{R}) = 3.19 \pm 0.04~\mathrm{(WMAP7+ACT)}, \ee
where the factor of $10^{10}$ is a numerical factor included to ensure robustness to numerical precision errors while sampling.
\subsubsection{The scalar spectral index}
A generic prediction of inflationary models is a nearly scale invariant spectrum; any deviations provide powerful tests of inflationary models \citep*{mukhanov/chibisov:1981,hawking:1982,starobinsky:1982,guth/pi:1982,bardeen/steinhardt/turner:1983,mukhanov/feldman/brandenberger:1992}. Deviations from scale invariance have been tested in a variety of models and contexts \citep[e.g.,][]{wang:etal/1999, tegmark/etal:2002Pk, bridle/etal:2003, hannestad:2003, martin/ringeval:2004a, martin/ringeval:2004b, sealfon/etal:2005, spergel/etal:2007, verde/peiris:2008, peiris/etal:2010, vazquez/etal:2011}, with the high-resolution of ACT expanding possibilities for direct measurements of the power spectrum \citep{hlozek/etal:2012}.  
The standard cosmological scenario in which no variation of the spectral index is allowed (i.e. $d n_s/d\ln k = 0$) provides an excellent fit to the data. 
We obtain a constraint on the scalar spectral index of 
\begin{equation}
n_s = 0.972\pm0.012~(\mathrm{WMAP7+ACT}).
\end{equation}
When  $d n_s/d\ln k$ is allowed to float, $n_s$ moves to slightly different values as shown in Figure~\ref{fig:ns_run}. 
When considering the ACT-S and ACT-E spectra separately (while still in combination with WMAP7), we find
\begin{eqnarray}
n_s \mathrm{(ACT-E)} &=& 0.981\pm  0.012;\nonumber \\
n_s \mathrm{(ACT-S)} &=& 0.966 \pm 0.012.
\end{eqnarray}

The ACT-E data specifically prefer a higher value of the ACT-S data, which is consistent with $n_s=0.962\pm0.013$, the value in \citet{dunkley/etal:2011}. However, the cosmological models that describe the two statistically independent data sets are consistent as shown in Appendix~\ref{dataconsistency}. The difference in the marginalized $n_s$ is indicative of residual correlations between parameters. Combining the two data sets increases the value of $n_s$ relative to WMAP7 alone. 
It is similarly higher than $n_s=0.9538\pm0.0081$ from WMAP7+SPT+BAO+H0 reported in \citet{story/etal:2012}.

Including BAO and $H_0$ data yields the constraint
\begin{equation}
n_s = 0.971\pm0.009~(\mathrm{WMAP7+ACT+BAO+H}_0),
\end{equation}
which rules out a scale-invariant $n_s=1$ spectrum at a significance of $3.2\sigma.$
The ACT+WMAP7 value of $n_s = 0.972\pm0.012$ is $\sim1.2\sigma$ higher than the recently released WMAP9 result $n_s = 0.9579 \pm0.0081$ \citep{bennett/etal:2012, hinshaw/etal:2012}, however this constraint includes SPT data and the improved WMAP data.

\subsubsection{$\Lambda$CDM + Running Index}
 \begin{figure}[htbp!]
\begin{center}
$\begin{array}{@{\hspace{-0.0in}}l}
\includegraphics[scale=0.4]{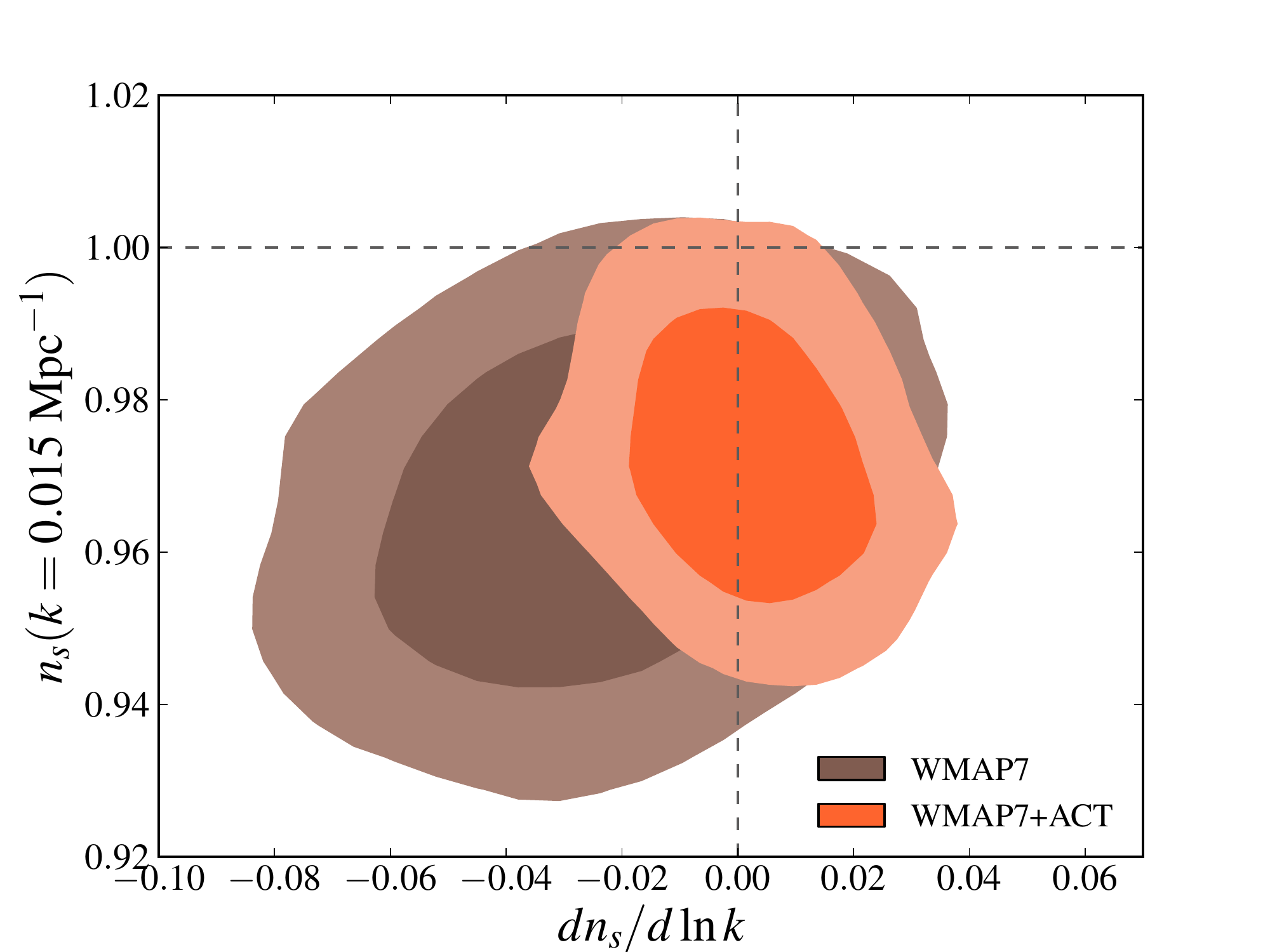}\\ [0.0cm]
 \end{array}$
 \caption{Two-dimensional marginalized limits (68\% and 95\%) for the spectral index, $n_s$, 
 plotted at the pivot point $k_0=0.015~\mathrm{Mpc}^{-1}$, and the running of the index $dn_s/d\ln k$, for ACT+WMAP7, compared to WMAP7 data alone. The data are consistent with the model $dn_s/d\ln k = 0.$ \label{fig:ns_run}}
\end{center}
 \end{figure}
In addition to the spectral index alone,  we test for deviations from a perfect power law, in the form of running of the spectral index. 
Our cosmological models are run around a pivot point of $k_0=0.002~{\rm Mpc}^{-1}$. Figure~\ref{fig:ns_run} shows the joint constraints on the spectral index and its running at a 
de-correlated pivot point $k_0=0.015~\mathrm{Mpc}^{-1}$, chosen to minimize the correlation between the two parameters \citep{cortes/liddle/mukherjee:2007}. We show the constraints for the ACT data combination with WMAP7, compared to the WMAP7 data alone. This relation between the indices at these two pivot points is given by:
\\
\\
\ba
n_s(k_0=0.015~\mathrm{Mpc}^{-1})&=&n_s(k_0=0.002~\mathrm{Mpc}^{-1}) \nonumber \\
&&+  \ln(0.015/0.002)\frac{dn_s}{d\ln k}. \nonumber \\
\ea
The ACT data are consistent with no running of the spectral index, with 
\begin{equation}
\frac{dn_s}{d\ln k} =-0.004 \pm 0.012~\mathrm{(WMAP7+ACT)}. 
\end{equation}

 \begin{table*}[htbp!]
\begin{center}
\caption{\small{Constraints on the spectral index and tensor-to-scalar ratio for different versions of the Recfast code.\label{table:ns_r}}}
\begin{tabular}{c|c|c|c}
\hline
\hline
Data set & Recfast version &$ n_s$&$r$ (95\% CL). \\
\hline
WMAP7 alone & v1.4.2 & $0.981\pm 0.020$& $< 0.32$\\
& v1.5&$0.990\pm 0.021$&$<0.39$\\
WMAP7 + ACT & v1.5&$0.980\pm0.017$&$<0.30$\\
\hline
\end{tabular}
\end{center}
\end{table*}
\begin{figure}[htbp!]
\begin{center}
$\begin{array}{@{\hspace{-0.0in}}l}
\includegraphics[scale=0.4]{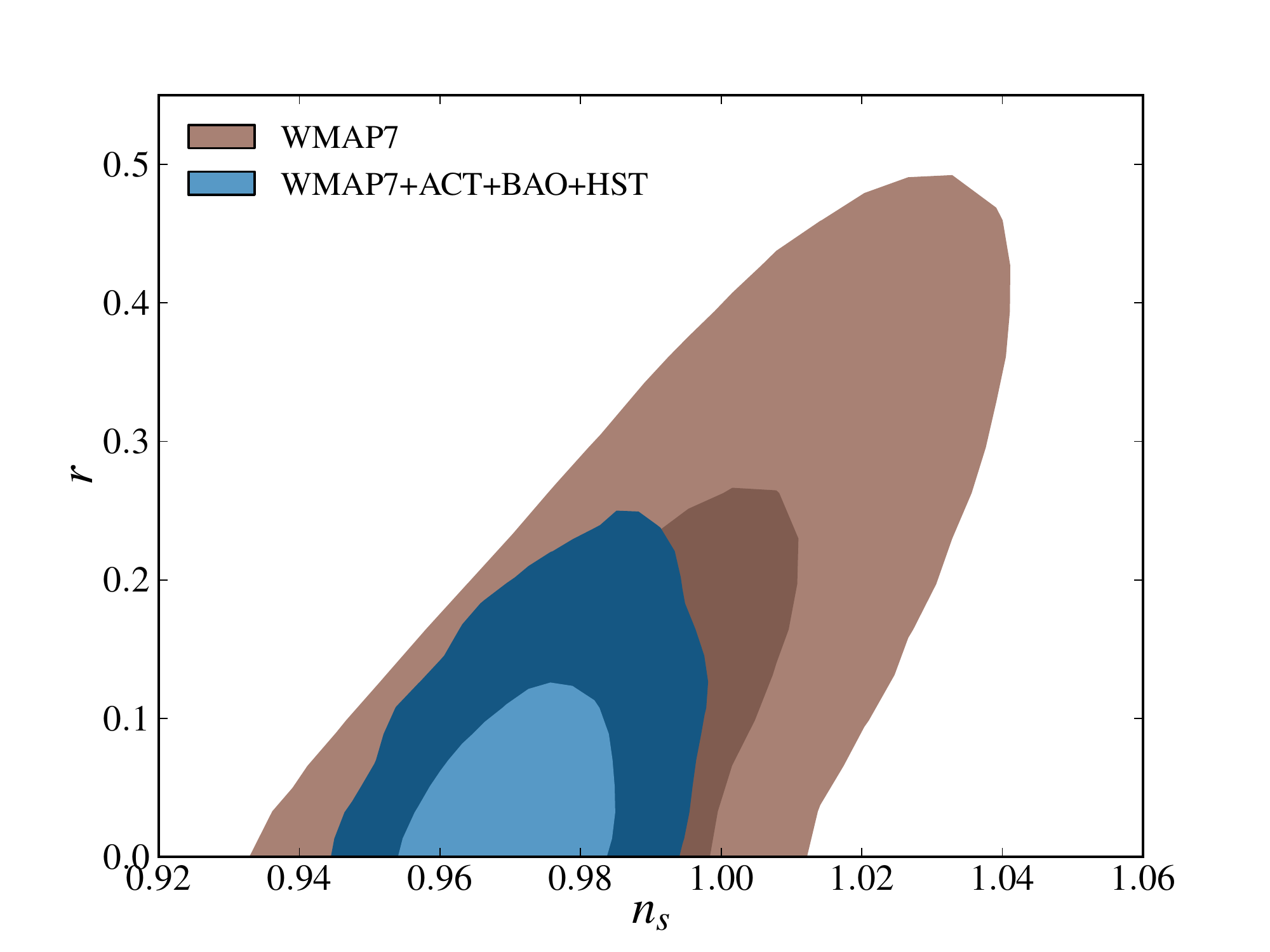}\\ [-0.2cm]
\includegraphics[scale=0.4]{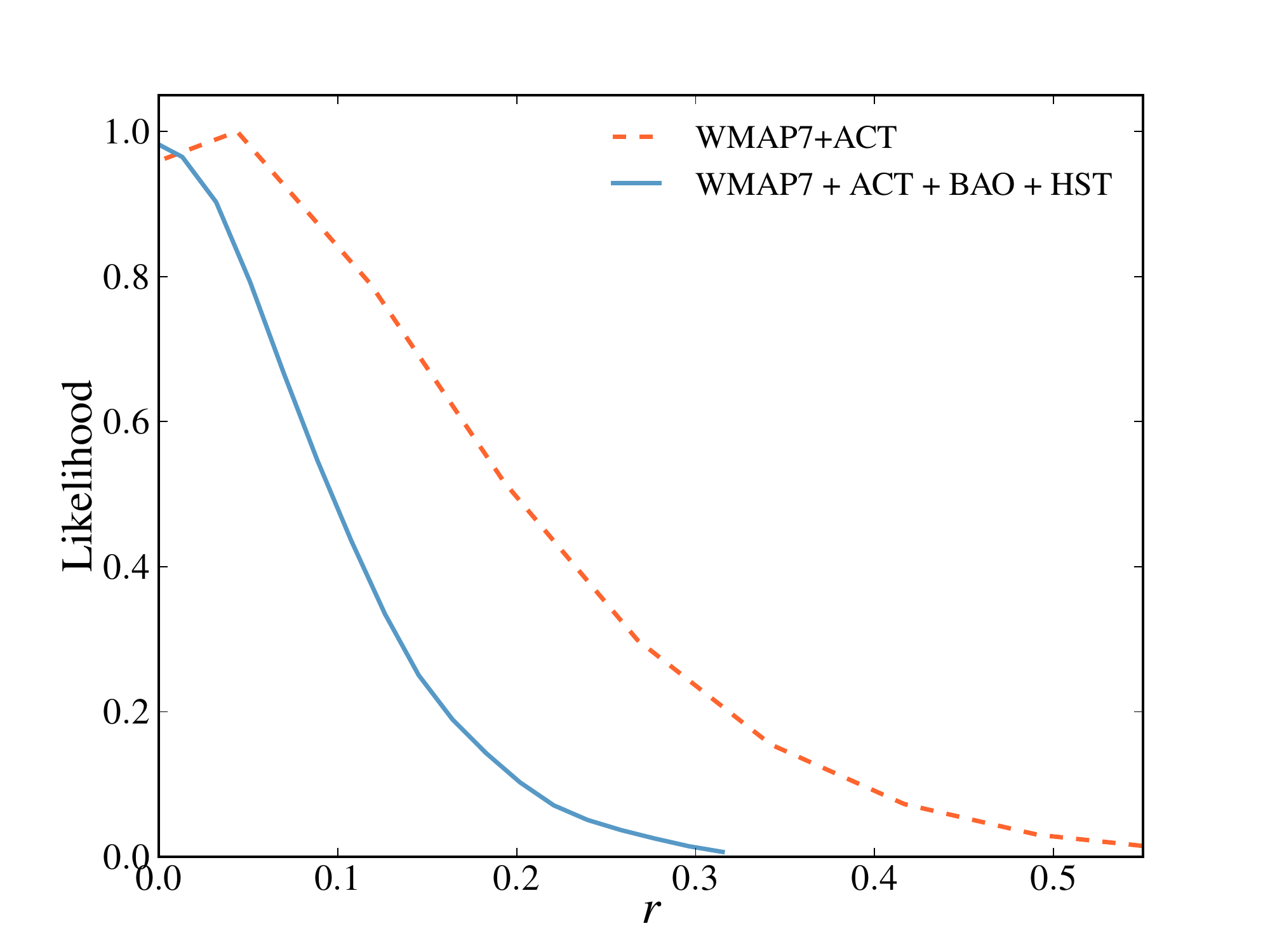}\\ [0.0cm]
 \end{array}$
 \caption{{\textit{Top panel:}}Two-dimensional marginalized distribution ($68\%$ and $95\%$ CL) for the tensor-to-scalar ration $r$ and the scalar spectral index $n_s$. The brown contours show the constraints using only WMAP7 data; the blue contours show the constraints when adding the 3-year ACT data, and BAO and $H_0$ measurements, and find the constraint $r<0.16$. Recfast v1.5 was used to compute the constraints. {\textit{Bottom panel}}: One-dimensional marginalized contours for the tensor-to-scalar ratio for the ACT data in combination with WMAP7 (dashed curve) and for the combination of WMAP7+ACT+BAO+$H_0$ (solid curve).  \label{fig:ns_tens}}
\end{center}
 \end{figure}

\subsubsection{$\Lambda$CDM+Tensor modes}
The standard $\Lambda$CDM model assumes that the initial fluctuations are purely scalar modes; however, tensor fluctuations can also be seeded at early times, generating  gravitational waves. While gravitational waves have not yet been detected, limits can be placed on the power of tensor fluctuations by constraining the so-called tensor-to-scalar ratio \be r =\Delta_h^2(k_0)/\Delta_{\cal R}^2(k_0),\ee 
where $\Delta_h^2(k_0)$ is  the amplitude of primordial gravitational waves, measured at a reference scale $k_0=0.002~\mathrm{Mpc}^{-1}$; these tensor fluctuations contribute to the CMB in both temperature and polarization.  Small-scale temperature measurements such as those from ACT help to constrain the amplitude $r$ by reducing its degeneracy with the scalar spectral index, $n_s$. 

This degeneracy arises as models with large values of $r$ have more power from tensor modes at small $\ell$ and therefore require smaller values of the spectral index to compensate. The analysis of the 300 square degree ACT 1-year data combined with WMAP7 data constrained 
$r<0.25~(95\%~{\rm CL})$ using only CMB data. 
In this 3-year analysis we find the limit on $r$ has increased, with
\be
r<0.30~(\mathrm{WMAP7+ACT}, 95\%~{\rm CL}).
\ee
We find an upper bound $\approx15\%$ higher for the tensor-to-scalar ratio than reported in \citet{dunkley/etal:2011}. This is driven by the $n_s-r$ degeneracy and the fact that the ACT 3-year constraint on $n_s$ is higher than the previously reported value. The curvature of the marginalized likelihood is shown in the bottom panel of Figure~\ref{fig:ns_tens} when only using CMB data. Once distance measurements are added the bound tightens. It is worth noting that the WMAP7-only constraints on the tensor-to-scalar ratio and scalar spectral index also differ from the published results \citep{komatsu/etal:2011}, as  can be seen by comparing Figure~\ref{fig:ns_tens} in this work to Figure 13 in \cite{larson/etal:2011}. These arise from changes in the version of Recfast used in the WMAP7 version of this particular parameter combination. These differences are summarized in Table~\ref{table:ns_r}.

When adding in data from BAO and the Hubble constant value, we obtain
\be
r<0.16~\mathrm{(WMAP7+ACT+BAO+}H_0, 95\%~{\rm CL}).
\ee
This value is generally consistent with the recently released WMAP9 results \citep{bennett/etal:2012, hinshaw/etal:2012}, which find a tighter bound of $r<0.13$ using WMAP9 data in combination with small-scale CMB spectra, BAO measurements and the Hubble constant measurement, of $r < 0.13.$
\bigskip
\bigskip
\subsection{Curvature}
One of the simplest extensions of the primary cosmological model is allowing the curvature of the universe to vary, through the relation 
\be
\Omega_k(z) = \Omega_k(1+z)^{2}, 
\ee
where $\Omega_k$ is the value of the curvature at $z = 0$. A negative sign for $\Omega_k$ implies a closed universe, while $\Omega_k > 0$ describes an open universe. \citet{guth/nomura:2012} describe how a measurement of non-zero curvature places limits not only on the geometry of the universe but also on the space of allowed models of inflation; should positive curvature be detected at the level of $\Omega_k \leq -10^{-4},$ the current paradigm of eternal inflation would be falsified, as this scenario generically predicts positive values of $\Omega_k$. A non-zero measurement of negative curvature ($\Omega_k > 0$) allows one instead to place constraints on the different possible pre-inflationary histories and the probability measures within those frameworks.

When combining the CMB data with measurements of the Hubble constant and BAO we constrain the curvature to be 
\begin{equation}
\Omega_k = -0.0020  \pm 0.0047~~\mathrm{(WMAP7+ACT+BAO+}H_0),
\end{equation}
\\
a value consistent with a flat universe, and the previously reported WMAP7 result of $\Omega_k = -0.0023^{+0.0054}_{-0.0056}$ \citep{komatsu/etal:2011}. 
 \begin{figure}[htbp!]
\begin{center}
$\begin{array}{@{\hspace{-0.0in}}l}
\includegraphics[scale=0.4]{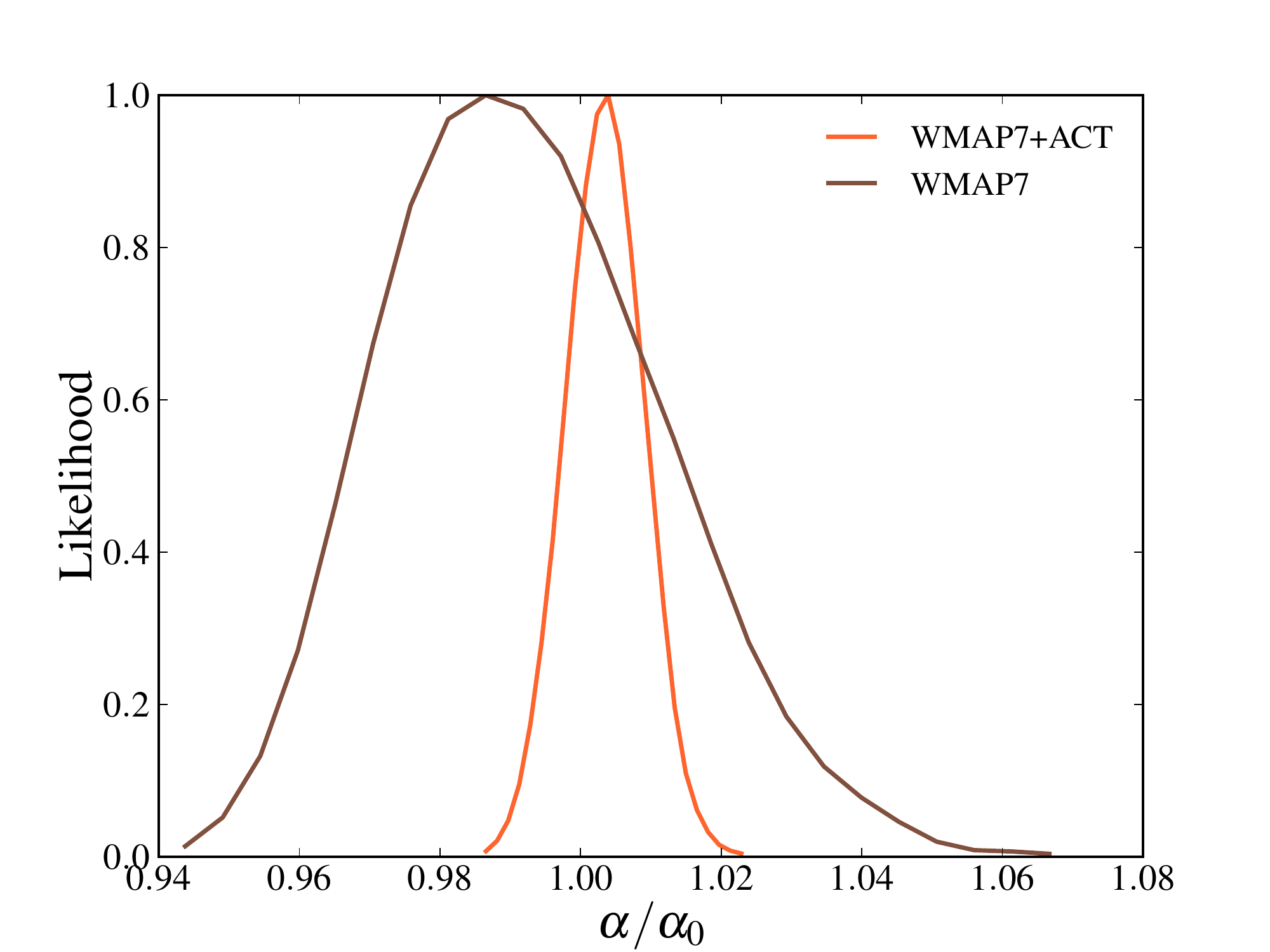}\\ [0.0cm]
 \end{array}$
 \caption{One-dimensional marginalized likelihood for $\alpha/\alpha_0$ from the ACT data in combination with WMAP7. The constraints are consistent with no variation in $\alpha$ at recombination. \label{fig:alpha}}
\end{center}
 \end{figure}

\subsection{$\Lambda$CDM + the fine structure constant $\alpha$}
Given the damping of the power spectrum observed at large multipoles, one might ask if this damping can be explained in the context of models which allow for the variation of fundamental constants, such as the fine structure constant, $\alpha = e^2/\hbar.$  Dark energy is a possible source of variation of the fine structure constant, and hence limits on the temporal variation in $\alpha$ put limits on the equation of state at late times \citep[e.g.,][]{parkinson/bassett/barrow:2003}. 
The parameter $\alpha$ mediates the strength of the electromagnetic interaction, and therefore it affects the formation of CMB anisotropies by changing the scattering rate of the Thomson scattering process as \be\dot{\tau}= x_en_ec\sigma_\mathrm{T},\ee where $\sigma_\mathrm{T} = 8\pi(\alpha/m_e)^2/3$ is the Thomson cross section for $\hbar = c = 1$ and $x_e$ is the electron fraction, itself dependent on the scale factor of the universe, $a(t)$ \citep[e.g.,][and references therein]{battye/etal:2000}. 

 We quantify this physical mechanism in terms of the ratio $\alpha/\alpha_0$, where $\alpha_0=1/(137.036)$ is the standard local value \citep{mohr/etal:2012} and $\alpha$ is the value during the recombination process.  A value $\alpha/\alpha_0 > 1$ will generate a shift in the acoustic peaks towards higher multipoles as
well as less damping. \citet{menegoni/etal:2012} allowed for a variation in $\alpha$ and illustrated the degeneracy between such variations and, for example, $N_\mathrm{eff}$, at early times. If variations in $\alpha$ arise because of some coupling of the photon to a scalar field \citep{bekenstein:2002}, in general one might expect fluctuations in $\alpha$ to be both temporal and spatial \citep{sigurdson/etal:2003}; however, for simplicity we ignore spatial variations here. In Figure~\ref{fig:alpha}, we show the constraints on the variation in $\alpha$ for the ACT+WMAP7 data combination, which give $\alpha/\alpha_0 = 1.004 \pm 0.005,$ consistent with the standard picture. 
 \begin{figure}[htbp!]
\begin{center}
$\begin{array}{@{\hspace{-0.0in}}l}
\includegraphics[scale=0.4]{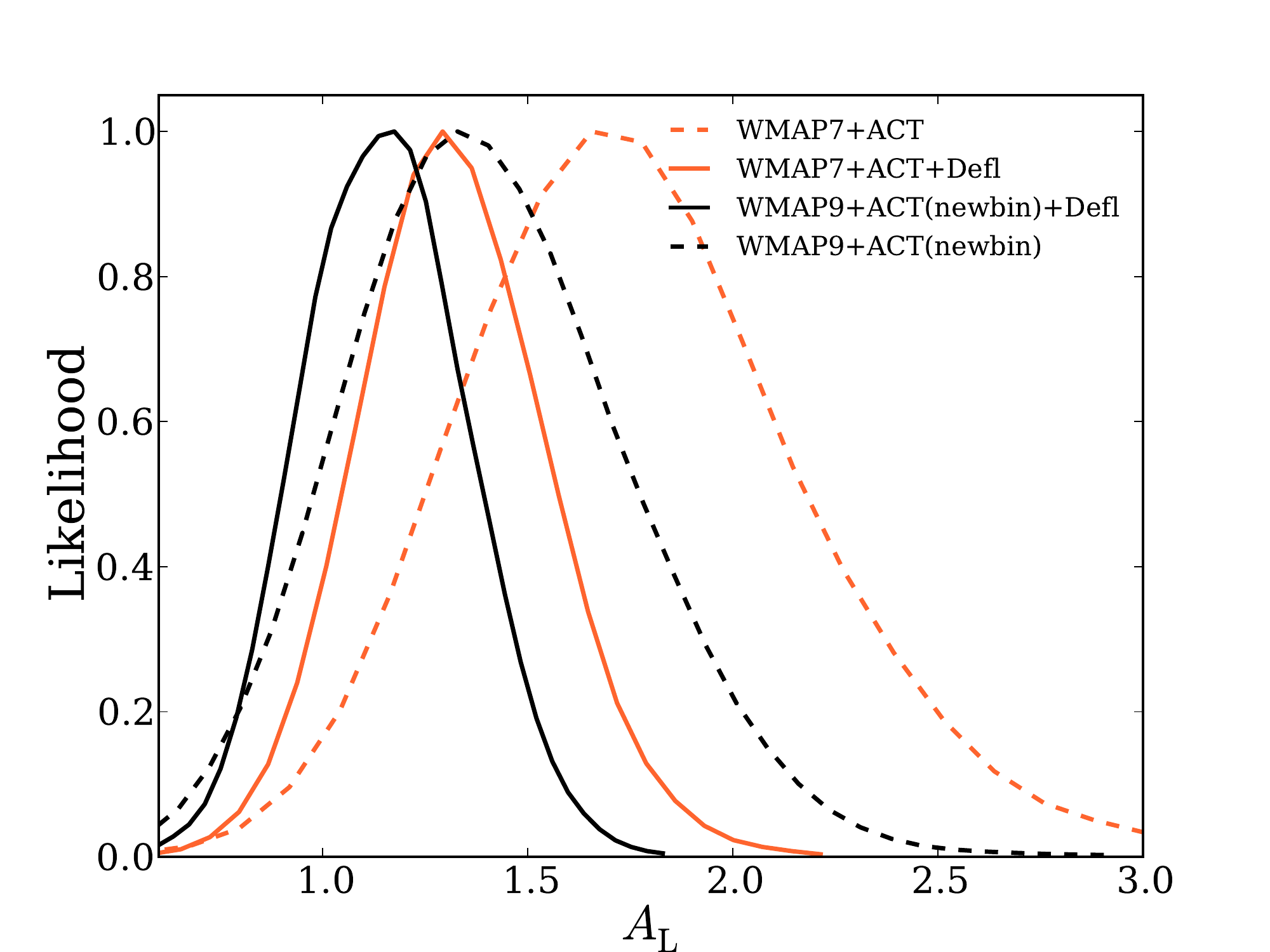}\\ [0.0cm]
 \end{array}$
 \caption{One-dimensional marginalized likelihood for the lensing parameter from WMAP7 and ACT temperature and deflection spectrum measurements. The black curves show the updated constraints when applying an alternative binning to the ACT data (designed so that the two ACT regions have similar band powers), using the beams presented in \citet{hasselfield/etal:2013b} and when including the WMAP9 data with ACT. \label{fig:lensing}}
\end{center}
 \end{figure}
\subsection{Lensing}
The lensing of the CMB photons by gravitational structure smears out the peaks and troughs of the CMB temperature spectrum, transferring power between modes. 
ACT reported the first detection of intrinsic CMB lensing \citep{das/sherwin/etal:2011} at $4\sigma$; in many models this provides evidence for dark energy from the CMB alone \citep{sherwin/dunkley/das/etal:2011}, adding further support to the $\Lambda$CDM model. The South Pole Telescope (SPT) presented a detection of the lensing deflection power spectrum at $6.3\sigma$ \citep{van_engelen/etal:2012}. The strength of the lensing can be simply parameterized through a lensing amplitude $A_L$ \citep{calabrese/etal:2008}
\begin{equation}
C_\ell^{\phi\phi,\mathrm{meas}} = A_LC_\ell^{\phi\phi,\mathrm{theory}},
\end{equation}
where $C_\ell^{\phi\phi}$ is the power spectrum of the lensing field $\phi,$ and $A_L = 1$ corresponds to the amount of lensing in the standard $\Lambda$CDM model. As this parameter multiplies the entire lensing power spectrum, it is defined over the same range in multipole space used for the theoretical temperature spectrum.
We measure the lensing amplitude from the smearing of the temperature power spectrum as
\begin{equation}
A_L = 1.7 \pm 0.38~(\mathrm{WMAP7+ACT}).
\end{equation}
When adding in the deflection spectrum measurement, this value is reduced to
\begin{equation}
A_L = 1.34 \pm 0.23~(\mathrm{WMAP7+ACT+ACTDefl}),
\end{equation}
as is shown in Figure~\ref{fig:lensing}. While this value is somewhat high, it is consistent with $A_L = 1$ at the $1.5\sigma$ level, where $A_L = 1$ is the amplitude of lensing in the $\Lambda$CDM scenario. We have tested that both the ACT-S and ACT-E patches yield the same value of $A_L$ at the $0.3\sigma$ level, and that the constraints are independent of whether a linear or logarithmic prior on $A_L$ is used. When using the WMAP9 data, and an alternate binning for the ACT data, the central value of $A_L$ shifts to $A_L=1.4\pm 0.34$ (WMAP9+ACTrebin) and $A_L = 1.15 \pm 0.21$ (WMAP9+ACTrebin+Defl), which is consistent with unity at the $1.2\sigma$ level for the TT only and the $0.75\sigma$ level including deflection. This alternative binning was designed to yield similar bandpower window functions to the ACT-S and ACT-E regions. The shift coming from the alternative binning procedure is $\sim 0.2\sigma$ for $A_L$ and less than $0.1\sigma$ for other cosmological models.

\begin{figure*}[htbp!]
\begin{center}
$\begin{array}{@{\hspace{-0.0in}}l@{\hspace{-0.0in}}l}
\includegraphics[scale=0.4]{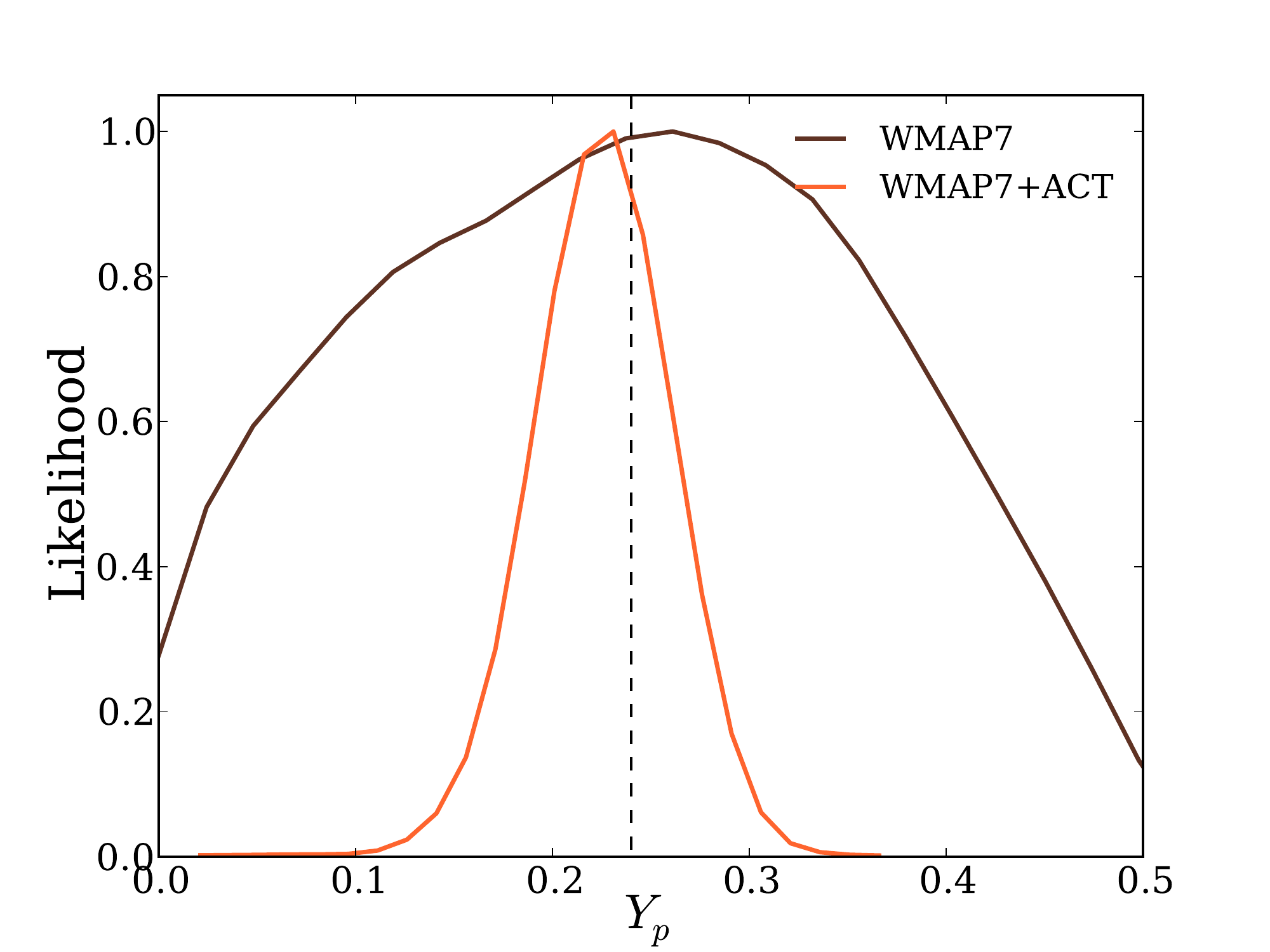}&
\includegraphics[scale=0.4]{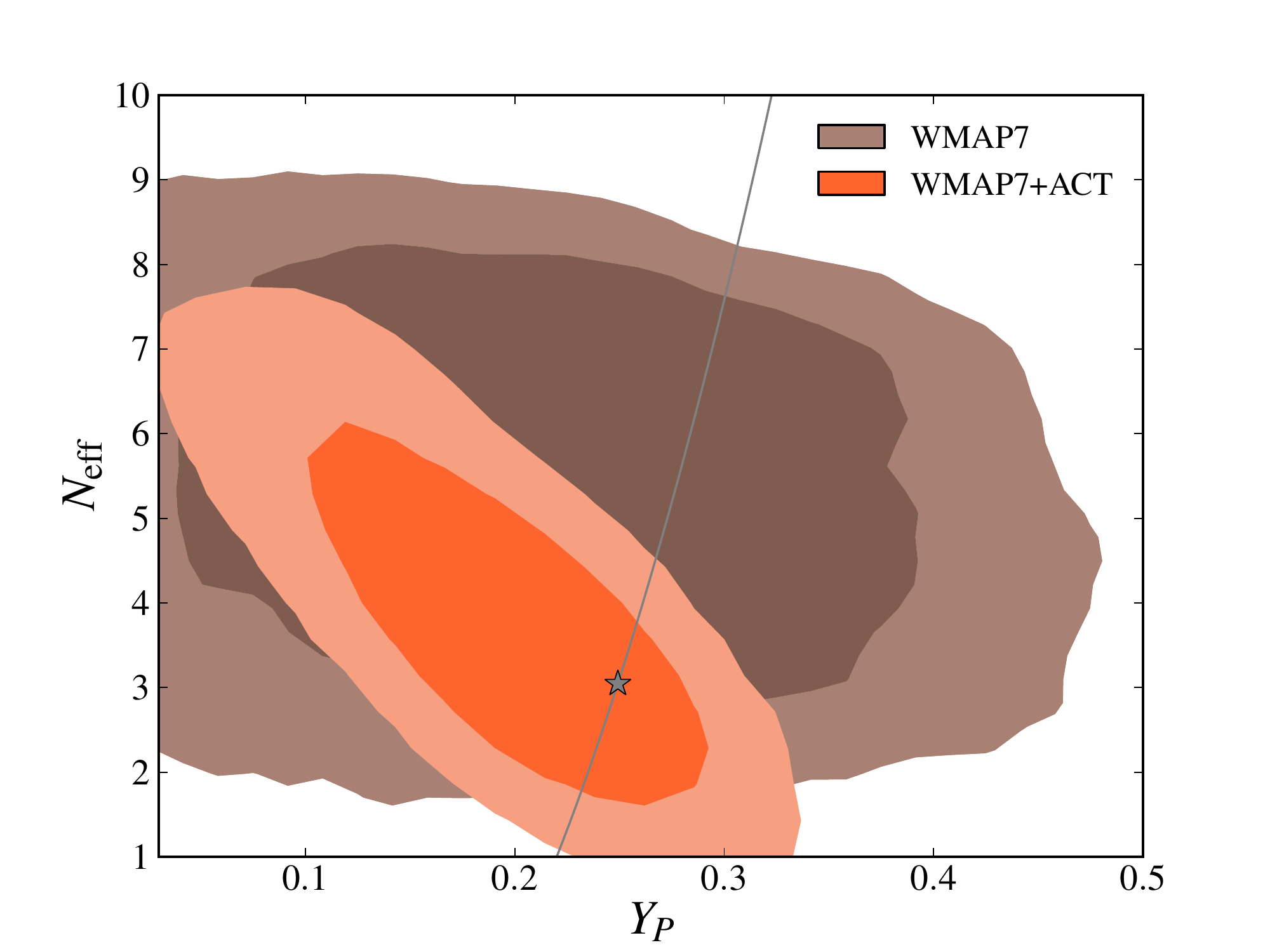}\\ [0.0cm]
 \end{array}$
 \caption{\textit{Left panel:} Marginalized one-dimensional constraints on the helium fraction $Y_p$ while fixing the number of effective relativistic degrees of freedom, $N_\mathrm{eff} =3.046.$ The dashed line indicates the standard value of $Y_p = 0.24$. \textit{Right panel:} Marginalized two-dimensional $68\%$ and $95\%$ contours for the effective number of relativistic degrees of freedom and the primordial helium abundance. The line indicates the BBN consistency relation (given in Eq.~(\ref{eq:bbn_consistency})), while the star labels the $\Lambda$CDM model with $N_\mathrm{eff}=3.046$ and $Y_p=0.24$. \label{fig:neff_yhe}}
\end{center}
 \end{figure*}

\subsection{Primordial Helium}
\label{sec:helium}
The conditions at decoupling are highly sensitive to the expansion rate of the universe at early times. Extra relativistic species change the expansion rate of the universe at a given temperature. An increased expansion rate cools the universe faster, and nuclear freeze-out occurs at an earlier time (recall that $H=1/(2t)$ at early times), when the concentration of neutrons is higher, changing the primordial abundances of light elements including $^4$He. We parameterize the primordial helium mass fraction as 
\be
Y_P = 2x,
\ee
where $x = {n_n}/(n_n+n_p)$ is the neutron abundance, $n_n$ is the number density of neutrons and $n_p$ the number density of protons \citep{peebles_book}. The abundance of primordial helium is degenerate with $N_\mathrm{eff}$ in a model with standard Big Bang Nucleosynthesis through Eq.~(\ref{eq:bbn_consistency}). The constraints from WMAP7 in combination with ACT data are:
\be
Y_p = 0.225\pm 0.034~\mathrm{(WMAP7+ACT)},
\ee
where $N_\mathrm{eff}$ is held fixed at the standard value of $N_\mathrm{eff} = 3.046$. The marginalized one-dimensional likelihood in this case is shown in the left-hand panel of Figure~\ref{fig:neff_yhe}.

When constraining $N_\mathrm{eff}$ in general, as in Section~\ref{sec:neff} and Tables~\ref{table:wa_damp_params} and~\ref{table:wabh_params}, we impose the BBN consistency relation on the helium abundance. Here we also consider the case where both parameters are allowed to vary freely. The constraints from this model are shown in the right-hand panel of Figure~\ref{fig:neff_yhe}, with the BBN consistency relation shown as a line in the plane. The model with $N_\mathrm{eff} = 3.046$ and $Y_p  =0.24$ is shown as a star.
\subsection{Cosmic strings}
While cosmic strings are no longer a viable seed of all cosmic structure, should a sub-dominant connected gas of cosmic strings exist, they would add power to the small-scale tail of the CMB power spectrum. The amplitude of diffuse power is proportional to the string tension, $G\mu$, and hence small-scale measurements provide a handle to constrain various string models. 
While direct searches for map linear string signatures in CMB maps provide interesting constraints on the string tension \citep{gott:1985, kaiser/stebbins:1984}, the simplest prescription for including the diffuse power is to model it as a network of string segments which produce loops that decay into radiation.
 
We consider a Nambu string template as used in \citet{battye/moss:2010}, modeled as a power law inversely proportional to multipole on especially small scales \citep{hindmarsh:1994}. Various authors model the string network using different approaches \citep[e.g.,][]{bennett/bouchet:1990,pogosian/vachaspati:1999,  fraisse/etal:2008,battye/moss:2010,bevis/etal:2010, mukherjee/etal:2010,urestilla/etal:2011}, where the equations of motion are solved using either Nambu, Unconnected String Model or Abelian-Higgs methods. 
Other approaches allow for more freedom \citep{foreman/etal:2011}, including the mean velocity of strings, correlation length of the string network and small-scale structure of the strings as parameters in the model.

Previous constraints on the string tension from the ACT 1-year data in combination with WMAP7 limited $G\mu < 1.6\times 10^{-7}$ \citep{dunkley/etal:2011}, while constraints from the WMAP7 data in combination with SPT \citep{dvorkin/etal:2011}  using a ``Nambu-Goto" method bound the allowed string tension to be $G\mu < 1\times10^{-7}~(95\%~{\rm CL}).$ 
In this analysis, we constrain \be
G\mu < 1.15\times10^{-7}~(\mathrm{CMB+Defl+BAO+H}_0, 95\%~\mathrm{CL}).\\
\ee 
\\
Early analyses of CMB data at ($\ell < 1000$) in the context of string models found that the model with a Harrison-Zeld'ovich spectrum with $n_s=1$ and cosmic strings was consistent with the data~\citep{bevis/etal:2008, battye/etal:2006} (and even that the six-parameter model with both strings and a scale-invariant perturbation spectrum fit the data as well as the $\Lambda$CDM model without strings), while more recent studies have shown that this model no longer provides an adequate fit to small-scale CMB data \citep{urestilla/etal:2011}. We find that fixing the primordial power to be scale invariant and combining all CMB data (WMAP7+ACT+SPT+ACTDefl) with BAO and the $H_0$ measurement yields a constraint on the string amplitude of $q_\mathrm{str} = 0.005,$ or $G\mu = 7.1\times10^{-8}~(95\%~\mathrm{CL}).$ The difference in effective $\chi^2$ values between the two models ($n_s=1 + q_\mathrm{str}$ versus $n_s$ free and no strings; both models have the same number of parameters) is $-2\ln {\mathscr L}_{\Lambda\mathrm{CDM}} +2\ln {\mathscr L}_{n_s=1,\mathrm{strings}} =32,$ indicating an $5.7\sigma$ preference for the $\Lambda$CDM model over a model with strings (using a Nambu string template) and a scale-invariant spectrum. Recent work by \citet{lizarraga/etal:2012} note that the string amplitude is highly correlated with  $N_\mathrm{eff},$ with $N_\mathrm{eff}$ being pushed towards much higher values when both the string tension and the number of effective degrees of freedom are varied together.

\section{Secondary parameters}
The assumed priors placed on the secondary parameters are discussed in detail in \citet{dunkley/etal:preplike}, where the values for assumed priors are presented and justified. We summarize and interpret our constraints here.
\label{sec:secondaries}
\subsection{Point source power}
The power spectrum includes a contribution from point sources that lie below our detection threshold.  At 148 GHz and 218 GHz, both synchrotron emission from radio galaxies and infrared emission from dusty galaxies contribute to the observed power.  Point sources detected in either band are masked down to a flux limit of 15 mJy.  We impose a prior of $a_s = 2.9\pm0.4$ at $\ell = 3000$ on the residual power from radio galaxies at 148 GHz from sources below this threshold \citep{gralla/etal:prep2013}, based on models from \citet{tucci/etal:2011}.  For comparison, \citet{reichardt/etal:2012} use a model from \citet{deZotti/etal:2005}, which would predict a residual power below 15 mJy that differs by about $5\%
$ from the prediction of the \citet{tucci/etal:2011} model, and so lies within the 1$\sigma$ limit of the prior we impose.  

For the $\Lambda$CDM model, the error on the unresolved radio galaxy power (at 150 GHz) is dominated by the prior on the amplitude of power at $\ell= 3000,$ with 
\begin{equation}
a_s = 3.1 \pm 0.4.
\end{equation}

We constrain the clustered ($a_c$) and Poisson $(a_p$) power from CIB sources respectively to be
\begin{equation}
a_c = 5.0 \pm 1.0; \qquad a_p = 7.0 \pm 0.5,
\end{equation}
where again, the amplitudes are for templates normalized to $1~\mu\mathrm{K}^2$ at $\ell=3000$ and 150 GHz. 

The frequency dependence of the CIB power is modeled as the square of a modified blackbody \citep[see Section 2.3 of][]{dunkley/etal:preplike}, with effective dust temperature fixed to $T_d=9.7$~K. The effective emissivity index is constrained to be $\beta_c=2.2\pm0.1$. The corresponding constraint on the CIB spectral index between 150 and 220~GHz is $\alpha_d=3.68\pm0.10$, where $\alpha_d$ is defined in flux density units as $S_{\nu}\propto\nu^{\alpha_d}$. This result is in good agreement with the earlier ACT 1-year constraint of $\alpha_d=3.69\pm0.14$ \citep{dunkley/etal:2011}, and somewhat higher than, though not inconsistent with, $\alpha_d=3.56\pm0.07$ (or $3.45\pm0.11$, when a tSZ-CIB correlation is allowed), reported by SPT \citep{reichardt/etal:2012}.

While $\beta_c$ cannot be directly associated with the emissivity of individual CIB sources, we expect the CIB anisotropy measured by ACT to improve dusty star-forming galaxy constraints when used in conjunction with data at higher frequencies \citep[as in][]{addison/dunkley/bond:2012}, and through ongoing cross-correlation analyses.

\bigskip
\bigskip
\subsection{Thermal SZ}
The amplitude of the thermal SZ (tSZ) power spectrum is a strong
function of the amplitude of matter
fluctuations, and is very sensitive to $\sigma_8$
\citep{komatsu/kitayama:1999}, scaling like $\sigma_8^{8.3}$ \citep{shaw/etal:2010,trac/bode/ostriker:2011}.  We model the tSZ power using a template
\citep{battaglia/etal:2011} from hydrodynamic simulations of
cosmological volumes that includes radiative cooling, star formation,
and AGN feedback. For $\sigma_8=0.8$ this template has a predicted
amplitude of $a_{\rm tSZ}=5.6$ at $\ell=3000$.
In this analysis of the ACT 3-year data we separate the thermal and
kinetic SZ components
\citep{dunkley/etal:preplike}
into two dimensionless spectra normalized to unity at $\ell_0=3000$ in
Eq.~\ref{eqn:spectra_th}.
Hence, the theoretical expectation for the tSZ amplitude at this
$\ell_0=3000$ is $5.6.$
We constrain the tSZ amplitude to be
\begin{equation}
a_\mathrm{tSZ} =  3.4\pm 1.4.
\end{equation}
Our current result corresponds to a measurement of $\sigma_8 = 0.75^{+0.03}_{-0.05}$ and is consistent with our previously reported values from $a_\mathrm{tSZ}$ of
 $\sigma_8 = 0.77 \pm 0.05$ in 
\citet{dunkley/etal:2011} (which assumed a $\sigma_8^7$ scaling), and from 
ACT measurements of the three point function, $\sigma_8 = 0.79\pm0.03$
\citep{wilson/etal:2012}.
Our tSZ constraint is also consistent with the SPT-reported value
of $a_\mathrm{tSZ} = 3.3\pm1.1$ \citep{reichardt/etal:2012}.

In the ACT analysis, using only the 1-year data, \citep{dunkley/etal:2011} we modeled the
SZ power spectrum using only a {\textit{total}} (dimensionless) amplitude of the
SZ signal (thermal plus kinetic), which was constrained relative to a theoretical spectrum with $\sigma_8 = 0.8$ as
$a_\mathrm{SZ}/a_\mathrm{SZ,\sigma_8 = 0.8} = 0.85 \pm 0.26$ using both 148 GHz and 218 GHz frequency bands, and which can
be interpreted as $\ell(\ell+1)/2\pi C_\ell^\mathrm{totSZ} =6.8 \pm
2.9~\mu\mathrm{K}^2$ at $\ell = 3000$ using the \cite{battaglia/etal:2010} template.

However, the connection between a tSZ amplitude and constraints on $\sigma_8$ depends on the modeling of the cluster physics. For a fixed
value of $\sigma_8$ the predicted amplitude of the tSZ power spectrum
at $\ell=3000$ varies by tens of percent depending on the tSZ model
\citep[e.g.,][]{battaglia/etal:2010,shaw/etal:2010,trac/bode/ostriker:2011}. All
the direct measurements of cluster physics are limited to high-mass, low-redshift clusters \citep[e.g.][]{planck:2012clusters}, which do not
constrain tSZ models, since around half of the contribution to tSZ
amplitude at $\ell = 3000$ comes from lower mass and higher redshift
clusters and groups \citep[e.g.,][]{trac/bode/ostriker:2011,battaglia/etal:2011}.

The uncertainty in the tSZ model leads to a systematic error in $\sigma_8$, which is not included in the $\sigma_8$
error bar. Combining the power spectrum measurements with higher order
correlations and cluster number counts will provide tighter constraints on
$\sigma_8$ and the tSZ models \citep{battacharya/etal:2012, hill/sherwin:2012}.

\subsection{${tSZ}$-CIB correlations}
\begin{figure}[htbp!]
\begin{center}
$\begin{array}{@{\hspace{-0.0in}}l}
\includegraphics[scale=0.4]{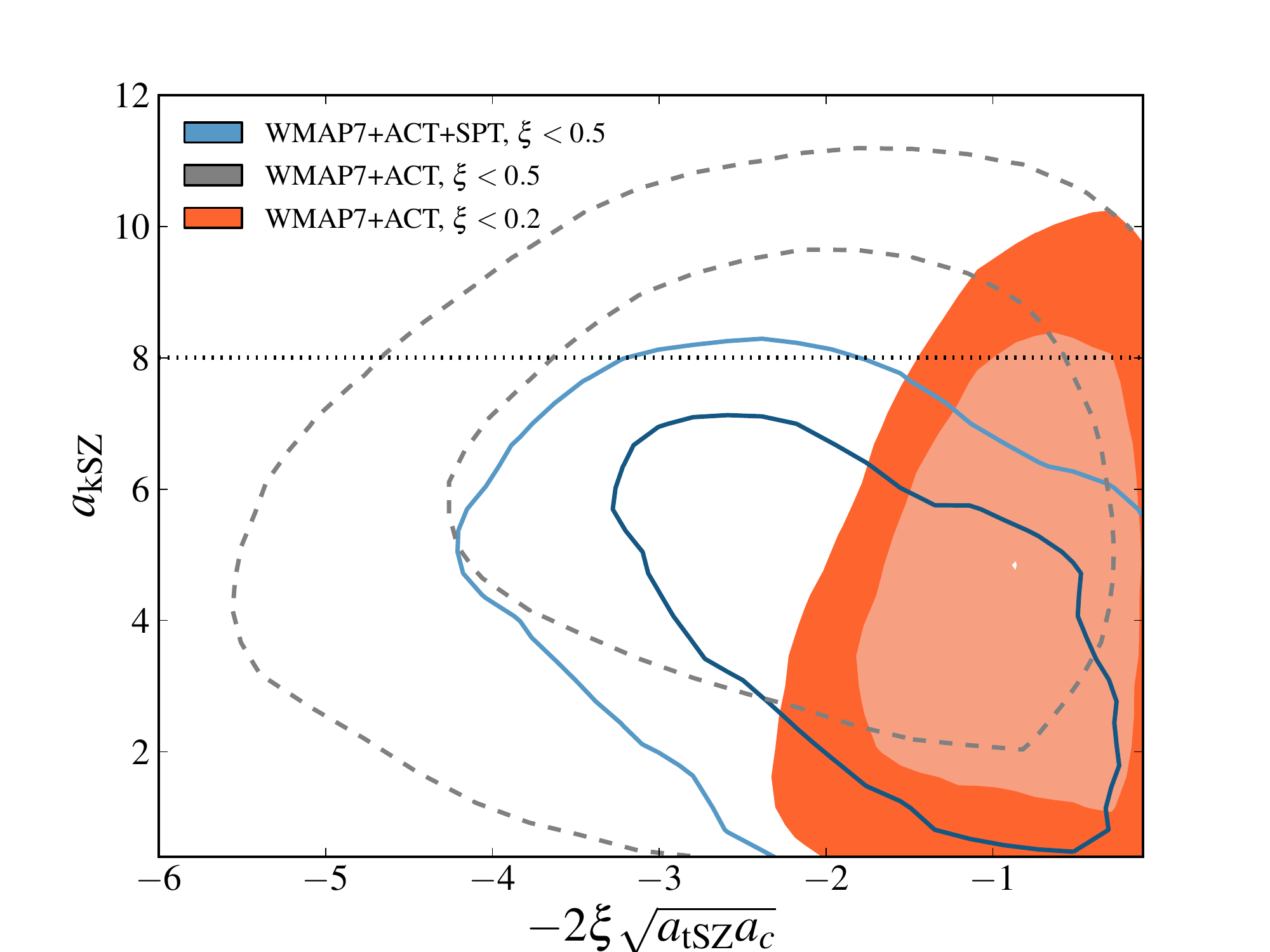}\\ [0.0cm]
 \end{array}$
 \caption{Marginalized two-dimensional $68\%$ and $95\%$ contours for
the amplitude of the kSZ power and the power in the tSZ-CIB
correlation for the combinations of ACT+WMAP7 and ACT+WMAP7+SPT. The filled contours are for the prior $\xi < 0.2,$ while the unfilled contours show the degrading of constraints when weakening the prior to $\xi < 0.5.$ The dotted line shows the limit on $a_\mathrm{kSZ}$ for $\xi = 0.$\label{fig:ksz_cib}}
\end{center}
 \end{figure}

The tSZ-CIB correlation term is important in interpreting the power from the
kSZ effect, as the frequency dependence of the tSZ-CIB and kSZ are
similar across the ACT and SPT frequency channels \citep[see e.g.,
Figure 5 of][]{addison/dunkley/spergel:2012}. A larger value of $\xi$
removes more power at $\ell\sim3000$, which can be compensated for by an
increase in the kSZ amplitude. As shown in
Figure~\ref{fig:lcdm_params}, $\xi$ is largely unconstrained by the
ACT data, and the choice of the $\xi < 0.2$ prior has little impact on
the cosmological analysis. Adopting a wider prior of $\xi<0.5$,
estimated as the upper limit from the modeling by
\cite{addison/dunkley/spergel:2012}, degrades the kSZ amplitude
constraint from $a_\mathrm{kSZ} < 8.6$ to $a_\mathrm{kSZ} <
9.4~(95\%~\mathrm{CL})$, shown by the unfilled, dashed contours in
Figure~\ref{fig:ksz_cib}. The kSZ constraint from fixing $\xi=0$ is
$a_\mathrm{kSZ} <8.0$, indicated by the dotted horizontal line in
Figure~\ref{fig:ksz_cib}. The anti-correlation between the tSZ-CIB and
kSZ power is more apparent in the joint fit with the SPT spectra.

\subsection{Kinematic SZ}

The kSZ power spectrum can be separated into two components.
The first is a low-redshift component (which we refer to as the homogenous kSZ) that is sourced by perturbations in the free
electron density after reionization with peculiar velocities;
on large scales this is known as the Ostriker-Vishniac (OV) effect
\citep{ostriker/vishniac:1986,ma/fry:2002,sunyaev/zeldovich:1970}.
  \begin{figure}[htbp!]
\begin{center}
$\begin{array}{@{\hspace{-0.0in}}l}
\includegraphics[scale=0.5]{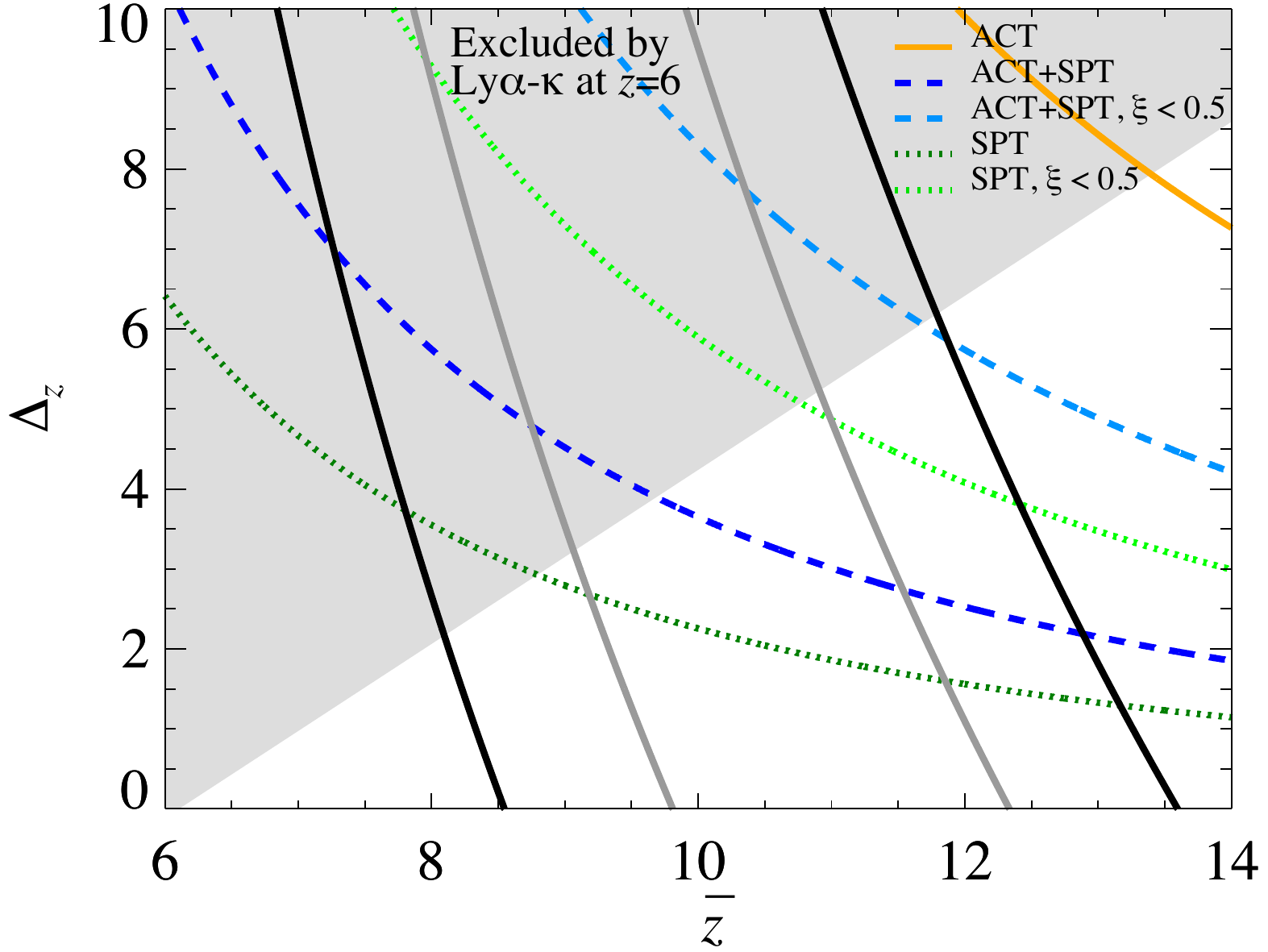}\\ [0.0cm]
 \end{array}$
 \caption{Constraints in the $\bar{z}-\Delta_z$ plane from the patchy
reionization signal at $\ell=3000$. The solid curves are for the ACT
data in combination with WMAP7 data, the light (dark) dotted green curves are for SPT+WMAP7, under the assumption of priors on the tSZ-CIB correlation of $\xi < 0.5$ ($\xi < 0.2$). The light (dark) dashed blue lines show the combination for WMAP7 data, ACT and SPT,  also under the priors of $\xi < 0.5$ ($\xi < 0.2$); the allowed
regions are below the colored lines. The
gray region is excluded by Gunn-Peterson \citep{gunn/peterson:1965}
observations and the WMAP7 limits on $\bar{z}$ are shown by the dark
gray $(1\sigma)$ and black $(2\sigma)$ lines. \label{fig:deltaz}}
\end{center}
 \end{figure}
The second is a high-redshift component that is sourced by
fluctuations in the ionized fraction as well as the electron density
during reionization,
which we refer to as patchy kSZ
\citep{knox/etal:1998,gruzinov/hu:1998}. In
general these two components have different shapes as a function of
$\ell$ \citep[e.g.,][]{mcquinn/etal:2007,zahn/etal:2012,mesinger/etal:2012}.

The upper limit on the kSZ for ACT data in combination with WMAP7 data is
\begin{equation}
a_{\mathrm{kSZ}} < 8.6~(\mathrm{WMAP7} + \mathrm{ACT}, \xi < 0.2,95\%~\mathrm{CL}).
\end{equation}

When relaxing the prior on the tSZ-CIB correlation to $0< \xi < 0.5$ and including SPT data, we obtain the bound
\begin{eqnarray}
a_{\mathrm{kSZ}} <6.9 &&(\mathrm{WMAP7} + \mathrm{ACT}+\mathrm{SPT}, \nonumber \\
&& \xi < 0.5,95\%~\mathrm{CL}). \nonumber \\
\end{eqnarray} We
approximate the homogeneous kSZ contribution at $\ell = 3000$,
$a_\mathrm{kSZ, hom} = 1.5 \pm 0.5$ using the scaling relation from
\citet{shaw/etal:2012}; however, we do not include the uncertainty in our analysis.
We remove this homogeneous kSZ contribution
leaving only the contribution from patchy kSZ, $a_\mathrm{kSZ, patchy}
< 7.1$. The amplitude of the
patchy kSZ places
constraints on the mean redshift ($\bar{z}$) and duration of ($\Delta_z$) reionization via the relation given by \citet{battaglia/etal:2012} as
 \begin{equation}
 a_\mathrm{kSZ,patchy} = 2.03
\left[\left(\frac{1+\bar{z}}{11}
-0.12\right)\right]\left(\frac{\Delta_z}{1.05}\right)_.^{0.51}\,
 \end{equation}

The curves in Figure~\ref{fig:deltaz} illustrate the constraints on
$\bar{z}$ and $\Delta_z$ derived from combinations of ACT, SPT, and
WMAP data assuming the \citet{battaglia/etal:2012}
scaling relation. The grey and black lines show the $68\%$ and $95\%$
confidence regions from large-scale WMAP 7-year CMB polarization
measurements \citep{larson/etal:2011}. The ACT data mildly prefer a
higher kSZ amplitude than SPT, which leads to an increase in the kSZ
upper limit for WMAP7+ACT+SPT (dashed line in
Figure~\ref{fig:deltaz}) compared to the WMAP7+SPT case (dotted
line). The choice of $\xi$ prior drives the constraint when SPT data
are included; constraints on the tSZ-CIB correlation and thus kSZ
power should be significantly improved with cross-correlations of the
ACT maps with CIB-dominated Herschel/SPIRE maps
\citep{addison/dunkley/spergel:2012}.

Recently \citet{zahn/etal:2012} obtained $\Delta_z < 7.9$ at $95\%$
confidence using SPT data, although their analysis used a somewhat
different treatment of the tSZ-CIB correlation, definition of
$\Delta_z$ and reionization model to that considered here.

\section{Conclusions}
\label{sec:conclusions}
We have presented parameter constraints from the ACT 3-year data in the ACT-S and ACT-E regions, over three seasons, at 148 GHz and 218 GHz. Using the multi-frequency likelihood presented in \citet{dunkley/etal:preplike}, the data provide constraints on both the primary cosmological model and a suite of secondary parameters. 
We confirm that the ACT data alone are fully consistent with the standard $\Lambda$CDM cosmological model. Our cosmological model is also consistent with data from the South Pole Telescope \citep{reichardt/etal:2012, keisler/etal:2011}. 
While this work was being completed, updated cosmological results were presented by the WMAP team \citep{bennett/etal:2012, hinshaw/etal:2012}. Our results, based on the ACT data in combination with WMAP7 data, are consistent with the WMAP9 results, which have been tightened through the addition of two years of additional data and improved analysis techniques. We discuss general consistency between data sets in Appendix~\ref{dataconsistency}.

We measure a power spectrum from the thermal Sunyaev-Zel'dovich effect that is consistent with $\Lambda$CDM, and place an upper bound on the power from the kinematic SZ effect. This amplitude provides constraints on the time and duration of reionization; future cross-correlation studies will improve this bound.

When combining the ACT data with data from the seven-year release from the WMAP satellite, we constrain a host of extended cosmological models. We see no evidence for mean spatial curvature, running of the scalar perturbation
spectral index, or extra relativistic degrees of freedom beyond three light
neutrinos. The amount of primordial $^4$He is also constrained by the ACT data, and is consistent with the standard prediction from Big Bang Nucleosynthesis.  

The ACT data tighten the bound on the sum of light neutrino masses to 0.39 eV,
using a measurement of $\sigma_8$ from the skewness of the ACT temperature map. 
We bound the allowed contribution of tensor modes to the temperature spectrum, and place limits on the amplitude of power coming from cosmic strings.
Small-scale temperature data from ACT constrain the density parameter associated
with early dark energy models to be less than 3\%.
 
\acknowledgements
This work was supported by the U.S. National Science
Foundation through awards AST-0408698 and AST-0965625 for the ACT
project, and PHY-0855887, PHY-1214379, AST-0707731 and PIRE-0507768
(award number OISE-0530095). Funding was also provided by Princeton University, the University of Pennsylvania, and a Canada Foundation for Innovation (CFI) award to UBC. ACT operates in the Parque Astron\'omico Atacama in northern Chile under the auspices of the Comisi\'on Nacional de Investigaci\'on Cient\'ifica y Tecnol\'ogica de Chile (CONICYT). Computations were performed on the GPC supercomputer at the SciNet HPC Consortium. SciNet is funded by the CFI under the auspices of Compute Canada, the Government of Ontario, the Ontario Research Fund -- Research Excellence; and the University of Toronto. Funding from ERC grant 259505 supports JD, EC, and TL. We especially wish to thank Norm Jarosik and 
Astro-Norte. RD received additional support from
a CONICYT scholarship, from MECESUP, from Fundaci—n
Andes and from FONDECYT-11100147. NS is supported
by the U.S. Department of Energy contract to SLAC no. DEAC3-
76SF00515 and by the NSF under Award No. 1102762.
ERS acknowledges support by NSF Physics Frontier Center
grant PHY-0114422 to the Kavli Institute of Cosmological
Physics. AK has been supported by NSF-AST-0807790 for
work on ACT. We are grateful for the assistance
we received at various times from the ALMA, APEX, ASTE,
CBI/QUIET, and NANTEN2 groups. RH thanks David Marsh, Shaun Bristow and Daan Meerburg for useful discussions.

\bibliography{bibtexparams/act_trim,bibtexparams/actnew,bibtexparams/wmap_jo,bibtexparams/wmap_supp,bibtexparams/act_renee,bibtexparams/pk,bibtexparams/act,bibtexparams/thesis}
\appendix
\section{ACT data consistency}
\label{dataconsistency}
\begin{figure*}[htbp!]
\begin{center}
$\begin{array}{@{\hspace{-0.2in}}l@{\hspace{-0.2in}}l}
\includegraphics[scale=0.4,trim = 10mm 15mm 0mm 15mm, clip]{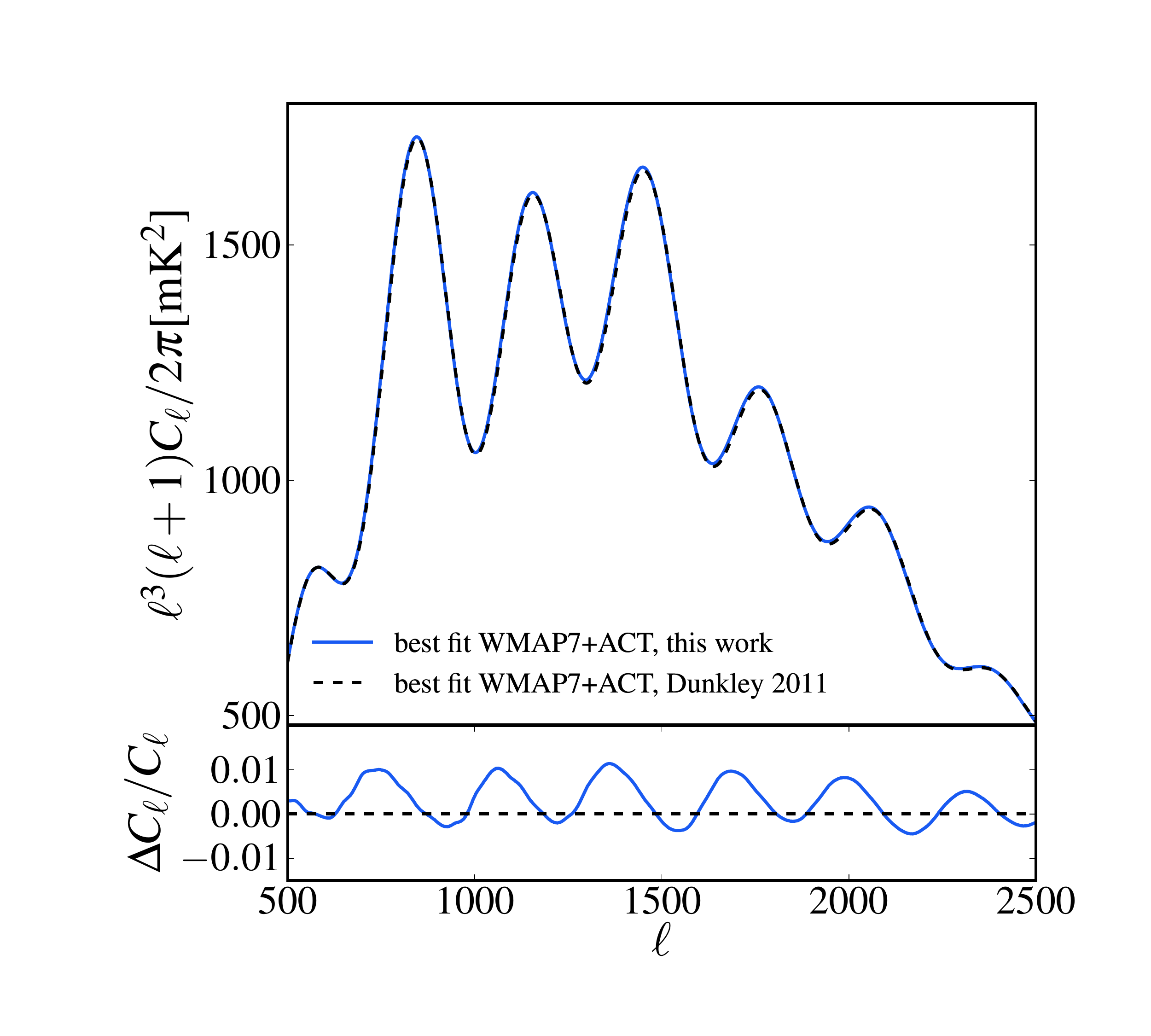}&
\includegraphics[scale=0.42,trim = 45mm 15mm 10mm 15mm, clip]{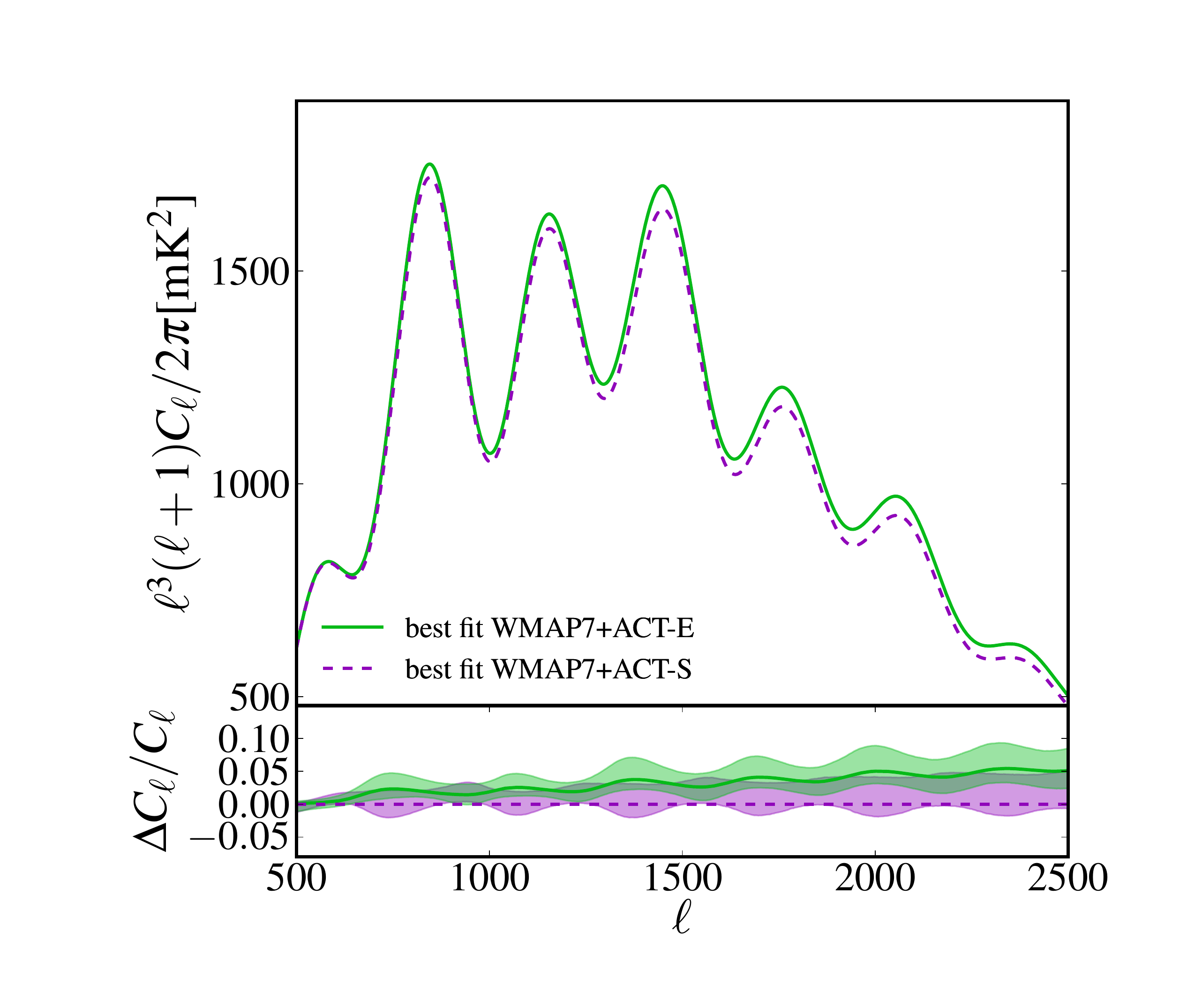}\\ [0.0cm]
 \end{array}$
 \caption{The best-fit spectra from the ACT data. \textit{Left panel:} the solid curve indicates the best-fit $C_\ell$ from the 3-year ACT data set (top) and difference from the best-fit model (bottom), while the dashed line shows the curves from the best-fit parameters for the ACT 1-year data, as presented in \citet{dunkley/etal:2011}. \textit{Right panel}: the best-fit power spectra for the WMAP7+ACT-E and WMAP7+ACT-S data sets. The bands around each best-fit spectrum are all models which lie within $1\sigma$ of the best-fit. The spectra are consistent at the $4\%$ level. At  $\ell=2500$, 4\% corresponds to about $4~\mu\mathrm{K}^2,$ which is about 25\% larger than our error bar at $\ell=2500.$ \label{fig:bf_compare}} 
\end{center}
 \end{figure*}

In order to check consistency of the current best-fit cosmological model with previous ACT constraints presented in \citet{dunkley/etal:2011}, it is useful to consider the $C_\ell$ plot derived from the best-fit model. This is shown in the left-hand panel of Figure~\ref{fig:bf_compare}. The reduced error bars at larger multipoles yield tighter constraints on the positions of the fifth and sixth acoustic peaks, providing tighter constraints on the baryon density, $\Omega_bh^2,$ relative to the WMAP7-only constraints. This comparison is shown directly in Figure~\ref{fig:lcdm_params}. 

\begin{figure}[htbp!]
\begin{center}
$\begin{array}{@{\hspace{-0.0in}}l}
\includegraphics[scale=0.75]{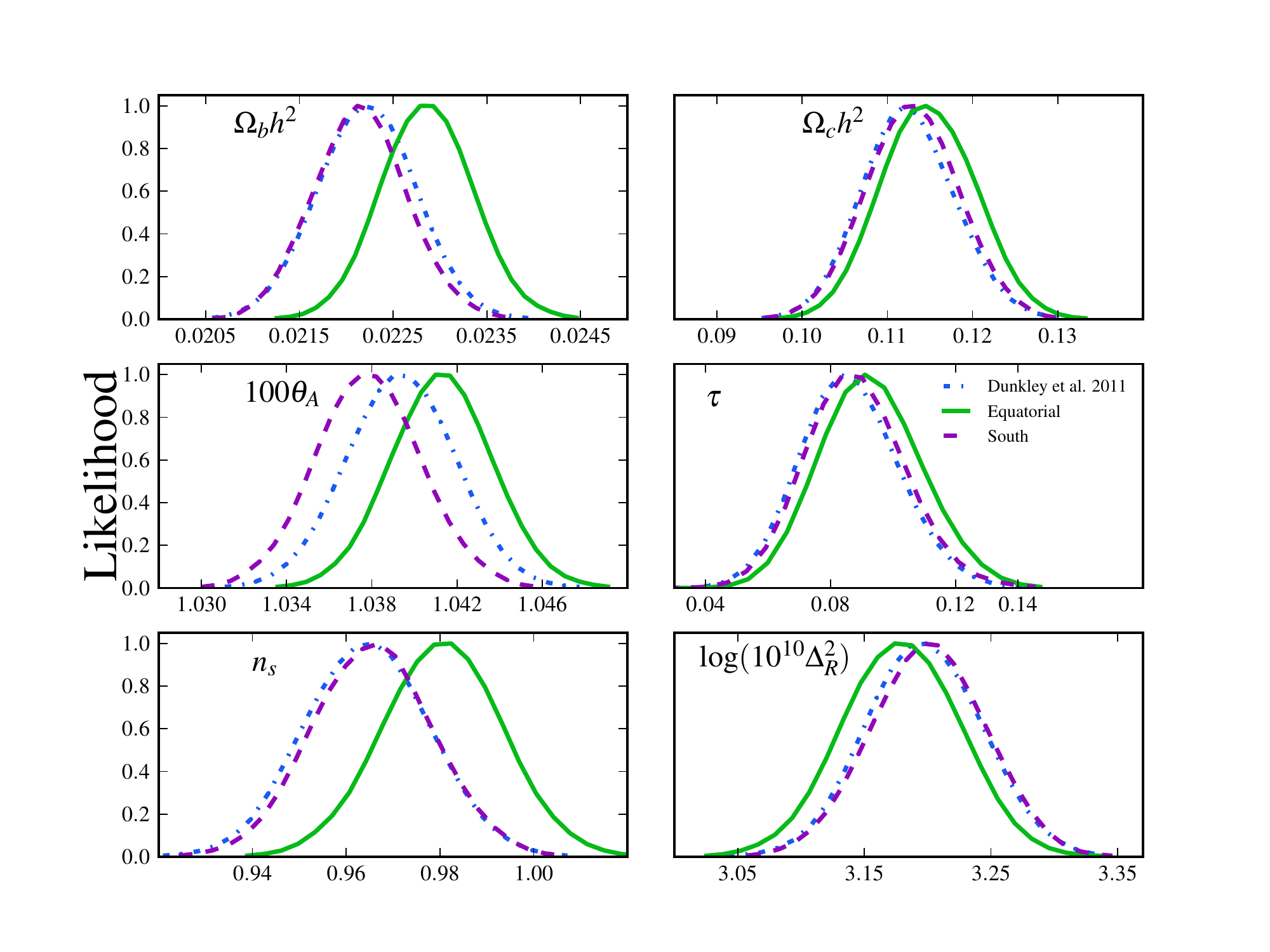}\\ [0.0cm]
 \end{array}$
 \caption{Marginalized 1-dimensional likelihoods for the $\Lambda$CDM cosmological parameters, computed for the ACT-E data in combination with WMAP7, the ACT-S+WMAP7 combination, and the constraints computed using only the ACT 1-year data, as presented in \citet{dunkley/etal:2011}. The dot-dashed curves show the parameters from the ACT 1-year analysis \citep{dunkley/etal:preplike}, while the solid and dashed lines are the parameters from the WMAP7+ACT-E and WMAP7+ACT-S constraints respectively. While there are small shifts between the two parameter sets, the best-fit spectra, shown by the curves in Figure~\ref{fig:bf_compare}, agree to the $4\%$ level, and the parameter sets are consistent when considering the full covariance between parameters. \label{fig:lcdm_compare}} 
\end{center}
 \end{figure}

The marginalized one-dimensional likelihoods for the $\Lambda$CDM parameters are shown in Figure~\ref{fig:lcdm_compare}. The $\simeq 0.7\sigma$ shift in $\theta_A$ from the \citet{dunkley/etal:2011} release relative to the 3-year ACT data constraints presented here can be explained by the similar shift in parameters of the first 1-year data maps used in \citet{dunkley/etal:2011} and \citet{das/etal:2011} relative to the maps released with \citet{dunner/etal:2012}. This shift was investigated thoroughly, and found to be caused by improved noise treatment and map-making pipeline. This shift in $\theta_A$ caused a negligible ($\simeq 0.03\sigma$) shift in the underlying cosmological parameter $\Omega_\Lambda.$
  
The best-fit curves for the ACT-E and ACT-S spectra in combination with WMAP7 data are shown in the right-hand panel of Figure~\ref{fig:bf_compare}, while the marginalized likelihoods are also shown in Figure~\ref{fig:lcdm_compare}. The ACT-E data on their own (in combination with WMAP7 data) prefer a higher value for $\Omega_bh^2$, $\theta_A$ and the scalar spectral index. Given the $\simeq 4\%$ differences in the best-fit $C_\ell$ curves, one might ask how well the combined fit compares to the fits obtained when considering only the ACT-E and ACT-S subsets individually in combination with WMAP7 data. Removing all information from WMAP7 other than a prior on $\tau$ yields differences of $< 8\%$ between the best-fit spectra obtained using only the ACT-E and ACT-S data.

The $-2\ln\mathscr{L}$ value of the best fit to ACT-S+WMAP7 data is  444.6 relative to 446.6 when combining ACT-S and ACT-E, for 480 data points and 10 fitted parameters, leading to 470 degrees of freedom. When considering WMAP7+ACT-E, the $-2\ln\mathscr{L}$ value is 227.7 compared to 229.0 when using the ACT 3-year data set in combination with WMAP7. The relative difference in $-2\ln\mathscr{L}$ between the south-only and equatorial-only parameters and the respective $-2\ln\mathscr{L}$ value for the combined fit is approximately $2;$ hence one might expect a roughly $1\sigma$ shift in cosmological parameters when considering the two data sets independently.

\begin{figure}[htbp!]
\begin{center}
$\begin{array}{@{\hspace{-0.0in}}l}
\includegraphics[scale=0.75]{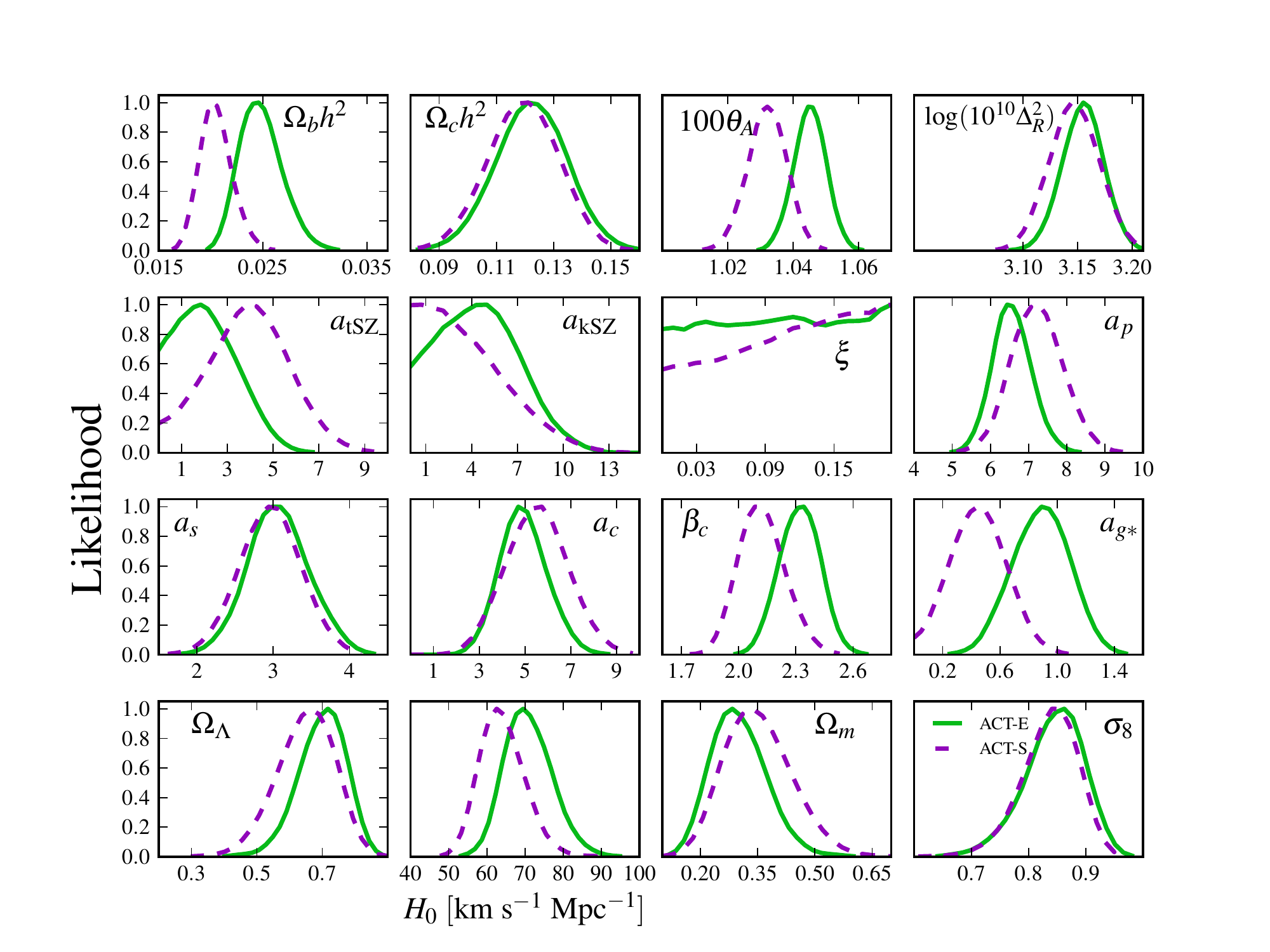}\\ [0.0cm]
 \end{array}$
 \caption{Marginalized 1-dimensional likelihoods for the $\Lambda$CDM cosmological parameters, computed for the ACT-E data (without WMAP7), and the ACT-S data. The spectral index $n_s$ and optical depth $\tau$ are held fixed, and all calibration parameters are set to their best-fit values from the ACT-E+WMAP7 and ACT-S+WMAP7 runs respectively. We find that the data sets are statistically consistent when considering the full covariance between the parameters. \label{fig:actse_compare}} 
\end{center}
 \end{figure}

One can go further and test for consistency between individual seasons
without including data from WMAP, which we show in
Figure~\ref{fig:actse_compare}. The roughly $1\sigma$ shifts in the
value of the baryon density and ratio of the acoustic horizon to the angular diameter distance at decoupling can be
understood from the fact that the data are taken from different
patches on the sky with different noise properties.  The two parameter
sets are statistically independent, and so ACT's internal consistency
can be quantitatively checked by approximating the likelihood as
Gaussian in the cosmological parameters.  
The covariance of the south
and equatorial parameter differences is then simply the sum of the
south and equatorial covariances, as reported by the MCMC chains.  We
excise $\xi$ from this test as it is clearly prior-driven.  For the
remaining 17 parameters, we find $\chi^2=8.78$ with a probability to
exceed this value (PTE) of 0.88.  The kinetic and thermal SZ amplitude distributions
are meaningfully affected by the constraint that they be non-negative;
excising these two parameters in addition has only a minimal impact,
with $\chi^2=8.14$ for 15 degrees of freedom (PTE=0.84).  We also find
that no particular parameter combinations stand out as being
inconsistent between the south and equatorial data.  

We divide each
parameter difference by its standard deviation and appropriately rescale the
combined covariance matrix (which now is unity along the diagonal);
this operation leaves $\chi^2$ unchanged.  We rotate into the
eigenvector space of the re-scaled covariance and find that the
eigenmode distribution is consistent with a Gaussian distribution; in particular
for both the 15- and 17-parameter case, the most discrepant mode falls
within the 68\% confidence interval expected from purely Gaussian
statistics.  We note that the posterior parameter distributions need
not be Gaussian, but that Gaussianity does appear to be a good
approximation, and that the cosmological parameters derived separately
from the ACT south and equatorial regions are fully consistent with
each other.
 
For the extended models, we see a shift of $\simeq 1\sigma$ in the preferred value for $N_\mathrm{eff}$ for ACT-E relative to ACT-S, but the lensing parameter $A_L$ shifts by no more than $0.2\sigma$ for the different patches.
 
\section{Galactic foreground treatment}
As discussed in \citet{dunkley/etal:preplike}, we marginalize over residual Galactic dust allowing for two different levels in ACT-S and ACT-E and imposing Gaussian priors on the dust amplitudes $a_{ge} = 0.8\pm0.2;  a_{gs} = 0.4\pm 0.2$, after masking bright dust sources in map space as reported in \citet{das/etal:2012prep}. We tested that using a bigger or smaller mask has a negligible effect on all the parameters. Moreover we masked out in the equatorial strip a``seagull"-like structure \citep[see Figure~7 in][]{das/etal:2012prep}, and again found that parameters do not move if we leave this region in the equatorial map.
Finally we cross correlate ACT and IRIS maps \citep{iris:2005}, as described in \citet{das/etal:2012prep}, and subtract a dust template (rescaled by the cross-correlation coefficient derived through this cross-correlation) to our maps, before computing the power spectrum. We compared parameters obtained with or without the dust template subtraction and found the impact on parameters to be negligible.
  \begin{figure}[htbp!]
\begin{center}
$\begin{array}{@{\hspace{-0.0in}}l}
\includegraphics[scale=0.55]{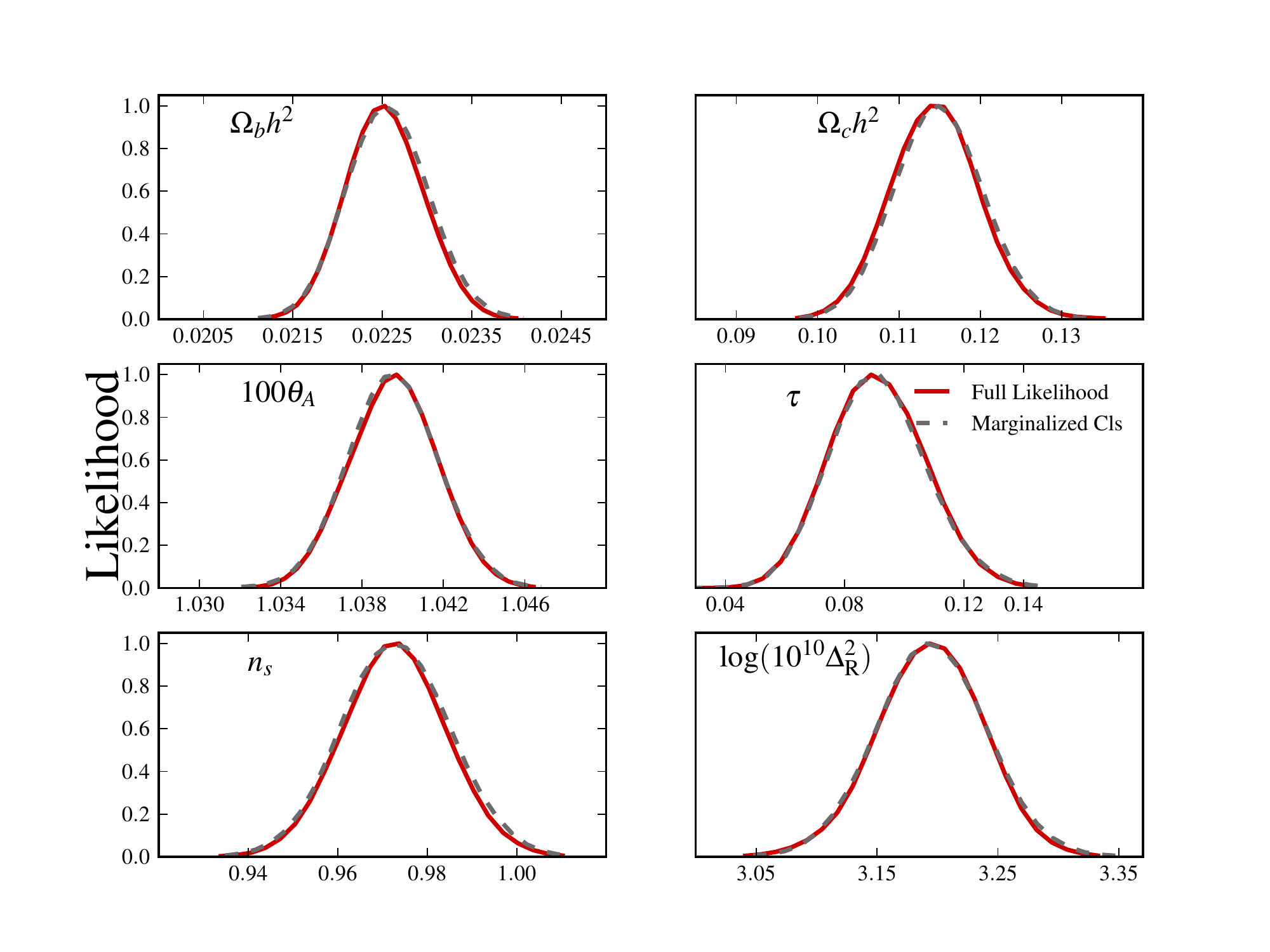} \\[0.0cm]
 \end{array}$
 \caption{$\Lambda$CDM parameter consistency between the full likelihood and the marginalized $C_\ell$ bandpower likelihood. The distributions are consistent between the two likelihood treatments. \label{fig:marg_lcdm_params}} 
\end{center}
 \end{figure}
\section{Calibration and beam uncertainty}
\label{beam_error}
The calibration factors of the 148 GHz and 218 GHz ACT-S and ACT-E spectra are allowed to vary in the cosmological analysis, yielding a $1~(2)\%$ calibration error for the 148 (218) GHz spectra respectively. This error is similar to or smaller than the calibration error obtained from calibrating the 148 GHz maps by cross-correlating with WMAP7 maps \citep{hajian/etal:2012}, which gives an error of $2\%.$ We apply a prior on the calibration parameters for the 148 GHz and 218 GHz spectra of $1.00 \pm 0.02.$ We test whether removing the prior has an effect on the parameters. This shift is negligible for the cosmological models considered here.

Similarly we test for a dependence of the cosmological parameters on the beam error as measured by planet observations, by doubling the assumed beam error and comparing this to the case where the beam error is ignored. We find that doubling the beam error shifts the peaks in the foreground parameters by $< 0.1 \sigma$ in all cases, relative to the constraints when no beam error is applied.

We present updated ACT beam parameters in \citet{hasselfield/etal:2013b}. With these changes, the difference in the window function rises up to a $2\%$ difference on $\cal{W}_\ell$ from $\ell=700$ to $\ell=1200$ but then remains constant. This changes the overall calibration in the spectrum, as the relative calibrations are taken at $\ell = 700.$ The largest shift in parameters when using these updated beams (as opposed to the ones employed in this analysis) occurs in the model where $N_\mathrm{eff}$ is a free parameter, and leads to a shift of $<0.2\sigma$ in $N_\mathrm{eff}.$ The overall $\chi^2$ of all models decreases by 8 when the updated beams are used.

 \section{Marginalized CMB likelihood}
 \label{likeconsistency}
To compare results to the full multi-frequency likelihood, we cross-check constraints on various models using the simpler `CMB-only' likelihood from \citet{dunkley/etal:preplike}, which first marginalizes over the secondary parameters to generate CMB band powers and a covariance matrix. This marginalized likelihood is then used to constrain cosmological models directly without requiring any further sampling of nuisance parameters.
   \begin{figure}[htbp!]
\begin{center}
$\begin{array}{@{\hspace{-0.0in}}c}
\includegraphics[scale=0.45]{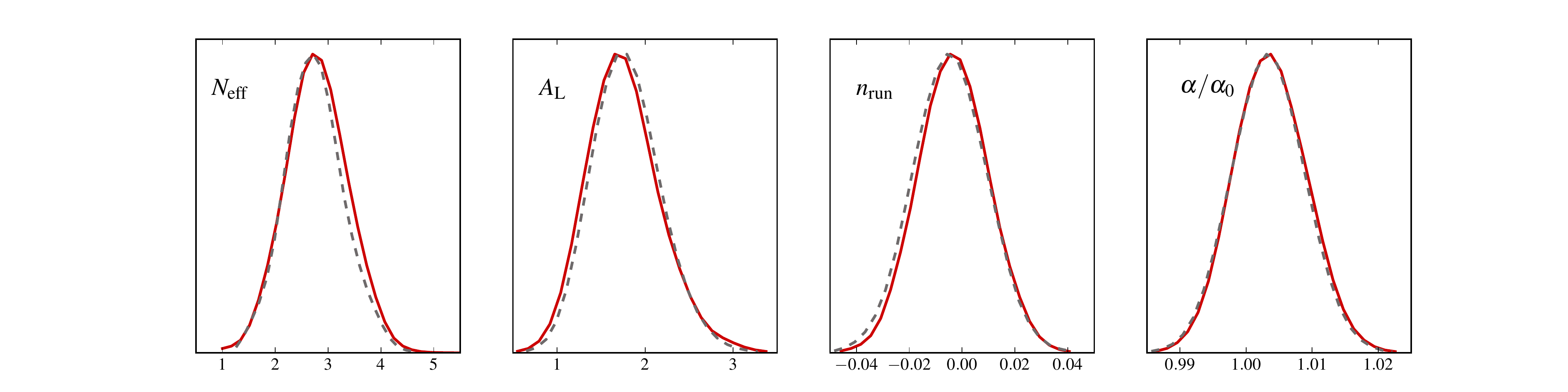} \\ [0.0cm]
 \end{array}$
 \caption{One-dimensional likelihoods for the extension parameters in various models. The two likelihoods  yield consistent results. The solid and dashed lines are the same as in Figure~\ref{fig:marg_lcdm_params}. \label{fig:2d_margcls}} 
\end{center}
 \end{figure}
 \subsection{$\Lambda$CDM model comparison}
In Figure 12 of \citet{dunkley/etal:preplike} we show that the secondary parameters estimated using this marginalized foreground likelihood and assuming the $\Lambda$CDM model are consistent with those obtained from the full multi-frequency case. There is a slight ($1\sigma$) shift in estimated kSZ amplitude, which shifts due to its correlation with the primary CMB; however, the constraints are still consistent with an upper limit on the kSZ amplitude. Figure~\ref{fig:marg_lcdm_params} shows the marginalized one-dimensional likelihoods for parameters in the $\Lambda$CDM cosmological model. The shift between the constraints from the full likelihood and those from the `CMB-only' likelihood are negligible.
 
\subsection{Extended model comparison}

In addition to the test for the $\Lambda$CDM model, we also test for the consistency between the full likelihood and the marginalized CMB-only likelihood for the extended models including the number of effective relativistic degrees of freedom, $N_\mathrm{eff},$ allowing for an estimated lensing amplitude, $A_L$, including the running of the spectral index, $d n_s/d\ln k,$ and variation in the fine-structure constant $\alpha/\alpha_0.$ The one-dimensional likelihoods are shown in Figure~\ref{fig:2d_margcls}, where the maximum likelihood values are consistent between methods; the shifts in the mean values are less then $0.05\sigma$ in all cases.
Both likelihood codes are available on LAMBDA; these tests indicate that the CMB-only likelihood returns robust results compared to the full likelihood, and may be used in cases where only the primary cosmological parameters are of interest.

\end{document}